\documentclass[onecolumn]{els-mrw_modified} 
\usepackage{amsmath,amssymb,amsfonts,amsthm,makeidx,graphicx}
\usepackage{txfonts,comment}
\usepackage{helvet}
\usepackage{upgreek}     % Added package for the beta symbol

\usepackage[T1]{fontenc} % Added package for accent symbols in the references
\usepackage{braket}      % Added package for dirac bra-ket notation
\usepackage{hyperref}    % Added package for hyper-referencing citations
\usepackage{bbm,bm}      % Added package (bbm) for identity symbol   
\usepackage{tikz}        % Added package to create tikz figure
\usepackage[bb=boondox,scr=boondox,cal=cm]{mathalfa} %Added

\newcommand{\ie}{{i.e.},\ }
\newcommand{\eg}{{e.g.},\ }

\newcommand{\Tr}{\operatorname{Tr}}

\def\m{\mathfrak{m}}      %Magnetization density=Magnetization/Volume
\def\n{\mathfrak{n}}      %Number density=Number/Volume
\def\E{E\!^{}_{\upbeta=0}}  %Energy at infinite temperature
\def\W{\Omega^{}_{\upbeta=0}}  %Density of states at infinite temperature
\def\O{\mathscr{O}}       % big O notation
\def\lO{O}                % local observable
\def\tO{\mathcal{O}}      % translationally invariant intensive observable
\def\lZ{Z}                % local Z
\def\tZ{\mathcal{Z}}      % translationally invariant Z
\def\lJ{J}                % local J
\def\tJ{\mathcal{J}}      % translationally invariant J
\def\bb{\upbeta}
\def\Nt{\mathsf{N}}      % symbol N for nearest- and next-nearest-neighbor operators, different from particle number N
\def\Nm{\mathcal{N}}     % Number of observables, different from number of particles N
\def\Z{\mathscr{Z}}      % Partition function
\def\s{\mathfrak{s}}                % Entropy per site
\begin{document}

\chapter{Quantum Chaos and Eigenstate Thermalization}\label{chap1}

\author[1]{Rohit Patil}%
\author[1]{Marcos Rigol}%

\address[1]{\orgdiv{Department of Physics}, \orgname{The Pennsylvania State University},  \orgaddress{University Park, Pennsylvania 16802, USA}}

%\articletag{Chapter Article tagline: update of previous edition, reprint..}
\articletag{Chapter Article tagline: \today}

\maketitle

\begin{glossary}[Keywords]
Eigenstate Thermalization Hypothesis, Quantum Chaos, Entanglement Entropy
\end{glossary}
 
\begin{abstract}[Abstract]
We provide a pedagogical introduction to eigenstate thermalization. This phenomenon, which occurs in generic quantum many-body systems, allows one to understand why thermalization takes place in isolated systems under unitary dynamics. We motivate eigenstate thermalization using random matrix theory and discuss recent complementary results for the volume-law entanglement entropy of Haar-random states. We discuss numerical results that highlight the corresponding behaviors in quantum many-body systems.
\end{abstract}

\section{Introduction}\label{sec:Intro}

Statistical mechanics is the field in physics whose goal is to understand the macroscopic behavior of systems with a large number of degrees of freedom in terms of the microscopic laws governing their dynamics. Rather than tracking the detailed evolution of a single system with many degrees of freedom like the one we may have in an experiment, which is often very difficult if not impossible and less informative, in statistical mechanics one considers an ensemble of systems with the same macroscopic constraints (or macrostates) but with different microscopic states (or microstates). When the underlying microscopic laws are classical in nature (such as Newton's laws of motion), the ensemble approach to equilibrium is justified based on arguments such as {\bf ergodicity}, i.e., given a sufficiently long evolution time, the system explores the phase-space region allowed by its conserved quantities according to the appropriate invariant measure, and {\bf typicality}, whereby most microstates compatible with the macroscopic constraints exhibit the corresponding equilibrium properties~\cite{dalessio_quantum_2016}. In generic interacting systems, such arguments are associated with nonlinear dynamics and classical chaos. On the other hand, if the microscopic laws follow quantum mechanics, then it is no longer possible to motivate chaos in terms of classical phase-space trajectories. Quantum chaos is instead motivated in terms of {\bf random matrix theory} (RMT).

Thermalization in generic isolated quantum many-body systems is currently understood to be a general consequence of the phenomenon known as {\bf eigenstate thermalization}~\cite{deutsch_1991, srednicki_1994, rigol_2008}. The corresponding mathematical ansatz, known as the {\bf eigenstate thermalization hypothesis} (ETH), has its roots in quantum chaos and RMT and explains why statistical mechanics emerges in generic isolated quantum systems under unitary evolution, thus providing a bridge between quantum and statistical mechanics. In this chapter, we give a brief overview of quantum chaos and RMT, followed by a pedagogical introduction to the ETH. We discuss numerical results for the nonintegrable spin-$1$ $XXZ$ model to illustrate the spectral statistics associated with quantum chaos and examine the various aspects of the ETH, contrasting them with results for the same model at a special {\bf integrable} point. That contrast, which we carry throughout this chapter, allows us to highlight both generic and nongeneric behaviors in quantum many-body systems. Since our focus is on quantum systems, we set $\hbar=1$.

\subsection{Model Hamiltonian}\label{sec:Model_and_observables}
To exemplify the different behaviors related to quantum chaos indicators, entanglement, and eigenstate thermalization discussed in this chapter, we consider the spin-$1$ $XXZ$ Hamiltonian in chains with $L$ sites and periodic boundary conditions
\vspace{-0.17cm}
\begin{equation} \label{eq:Spin1_Hamiltonian}
    \hat H= -\sum_{j=1}^L \left(\hat{S}^{x}_{\!j} \hat{S}^x_{\!j+1} + \hat{S}^y_{\!j} \hat{S}^y_{\!j+1} +\Delta\hat{S}^{z}_{\!j} \hat{S}^z_{\!j+1}\right) +\lambda \sum_{j=1}^L \left((\hat{\vec{S}}^{}_{\!j}\cdot \hat{\vec{S}}^{}_{\!j+1})^2 
     -  \mu\left[2(\hat{S}^z_{\!j})^2-(\hat{S}^z_{\!j} \hat{S}^z_{\!j+1})^2\right]-\nu\left[(\hat{S}^x_{\!j} \hat{S}^x_{\!j+1} + \hat{S}^y_{\!j} \hat{S}^y_{\!j+1})\hat{S}^z_{\!j} \hat{S}^z_{\!j+1} + \text{H.c.}\right]\right) \,,
\end{equation}
\vspace{-0.25cm}

\noindent where $\mu=\Delta-1, \nu=2-\sqrt{2(1+\Delta)}$, and $\hat{\vec{S}}_{\!j}=(\hat{S}^x_{\!j},\hat{S}^y_{\!j},\hat{S}^z_{\!j})$ is the spin-$1$ operator at site $j$. For $\lambda=1$, this model is integrable and is known as the Zamolodchikov-Fateev model~\cite{zamolodchikov_1980, bytsko_2003}. For $\lambda=0$, it is quantum chaotic for generic values of the anisotropy parameter $\Delta$. All the numerical results shown in what follows, both at and away from integrability, are obtained for $\Delta=0.55$ for which quantum-chaos indicators are closest to the RMT predictions for $\lambda=0$ in our finite systems.

The model~\eqref{eq:Spin1_Hamiltonian} exhibits U(1) symmetry, so it conserves the total magnetization $\hat M=\sum_{j=1}^L\hat S^z_{\!j}$, whose eigenvalue $M$ plays a role analogous to the total particle number $N$ in itinerant models with at most two particles per lattice site (the two are related via $N=M+L$). In addition, the model~\eqref{eq:Spin1_Hamiltonian} exhibits several discrete symmetries including lattice translation, space reflection (parity), and spin inversion symmetries. The lattice translation symmetry results in the conservation of total quasimomentum $k$ with $k\in\Big\{\tfrac{2\pi \eta}{L}\,|\,\eta=\eta_{\min},\eta_{\min}+1,\ldots,\eta_{\max}\Big\}$ with $\eta_{\max}=-\eta_{\min}+1=L/2$ ($\eta_{\max}=-\eta_{\min}=\lfloor L/2 \rfloor$) for $L$ even (odd). The quasimomentum sectors with $k=0$ and, when present, $k=\pi$ further split into two subsectors (even and odd) under the space reflection symmetry $P$. The zero total magnetization sector ($M=0$) also exhibits a spin-$z$ inversion symmetry $\mathbb{Z}^{}_2$. Considering a fixed total magnetization sector, we report results for the weighted averages over the relevant symmetry sectors. Finally, it will be important for our comparisons with the RMT predictions [\ie the GOE ($\beta=1$) predictions introduced in the next section] that the model~\eqref{eq:Spin1_Hamiltonian} exhibits time-reversal symmetry and, specifically, that the quasimomentum sectors with $k\neq0,\pi$ (on which we focus in our calculations) are invariant under the combination of time-reversal and space-reflection symmetries.

\begin{figure}[!t]
    \centering
    \includegraphics[width=0.8\linewidth]{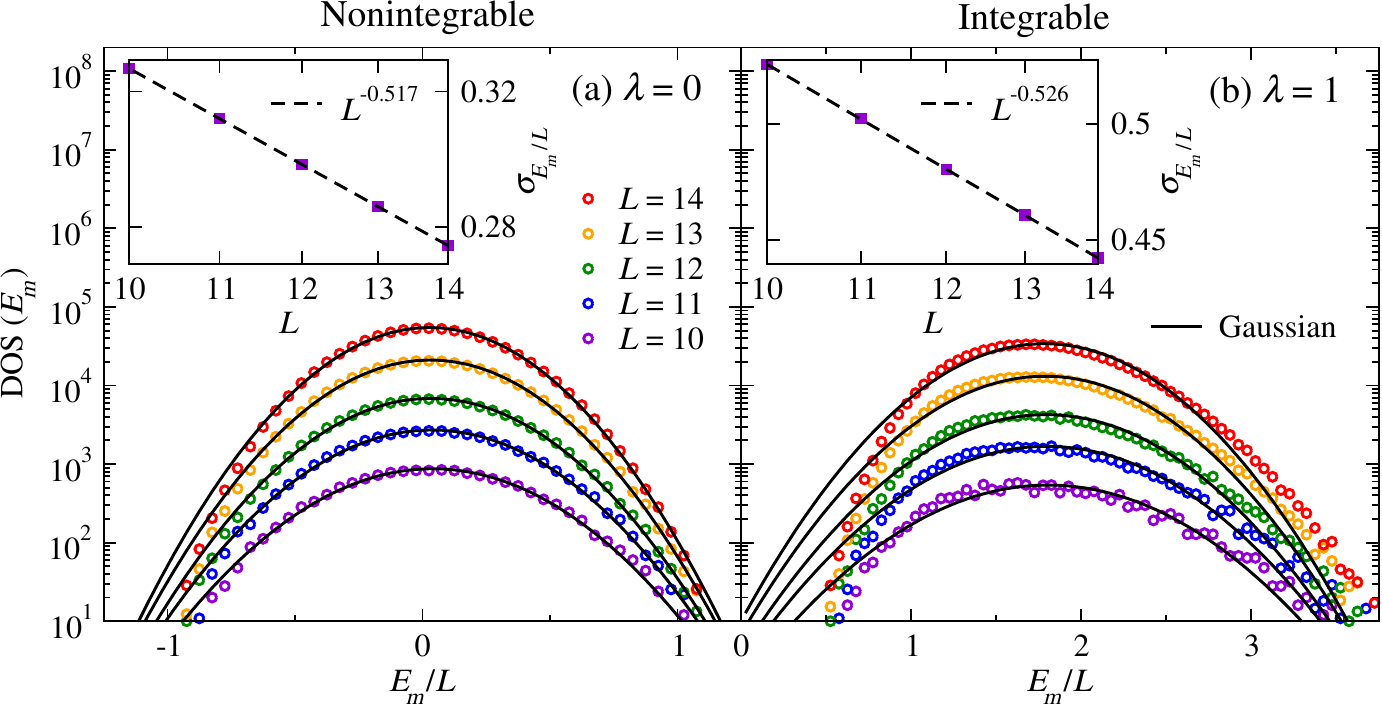}
    \caption{{\bf Density of states}: The $\text{DOS}(E^{}_m)$ is plotted as a function of the energy density $E^{}_m/L$ for the spin-$1$ $XXZ$ model at (a) the nonintegrable point ($\lambda=0$) and (b) the integrable point ($\lambda=1$). The solid lines show Gaussians with the same mean and standard deviation as the numerical data for chains with $L=10$--$14$. The insets show the scaling of the standard deviation $\sigma^{}_{\!E^{}_m/L}\propto L^{-\gamma}$ with fitted exponent $\gamma\approx0.5$, as predicted.}
    \label{fig:DOS}
\end{figure}

We highlight that, in contrast to the random matrices to be introduced and discussed in the next section, the {\bf density of states} (DOS) of many-body Hamiltonians with local interactions, such as ours~\eqref{eq:Spin1_Hamiltonian}, approaches a {\bf Gaussian} with increasing system size~\cite{brody_81} regardless of whether the Hamiltonian is integrable or not. In Fig.~\ref{fig:DOS} we show the DOS associated with the energy eigenvalues $E^{}_m$ of Hamiltonian $\hat H$~\eqref{eq:Spin1_Hamiltonian} across the symmetry sectors with $\{M=0;\mathbb{Z}^{}_2=\pm1;k\neq0,\pi\}$ as a function of energy density $E^{}_m/L$, at the nonintegrable $\lambda=0$ [Fig.~\ref{fig:DOS}(a)] and integrable $\lambda=1$ [Fig.~\ref{fig:DOS}(b)] points. The DOS is close to Gaussian about the center of the energy spectrum and deviates from Gaussian towards the edges. The deviation is stronger and exhibits clear skewness toward higher energy states at the integrable point, which suffers from stronger finite-size effects. Important for many of our discussions in what follows is the fact that the standard deviation of the energy density $E^{}_m/L$ vanishes as $\sigma^{}_{\!E^{}_m/L}\sim1/\sqrt{L}$ with increasing system size $L$, see the insets in Fig.~\ref{fig:DOS}. This means that the typical energy eigenstates in the thermodynamic limit ($L\rightarrow \infty$) are those that lie near the center of the spectrum, i.e., the energy eigenstates whose energy density is the same as the mean energy density $\E/L=\Tr(\hat H)/(LD)$ at {\bf infinite temperature} ($k^{}_BT=1/\upbeta\rightarrow \infty$), where $D=\Tr(\mathbbm{1})$ is the dimension of the Hilbert space and $k^{}_B$ is the Boltzmann constant. Many of our results in this chapter focus on such {\bf typical} energy eigenstates.

We present numerical results for three translationally invariant intensive observables, the nearest- and next-nearest-neighbor $\hat S^{z}$-$\hat S^{z}$ correlations $\hat \tZ^{}_{\Nt}$ and $\hat \tZ^{}_{\Nt\Nt}$, and the spin-current operator $\hat \tJ^{}_{\Nt}$ (which we write in terms of the raising and lowering operators $\hat{S}^\pm_{\!j}=\hat{S}^x_{\!j}\pm i\hat{S}^y_{\!j}$):
\vspace{-0.2cm}
\begin{equation}\label{eq:observables}
    \hat \tZ^{}_{\Nt} \equiv\frac{1}{L}\sum_{j=1}^L\hat\lZ^{j}_{\Nt}, \quad \hat\lZ^{j}_{\Nt}\equiv\hat{S}^z_{\!j} \hat{S}^z_{\!j+1}\,,\qquad
    \hat \tZ^{}_{\Nt\Nt} \equiv\frac{1}{L}\sum_{j=1}^L\hat\lZ^{j}_{\Nt\Nt}, \quad \hat\lZ^{j}_{\Nt\Nt}\equiv\hat{S}^z_{\!j} \hat{S}^z_{\!j+2}\,,\qquad
    \hat \tJ^{}_{\Nt} \equiv\frac{1}{L}\sum_{j=1}^L\hat\lJ^{j}_{\Nt}, \quad \hat\lJ^{j}_{\Nt}\equiv-\frac{i}{2}(\hat{S}^+_{\!j} \hat{S}^-_{\!j+1}-\hat{S}^-_{\!j} \hat{S}^+_{\!j+1})\,.
\end{equation}
\vspace{-0.3cm}

\noindent In Sec.~\ref{sec:outlook}, we discuss fundamental differences between the local observables $\hat \lO^j$ and their translationally invariant intensive counterparts $\hat \tO$.

\section{Quantum Chaos and Random Matrix Theory}\label{sec:Chaos}

In classical mechanics, chaos is characterized by a strong sensitivity to initial conditions, whereby trajectories with almost identical initial conditions separate exponentially from each other in phase space with time (as characterized by the Lyapunov exponent). Chaotic systems have fewer, generally $\O(1)$, mutually Poisson-commuting independent integrals of motion than the number of degrees of freedom. Classically integrable systems, on the other hand, exhibit dynamics with regular trajectories (on an $N$-dimensional torus in the action-angle formalism) as they possess as many independent integrals of motion in involution as the number of degrees of freedom $N$.

In contrast, quantum chaos is associated with RMT. The pioneering works of Wigner~\cite{wigner_55, wigner_56, wigner_58}, Dyson~\cite{dyson_62} and others in RMT aimed at understanding the spectral statistics of complex atomic nuclei. Wigner's key insight was that the spectral properties of complicated quantum systems such as complex atomic nuclei can be described in a statistical framework, i.e., within a sufficiently narrow energy window in which the density of states is approximately constant, the Hamiltonian in a generic basis resembles a random matrix, so that, by studying random matrices (subject to appropriate symmetries of the Hamiltonian), one expects to capture the universal spectral properties of the Hamiltonian.   

\subsection{Eigenvalues of random matrices}

In analogy with sampling a random variable $x$ from a {\bf probability density function} (pdf) $\rho(x)$, one can sample a random $D\times D$ matrix ${\bf M}$ from a pdf $\rho({\bf M})$, which is the joint pdf (jpdf) of its independent matrix elements $M^{}_{\!jk}$. When representing Hamiltonians using random matrices ${\bf H}$, we require that the pdf $\rho({\bf H})$ be invariant under changes of basis that preserve the {\bf symmetry class} of the Hamiltonian. For instance, a Hamiltonian $\hat H$ with an antiunitary symmetry that squares to $+1$ within the symmetry sector under consideration, such as time reversal or the combination of time reversal and parity, can be represented by a real-symmetric matrix. Therefore, we require that the pdf be invariant under orthogonal basis transformations implemented by ${\bf R}$, $\rho({\bf H})=\rho({\bf R}^T {\bf H} {\bf R})$. In the absence of an antiunitary symmetry, the Hamiltonian is generically represented by a complex-Hermitian matrix, and we require that the pdf be invariant under unitary basis transformations implemented by ${\bf U}$, $\rho({\bf H})=\rho({\bf U}^\dagger {\bf H} {\bf U})$. 

Among the random-matrix ensembles satisfying these requirements, we focus on the {\bf Gaussian orthogonal ensemble (GOE)} for real-symmetric matrices and on the {\bf Gaussian unitary ensemble (GUE)} for complex-Hermitian matrices. Their pdf can be written as~\cite{alhassid_2000}
\begin{equation}\label{eq:jpdf_matrix_elements_Gaussian}
    \rho({\bf H})\propto \exp\left[-\beta\frac{\Tr({\bf H}^2)}{2\sigma^2}\right]\,,
\end{equation}
where $\beta=1$ ($2$) labels the Dyson index corresponding to GOE (GUE) and $\sigma^2/\beta$ is the variance of the diagonal matrix elements. For the GOE, for which ${\bf H}$ is real symmetric, we have $\Tr({\bf H}^2)=\sum_{j=1}^D H_{\!jj}^2+2\sum_{k>j}H_{\!jk}^2$, so that Eq.~\eqref{eq:jpdf_matrix_elements_Gaussian} can be written as
\begin{equation}\label{eq:jpdf_matrix_elements_GOE}
    \rho({\bf H})=\prod_{j=1}^D \frac{1}{\sqrt{2\pi\sigma^2}}\exp\left(-\frac{H_{\!jj}^2}{2\sigma^2}\right)\,\prod^{}_{k>j}\frac{1}{\sqrt{\pi\sigma^2}}\exp\left(-\frac{H_{\!jk}^2}{\sigma^2}\right)\,,
\end{equation}
where the normalization has been included explicitly. Hence, $H^{}_{\!jk}$ with $k\geq j$ are independent Gaussian random variables with zero mean and with the variance of the diagonal elements $H^{}_{\!jj}$ being twice that of the off-diagonal elements $H^{}_{\!jk}$.

The jpdf of eigenvalues of $D\times D$ matrices belonging to a Gaussian ensemble is given by~\cite{mehta2004, livan_18}
\begin{equation}\label{eq:jpdf_Gaussian_eigenvalues}
    \rho^{}_{D;\,\beta}\left(x^{}_1,\ldots,x^{}_D\right)=\frac{1}{Z^{}_{D;\,\beta}}\prod^{D}_{m=1}\exp\left(-\frac{1}{2}x_m^2\right) \,\prod^{}_{n>m}\left|x^{}_m-x^{}_n\right|^{\,\beta}\,,\quad \text{where}\quad Z^{}_{D;\,\beta}=(2\pi)^{D/2}\prod_{m=1}^D\frac{\Gamma(1+m\beta/2)}{\Gamma(1+\beta/2)}\,,
\end{equation}
and the eigenvalues are expressed in units such that the standard deviation of the diagonal matrix elements is $\sigma/\sqrt{\beta}=1$. Note that the jpdf vanishes whenever $x^{}_m=x^{}_n$ for any $m\neq n$ due to the factors of the form $|x^{}_m-x^{}_n|^{\,\beta}$ encoding the repulsion between the eigenvalues. Interestingly, the jpdf in Eq.~\eqref{eq:jpdf_Gaussian_eigenvalues}, encoding the repulsion between eigenvalues and their confinement [due to the $\exp(-x^2_m/2)$ factors], also has a natural statistical interpretation as an equilibrium probability distribution of a two-dimensional Coulomb gas in a harmonic trap with $D$ point charges \big(eigenvalues $x^{}_m$\big) confined to a line~\cite{mehta2004, livan_18}. This can be used to derive the level density of the eigenvalues $p^{}_{D;\,\beta}(x)=\int dx^{}_2\ldots dx^{}_D\, \rho^{}_{D;\,\beta}\big(x,x^{}_2,\ldots,x^{}_D\big)$ and one can show that it follows the so-called Wigner semicircle law~\cite{mehta2004, livan_18}
\begin{equation} 
    \lim_{D\rightarrow\infty} \sqrt{\beta D}\ p^{}_{D;\,\beta}\!\left(\!\sqrt{\beta D}x\right)=\frac{1}{\pi}\sqrt{2-x^2}\,\quad\text{for}\quad |x|\leq\sqrt{2},\quad\text{and zero otherwise}.
\end{equation}
The {\bf semicircular level density} for Gaussian random matrix ensembles in RMT is to be contrasted with the Gaussian density of states for generic many-body local Hamiltonians as discussed in Sec.~\ref{sec:Model_and_observables}. This constitutes an important difference between the two.

Essential features of RMT level statistics, such as level repulsion, are manifest in the distributions $P^{}_{D;\,\beta}(s)$ of nearest-neighbor level spacings $s$. While Eq.~\eqref{eq:jpdf_Gaussian_eigenvalues} allows one to calculate $P^{}_{D;\,\beta}(s)$ for arbitrary values of $D$, a simple elementary closed form is available only for $D=2$. It can be directly obtained by considering $2\times2$ random matrices drawn from the respective Gaussian ensembles~\cite{dalessio_quantum_2016,livan_18}. Substituting $D=2$ and $\beta=1$ ($2$) for the GOE (GUE) in Eq.~\eqref{eq:jpdf_Gaussian_eigenvalues}, we find  
\begin{equation}
      \rho^{}_{2;1}\left(x^{}_1,x^{}_2\right)=\frac{1}{4\sqrt{\pi}}\left|x^{}_1-x^{}_2\right|\,\exp\left[-\frac{1}{2}\left(x_1^2+x_2^2\right)\right]\,,\quad \rho^{}_{2;2}\left(x^{}_1,x^{}_2\right)=\frac{1}{4\pi}\left|x^{}_1-x^{}_2\right|^2\,\exp\left[-\frac{1}{2}\left(x_1^2+x_2^2\right)\right]\,.
\end{equation}
We can then compute $P^{}_\beta(s)\equiv\iint dx^{}_1dx^{}_2\, \rho^{}_{2;\,\beta}\big(x^{}_1,x^{}_2\big)\,\delta\big(|x^{}_1-x^{}_2|-s\big)$, which for the GOE gives
\begin{equation}
    P^{}_1(s)=\frac{2}{4\sqrt{\pi}}\int_{-\infty}^\infty dx^{}_2\int_{x^{}_2}^\infty dx^{}_1\, \left(x^{}_1-x^{}_2\right)\, \exp\left[-\frac{1}{2}\left(x_1^2+x_2^2\right)\right]\delta\left(x^{}_1-x^{}_2-s\right)=\frac{se^{-s^2/4}}{2\sqrt{\pi}}\int_{-\infty}^\infty dx^{}_2\, \exp\left[-\left(x^{}_2+\frac{s}{2}\right)^2\right]=\frac{s}{2}\exp\left(-\frac{s^2}{4}\right). 
\end{equation}
A similar calculation for the GUE yields
\vspace{-0.1cm}
\begin{equation}
    P^{}_2(s)=\frac{s^2}{2\sqrt{\pi}}\exp\left(-\frac{s^2}{4}\right).
\end{equation}
Rescaling the spacings so that the mean level spacing is one, one has
\begin{equation}\label{eq:level_spacing_gaussian}
     P^{}_1(s)=\frac{\pi}{2}s\exp\left(-\frac{\pi}{4}s^2\right)\,,\quad  P^{}_2(s)=\frac{32}{\pi^2}s^2\exp\left(-\frac{4}{\pi}s^2\right).
\end{equation}
The level-spacing distributions in Eq.~\eqref{eq:level_spacing_gaussian} are of the general form
\vspace{-0.1cm}
\begin{equation}\label{eq:wigner_surmise}
    P^{}_\beta(s)=A^{}_\beta s^{\,\beta}\exp(-B^{}_\beta s^2)\,,
\end{equation}
which is known as the {\bf Wigner surmise}. The coefficients $A^{}_\beta$ and $B^{}_\beta$ are fixed by the normalization of $P^{}_\beta(s)$ and by requiring that the mean level spacing be one. $P^{}_\beta(s)$, derived for $2\times 2$ Gaussian random matrices, captures important features of RMT. It vanishes as $s\rightarrow0$, making it apparent that the levels repel each other, and large spacings are suppressed by the Gaussian in $s$. The level-spacing distributions of large Gaussian random matrices, known as {\bf Wigner-Dyson distributions}, are well approximated by the Wigner surmise~\eqref{eq:wigner_surmise}.

In a sufficiently narrow energy window in which the density of states is roughly constant, the level-spacing statistics of quantum systems with a classically chaotic counterpart were conjectured by Bohigas, Giannoni, and Schmit to be described by RMT~\cite{bohigas_giannoni_schmit_84}. This means that such systems are expected to exhibit Wigner-Dyson level-spacing distributions, which are well approximated by the Wigner surmise~\eqref{eq:wigner_surmise}. Over the years, Wigner-Dyson level-spacing statistics have also been found in quantum-chaotic systems without a classical counterpart, see, e.g., Refs.~\cite{Santos_2004, santos_rigol_10a}, so the emergence of Wigner-Dyson level-spacing statistics is considered to be a hallmark of quantum chaos~\cite{dalessio_quantum_2016}. On the other hand, the energy eigenvalues of integrable quantum systems generally exhibit a Poisson level-spacing distribution, as expected for uncorrelated levels and conjectured by Berry and Tabor for quantum systems whose classical counterparts are integrable~\cite{berry_tabor_77}. The Poisson probability of finding $n$ levels in an energy interval $[x,x+\Delta x]$ can be written as $\mathsf{P}^{}_n=e^{-\mu \Delta x}\frac{(\mu \Delta x)^n}{n!}$, where $\mu$ is the average level density, so that the average number of levels in the interval is $\mu \Delta x$. The probability of finding no levels between energies $x$ and $x+s$ is therefore $\mathsf{P}^{}_0(s)=e^{-\mu s}$. As a result, the probability density for the nearest-neighbor spacing is $P(s)=-\frac{d\mathsf{P}_0(s)}{ds}=\mu e^{-\mu s}$. Rescaling the spacings so that the mean spacing is one, one obtains the {\bf Poisson level-spacing distribution}~\cite{livan_18}
\begin{equation}\label{eq:poisson_level_spacing}
    P^{}_{\text{Poisson}}(s)=\exp(-s)\,.
\end{equation}

When comparing the level-spacing distribution of a Hamiltonian to the Wigner surmise~\eqref{eq:wigner_surmise} or Poisson~\eqref{eq:poisson_level_spacing} distributions to diagnose quantum chaos and integrability, the (energy) density of states needs to be approximately constant. However, this is not the case when considering the full spectrum of many-body local Hamiltonians, whose density of states is Gaussian (see Sec.~\ref{sec:Model_and_observables}), so one generally needs to perform an unfolding of the spectrum before comparing with the RMT/Poisson statistics~\cite{mehta2004, atas_level_spacing_13}. Another possibility, when sufficiently large Hamiltonian matrices can be diagonalized, is to restrict the analysis of the level-spacing distribution to a narrow energy window in which the density of states is approximately constant. A common alternative to both approaches is to use the {\bf ratio of consecutive level spacings}, which does not require unfolding~\cite{huse_2007}. Given a set of eigenvalues $\{x_m\}$ arranged in increasing order with level spacings $s_m=x_{m+1}-x_{m}$, one defines the ratios of consecutive spacings as $r_m=\min(s_m,s_{m-1})/\max(s_m,s_{m-1})$. Their distribution is known as the ratio-of-level-spacings distribution $\tilde P^{}_\beta(r)$. In a similar spirit to that of the Wigner surmise $P^{}_\beta(s)$, which was derived using the eigenvalues of $2\times2$ matrices, one can start with the jpdf~\eqref{eq:jpdf_Gaussian_eigenvalues} for the eigenvalues of a $3\times3$ Gaussian random matrix to obtain~\cite{atas_level_spacing_13}
\begin{equation}\label{eq:ratio_gaussian_RMT}
    \tilde P^{}_\beta(r)=\frac{2}{Z^{}_\beta}\frac{(r+r^2)^{\,\beta}}{(1+r+r^2)^{1+(3/2)\,\beta}}\,,
\end{equation}
\vspace{-0.25cm}

\noindent where $Z^{}_1=8/27$ [$Z^{}_2=4\pi/(81\sqrt{3})$] for GOE (GUE). Note that for Gaussian random matrices $\tilde P^{}_\beta(r)$~\eqref{eq:ratio_gaussian_RMT} vanishes as $r\rightarrow0$, just as $P^{}_\beta(s)$~\eqref{eq:wigner_surmise} vanishes as $s\rightarrow0$ due to level repulsion. Similarly, in the Poisson case, the ratio-of-level-spacings distribution is given by~\cite{atas_level_spacing_13}
\vspace{-0.1cm}
\begin{equation}\label{eq:ratio_Poisson}
    \tilde P^{}_{\text{Poisson}}(r)=\frac{2}{(1+r)^2}\,.
\end{equation}
\vspace{-0.3cm}

\noindent The mean ratios corresponding to $\tilde P^{}_\beta(r)$~\eqref{eq:ratio_gaussian_RMT} and $\tilde P^{}_{\text{Poisson}}(r)$~\eqref{eq:ratio_Poisson} take the values $\braket{r}^{}_{\beta=1}\approx0.536$ ($\braket{r}^{}_{\beta=2}\approx0.603$) and $\braket{r}^{}_{\mathrm{Poisson}}\approx0.386$, respectively~\cite{atas_level_spacing_13}. For large GOE (GUE) matrices, studied numerically, $\braket{r}^{}_{\beta=1} \approx 0.531$ ($\braket{r}^{}_{\beta=2} \approx 0.600$)~\cite{atas_level_spacing_13}. These mean ratios are also often used to diagnose quantum chaos and integrability. If the Hamiltonian of interest has symmetries, one also needs to resolve them in order to compare the energy eigenvalue distribution of the Hamiltonian with the RMT and Poisson predictions. This is because in the quantum-chaotic setting, although the energy eigenvalues within each symmetry sector exhibit RMT level statistics, the energy eigenvalues belonging to different symmetry sectors do not exhibit level repulsion with one another. Therefore, the spectral statistics of a quantum-chaotic Hamiltonian with unresolved symmetries may exhibit level statistics similar to a Poisson distribution, see, e.g., Ref.~\cite{santos_rigol_10b}.

The distributions $P^{}(s)$ and $\tilde P^{}(r)$ discussed so far are diagnostics of quantum chaos that probe short-range spectral correlations on the scale of the mean level spacing $\omega^{}_H\equiv \braket{s}$, which sets the Heisenberg time $t^{}_H=2\pi/\omega^{}_H$. Another important quantity, which captures longer-range spectral correlations, is the {\bf spectral form factor (SFF)} $\mathcal{K}(t)\equiv\tfrac{1}{D}\overline{|\sum_m e^{-iE^{}_mt}|^2}=\tfrac{1}{D}\overline{\sum_{m,n}e^{-i(E^{}_m-E^{}_n)t}}$~\cite{cotler_17, kos_ljubotina_18, gharibyan_hanada_2018, chan_deluca_18, bertini_kos_18}, where $\{E^{}_m\}$ is the energy spectrum and $\overline{(\ldots)}$ is an average over results computed in different symmetry sectors~\cite{ceven_patil_26} (disorder realizations~\cite{ceven_patil_26, sierant_delande_20, suntajs_bonca_20, sierant_25, ceven_peinemann_26}) in clean (disordered) systems. Unlike the level-spacing distributions, the SFF is not self-averaging~\cite{prange_97}, so its extraction generally requires averaging. In practice, to compute the SFF, one considers an unfolded energy spectrum $\{\varepsilon^{}_m\}$, for which the mean level spacing is unity, and introduces the normalized time $\tau=2\pi t/t^{}_H$ (so that $\tau=2\pi$ corresponds to $t=t^{}_H$), and calculates a filtered SFF~\cite{gharibyan_hanada_2018, suntajs_bonca_20}
\begin{equation}\label{eq:K-Gaussian-Filtered}
    \tilde K(\tau)=\overline{\left[\frac{|\sum_m g^{}_\eta(\varepsilon^{}_m) e^{-i\varepsilon^{}_m \tau}|^2}{\sum_m |g^{}_\eta(\varepsilon^{}_m)|^2}\right]}\,,\quad\text{where}\quad g^{}_\eta(\varepsilon^{}_m)=\exp\left[-\frac{1}{2}\left(\frac{\varepsilon^{}_m-\braket{\varepsilon}}{\eta\, \sigma^{}_\varepsilon}\right)^2\right]
\end{equation}
is a Gaussian filter of width $\eta\, \sigma^{}_\varepsilon$ (with $\sigma^{2}_\varepsilon$ the variance of the unfolded spectrum) centered at the mean unfolded energy $\braket{\varepsilon}$, introduced to reduce edge effects. In addition, in our calculation of $\tilde K(\tau)$ we remove $5\%$ from both edges of the unfolded spectrum, which suffer from poor unfolding. Within RMT, universal spectral correlations manifest in the SFF as an approximately linear-in-$\tau$ growth known as the {\bf ramp}, which crosses over around $\tau=2 \pi$ into an approximately flat regime known as the {\bf plateau}. For an ensemble of random matrices drawn from the GOE, in the large-$D$ limit, the universal connected contribution to the SFF $\tilde K^{}_{\text{GOE}}(\tau)$ has the following analytical form~\cite{mehta2004}
\vspace{-0.1cm}
\begin{equation}\label{eq:K_GOE}
    \tilde K^{}_{\text{GOE}}(\tau)=
    \begin{cases}
    \frac{\tau}{\pi}-\frac{\tau}{2\pi}\ln\left(\frac{\tau}{\pi}+1\right)\,, & \tau\leq2\pi\\
    2-\frac{\tau}{2\pi}\ln\left(\frac{\frac{\tau}{\pi}+1}{\frac{\tau}{\pi}-1}\right)\,, & \tau >2\pi
    \end{cases}.
\end{equation}
The SFF for a local chaotic quantum many-body system is expected to exhibit a similar RMT ramp-plateau behavior after sufficiently long times $t\geq t^{}_{\text{SFF}}$, where it is described by RMT. The time $t^{}_{\text{SFF}}$ marking the onset of RMT behavior in the SFF is often identified with, or referred to as, the Thouless time $t^{}_{\text{Th}}$. In systems exhibiting diffusive transport, $t^{}_{\text{SFF}}$ is expected to scale as $t^{}_{\text{SFF}}\propto L^2$.

\begin{figure}[t]
    \centering
    \includegraphics[width=\linewidth]{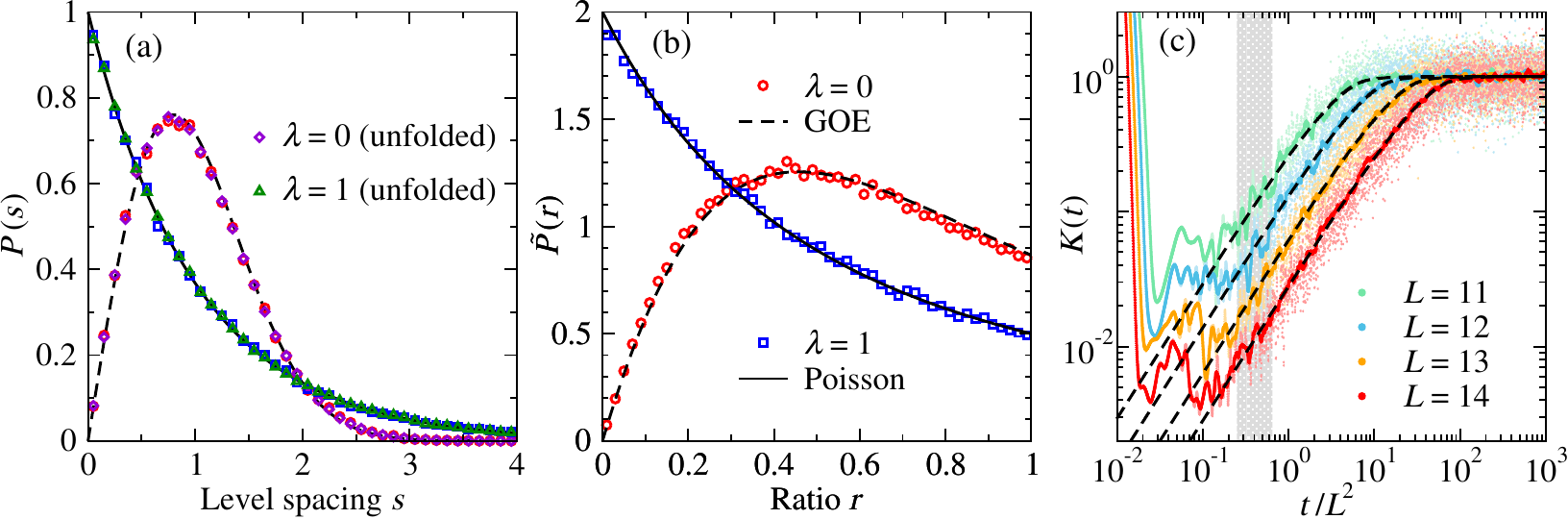}
    \caption{{\bf Level statistics}: Histograms of (a) the level-spacing distributions $P(s)$ calculated using the central $90\%$ of the unfolded spectrum and the central $20\%$ of the spectrum with the mean level spacing rescaled to 1, and (b) the distribution of ratios $\tilde P(r)$ calculated using the central $90\%$ of the spectrum, for the spin-$1$ $XXZ$ model~\eqref{eq:Spin1_Hamiltonian} at the quantum-chaotic ($\lambda=0$) and integrable ($\lambda=1$) points for the largest diagonalized chain ($L=14$). The dashed and solid lines show the corresponding predictions for the GOE ($\beta=1$) and Poisson distributions, respectively. (c) The spectral form factor $K(t)\equiv\tilde K(2\pi t/t^{}_H)$, see Eq.~\eqref{eq:K-Gaussian-Filtered}, calculated using the central $90\%$ of the unfolded spectrum obtained by fitting a degree-10 polynomial, with a Gaussian filter of width $\eta=0.4$~\cite{ceven_patil_26}, and the mean level spacing $\omega^{}_H$ calculated in the central $20\%$ of the spectrum to compute $t^{}_H=2\pi/\omega^{}_H$ for chains with $L=11$--$14$ at $\lambda=0$. The larger, darker (smaller, lighter) points show the moving averages over time windows of size $\delta(\log_{10}t)=0.051$ (raw data), and the dashed lines show the GOE prediction $K^{}_{\text{GOE}}(t)\equiv\tilde K^{}_{\text{GOE}}(2\pi t/t^{}_H)$, see Eq.~\eqref{eq:K_GOE}. The gray shaded band indicates the rescaled-time interval in which, for the chain sizes considered, the moving averages (larger, darker points) begin to resemble and fluctuate about $K^{}_{\text{GOE}}(t)$ and the fluctuations of the raw data (smaller, lighter points) begin to resemble those at long times. We expect $t^{}_{\text{SFF}}/L^2$ to lie within this interval.}
    \label{fig:level-statistics}
\end{figure}

In Fig.~\ref{fig:level-statistics} we compare the level-spacing [Fig.~\ref{fig:level-statistics}(a)] and ratio-of-level-spacings [Fig.~\ref{fig:level-statistics}(b)] distributions for the Hamiltonian in Eq.~\eqref{eq:Spin1_Hamiltonian} at the quantum-chaotic ($\lambda=0$) and integrable ($\lambda=1$) points with the analytical predictions for the GOE [Eqs.~\eqref{eq:level_spacing_gaussian} and~\eqref{eq:ratio_gaussian_RMT}] and the Poisson [Eqs.~\eqref{eq:poisson_level_spacing} and~\eqref{eq:ratio_Poisson}] distributions, respectively. The reported level statistics correspond to a weighted average over the symmetry sectors with $\{M=0;\mathbb{Z}^{}_2=\pm1;k\neq0,\pi\}$. In Fig.~\ref{fig:level-statistics}(a), we show results for the distribution of level spacings both after unfolding (for the central $90\%$ of the spectrum) and after a direct rescaling of the mean to one (for the central $20\%$ of the spectrum). The level-spacing statistics for the Hamiltonian at the quantum-chaotic (integrable) point agree well with the GOE (Poisson) level-spacing distribution. Note that no unfolding procedure is needed for the central $20\%$ of the spectrum at the system size considered. In Fig.~\ref{fig:level-statistics}(b), we compare the ratio-of-level-spacings distribution for the energy eigenvalues over the central $90\%$ of the spectrum (neither unfolding nor rescaling is required) with the corresponding GOE and Poisson predictions, which are again in good agreement. In Fig.~\ref{fig:level-statistics}(c) we plot $K(t)\equiv\tilde K(2\pi t/t^{}_H)$ as a function of the rescaled time $t/L^2$ for the four largest chains that we diagonalized. For all chain sizes, one can see a nonuniversal regime at short times, followed by the expected RMT ramp and plateau corresponding to $K^{}_{\text{GOE}}(t)\equiv\tilde K^{}_{\text{GOE}}(2\pi t/t^{}_H)$, about which our results for $K(t)$ fluctuate. The gray shaded band shows that $t^{}_{\text{SFF}}$ in our chains is consistent with the expected diffusive scaling $t^{}_{\text{SFF}}\propto L^2$. The onset of the plateau, on the other hand, moves to larger rescaled times because $t^{}_H$ grows exponentially with system size. 

Figures~\ref{fig:level-statistics}(a)--\ref{fig:level-statistics}(c) therefore show that the plotted short- and long-range spectral correlations of nonintegrable quantum many-body lattice Hamiltonians are well described by the predictions of RMT in the chaotic regime even when those Hamiltonians do not have a classical counterpart. In the integrable regime, on the other hand, short-range spectral correlations [Figs.~\ref{fig:level-statistics}(a) and~\ref{fig:level-statistics}(b)] are well described by the predictions of uncorrelated random variables. All this occurs despite the fact that there is nothing random about the matrices representing the corresponding Hamiltonians in the bases used for the numerical calculations. In fact, those matrices can be rather {\bf sparse} in computational bases, in stark contrast to the corresponding random matrices.

\subsection{Eigenvectors of random matrices}

For $D\times D$ real-symmetric (complex-Hermitian) random matrices drawn from an orthogonal (unitary) ensemble, the joint probability distribution $P^{}_{\!O}$ ($P^{}_{\!U}$) of the components $c^{}_j=\braket{a^{}_j|\psi}$ of an eigenvector $\ket{\psi}=\sum_{j=1}^Dc^{}_j\ket{a^{}_j}$ represented in a fixed orthonormal basis $\{\ket{a^{}_j}\}$ can be written as~\cite{brody_81}:
\begin{equation}
    P^{}_{\!O}(c^{}_1,c^{}_2,\ldots,c^{}_D)=\mathcal{N}\delta\left(\sum_{j=1}^Dc_j^2-1\right), \qquad P^{}_{\!U}(c^{}_1,c^{}_2,\ldots,c^{}_D)=\mathcal{N'}\delta\left(\sum_{j=1}^D|c^{}_j|^2-1\right)\,,
\end{equation}
where $\mathcal{N}$ ($\mathcal{N'}$) is the normalization constant for $P^{}_{\!O}$ ($P^{}_{\!U}$). This is a consequence of the orthogonal (unitary) invariance of the ensemble, which implies that the eigenvectors are uniformly distributed on the real (complex) unit sphere. Furthermore, the different eigenvectors of a given matrix are constrained to be mutually orthogonal.

Focusing on the GOE, which is the ensemble relevant to our model Hamiltonian~\eqref{eq:Spin1_Hamiltonian} (the equivalent results for the GUE can be found in Ref.~\cite{livan_18}), the probability distribution $\mathcal{P}^{}_{\!O}$ of a single squared amplitude, say $c_1^2$, can be obtained by computing the marginal distribution
\begin{equation}
    \mathcal{P}^{}_{\!O}(x)=\int_{-\infty}^{\infty}\prod_{j=1}^Ddc^{}_j\, P^{}_{\!O}(c^{}_1,\ldots,c^{}_D)\,\delta(x-c_1^2)=\mathcal{N}\int_{-\infty}^{\infty} \prod_{j=1}^Ddc^{}_j\, \delta\left(1-\sum_{k=1}^Dc_k^2\right)\delta(x-c_1^2)\,.
\end{equation}
One can use a Laplace transform to ``split'' the joint $\delta$ function involving all $D$ integration variables, and factorize the integral~\cite{livan_18}. To do so, one starts by introducing a parameter $t$ for the squared norm and defining
\begin{equation}
    \mathcal{P}^{}_{\!O}(x;t)=\mathcal{N}\int_{-\infty}^{\infty} \prod_{j=1}^Ddc^{}_j\, \delta\left(t-\sum_{k=1}^Dc_k^2\right)\delta(x-c_1^2)\,,
\end{equation}
where $\mathcal{P}^{}_{\!O}(x;t=1)=\mathcal{P}^{}_{\!O}(x)$. The Laplace transform with respect to $t$ yields
\vspace{-0.1cm}
\begin{equation}
  \mathcal{P}^{}_{\!O}(x;s)\equiv\mathcal{L}\{\mathcal{P}^{}_{\!O}(x;t)\}=\int_0^\infty dt e^{-st}\,\mathcal{P}^{}_{\!O}(x;t)=\mathcal{N}\left(\int_{-\infty}^{\infty} dc^{}_1 e^{-sc_1^2} \delta(x-c_1^2)\right)\left(\int_{-\infty}^{\infty} dc e^{-sc^2}\right)^{D-1} \propto x^{-1/2}e^{-sx}s^{-(D-1)/2}\,,
\end{equation}
where the integral over $c^{}_1$ can be performed by a change of variables $y=c_1^2$ and corresponds to the factor $x^{-1/2}e^{-sx}$, while the factor $s^{-(D-1)/2}$ appears from the $D-1$ Gaussian integrals. Inverting the Laplace transform using $\mathcal{L}\{t^n\}=\Gamma(n+1)/s^{n+1}$ and the translation property $\mathcal{L}^{-1}\{e^{-sx}F(s)\}=\Theta(t-x)f(t-x)$, where $f(t)=\mathcal{L}^{-1}\{F(s)\}$,\footnote{The first identity can be shown by changing variables $y=st$ in $\mathcal{L}\{t^n\}=\int_0^\infty dt\,e^{-st}\,t^n$ and using $\Gamma(n+1)=\int_0^\infty dy\,e^{-y}\,y^n$; see, \eg Ref.~\cite{arfken_weber_13}.} one has 
\begin{equation}
    \mathcal{P}^{}_{\!O}(x;t)\propto x^{-1/2}(t-x)^{(D-3)/2}\,\Theta(t-x)\,,
\end{equation}
where $\Theta$ is the Heaviside function. After setting $t=1$, one has $\mathcal{P}^{}_{\!O}(x)=\mathcal{N}''x^{-1/2}(1-x)^{(D-3)/2}$ with the normalization $\mathcal{N}''$ given by
\begin{equation}
   \int_0^1\!dx\,\mathcal{P}^{}_{\!O}(x)=\mathcal{N}''\!\int_{0}^{1}\!dx\,x^{-1/2}(1-x)^{\frac{D-3}{2}}=\mathcal{N}''\!\int_0^1 \!d(\sin^2\theta)\,(\sin^2\theta)^{-1/2}(1-\sin^2\theta)^{\frac{D-3}{2}}=2\mathcal{N}''\!\int_0^{\pi/2}\!d \theta\, (\cos\theta)^{D-2}=2\mathcal{N}''\frac{\Gamma\Big(\frac{D-1}{2}\Big)\,\Gamma\Big(\frac{1}{2}\Big)}{2\,\Gamma\Big(\frac{D}{2}\Big)}:=1, 
\end{equation}
where we used the Wallis integral $W^{}_n=\int_0^{\pi/2}d\theta \cos^n \theta=\Gamma\Big(\frac{n+1}{2}\Big)\,\Gamma\Big(\frac{1}{2}\Big)\,\Big/\,\Big[2\,\Gamma \Big(\frac{n}{2}+1\Big)\Big]$. Therefore, the normalized probability distribution for a single squared amplitude $x$ is given by
\vspace{-0.2cm}
\begin{equation}
    \mathcal{P}^{}_{\!O}(x)=\frac{\Gamma\Big(\frac{D}{2}\Big)}{\Gamma\Big(\frac{D-1}{2}\Big)\Gamma\Big(\frac{1}{2}\Big)}x^{-1/2}(1-x)^{(D-3)/2}\,,\quad 0\leq x\leq 1\,.
\end{equation}

Calculating the mean squared amplitude $\braket{ x}\equiv\int_0^1 dx\, \mathcal{P}^{}_{\!O}(x) x$, one finds
\vspace{-0.1cm} 
\begin{equation}
    \braket{x}= \frac{\Gamma\Big(\frac{D}{2}\Big)}{\Gamma\Big(\frac{D-1}{2}\Big)\Gamma\Big(\frac{1}{2}\Big)}\int_0^1 dx x^{1/2}(1-x)^{(D-3)/2}=\frac{\Gamma\Big(\frac{D}{2}\Big)}{\Gamma\Big(\frac{D-1}{2}\Big)\Gamma\Big(\frac{1}{2}\Big)}2\left[W^{}_{D-2}-W^{}_D\right]=1-\frac{\Gamma\Big(\frac{D}{2}\Big)}{\Gamma\Big(\frac{D-1}{2}\Big)}\frac{\Gamma\Big(\frac{D+1}{2}\Big)}{\Gamma\Big(\frac{D}{2}+1\Big)}=\frac{1}{D}\,.
\end{equation}
Since the mean $\braket{x}$ depends on $D$, one can instead consider the probability density in the rescaled variable $X=x/\!\braket{x}=xD$, which in the $D\rightarrow\infty$ limit gives
\vspace{-0.2cm}
\begin{equation}\label{eq:amplitude_orthogonal}
    \tilde{\mathcal{P}}^{}_{\!O}(X)=\lim_{D\rightarrow\infty}\frac{1}{D}\mathcal{P}^{}_{\!O}\left(\frac{X}{D}\right)=\frac{1}{\sqrt{2\pi X}}\exp\left(-\frac X2\right)\,,
\end{equation}
and similarly in the unitary case~\cite{livan_18}, defining $\mathcal{P}^{}_{\!U}(x)$ as the probability distribution of $x=|c^{}_j|^2$ and $X=|c^{}_j|^2/\!\braket{|c^{}_j|^2}=D|c^{}_j|^2$,
\begin{equation}\label{eq:amplitude_unitary}
    \tilde{\mathcal{P}}^{}_{\!U}(X)=\lim_{D\rightarrow\infty}\frac{1}{D}\mathcal{P}^{}_{\!U}\left(\frac{X}{D}\right)=\exp\left(-X\right)\,.
\end{equation}
The marginal distribution $\tilde{\mathcal{P}}^{}_{\!O}(X)$ in Eq.~\eqref{eq:amplitude_orthogonal} for the rescaled squared amplitude $X=c_j^2/\!\braket{c_j^2}$ is known as the {\bf Porter-Thomas distribution}~\cite{porter_thomas_56}. Note that, in the orthogonal (unitary) ensemble, assuming that, in the $D\to\infty$ limit, the rescaled coefficients $c^{}_j/\!\sqrt{\braket{c_j^2}}$ $\left[c^{}_j/\!\sqrt{\braket{|c^{}_j|^2}}\right]$ are real (complex) Gaussian random variables, as is sometimes done in the literature~\cite{dalessio_quantum_2016}, reproduces the squared-amplitude distribution in Eq.~\eqref{eq:amplitude_orthogonal} [Eq.~\eqref{eq:amplitude_unitary}]. We stress that the orthogonality between different eigenvectors plays no role in deriving the marginal distribution of a single squared amplitude. Correlations due to the orthogonality between different eigenvectors play an important role when computing the moments of the matrix elements of operators. They will be discussed in the following section.

\subsection{Matrix elements of Hermitian operators in the eigenstates of random matrices}\label{sec:matelemobsRMT}

Given a Hermitian operator $\hat O=\sum_jO^{}_j\ket{O^{}_j}\bra{O^{}_j}$ (in the GOE, taken to be real symmetric in the basis in which the random Hamiltonians are real symmetric) with eigenbasis $\{\ket{O^{}_j}\}$, its matrix elements with respect to the eigenstates $\{\ket{\psi^{}_m}\}$ of a random Hamiltonian can be written as 

\vspace{-0.4cm}
\begin{equation}\label{eq:RMT_matrix_elements}
    O^{}_{mn}\equiv\braket{\psi^{}_m|\hat O|\psi^{}_n}=\sum_{j}O^{}_j\braket{\psi^{}_m|O^{}_j}\braket{O^{}_j|\psi^{}_n}=\sum_{j} O^{}_j (c^m_j)^*c^n_j\,,
\end{equation}
where $c^m_j\equiv \braket{O^{}_j|\psi^{}_m}$ can be thought of as the matrix elements of a random unitary (orthogonal) matrix encoding the change of basis from the eigenstates $\{\ket{\psi^{}_m}\}$ of a Hamiltonian drawn from the GUE (GOE) to the eigenstates $\{\ket{O^{}_j}\}$ of $\hat O$. The eigenstates of a random matrix are random orthonormal vectors in a large Hilbert space of dimension $D$, which satisfy
\begin{equation}\label{eq:RMT_orthonormality}
    \overline{(c^m_j)^*c^n_k}=\frac{1}{D}\delta^{}_{jk}\delta^{}_{mn}\,,
\end{equation}
where $\overline{(\ldots)}$ represents an ensemble average over the random eigenstates $\ket{\psi^{}_m}$ and $\ket{\psi^{}_n}$. Using Eq.~\eqref{eq:RMT_orthonormality}, the ensemble-averaged matrix elements of the operator $\hat O$ in Eq.~\eqref{eq:RMT_matrix_elements} are given by
\begin{equation}
    \overline{O^{}_{mm}}=\frac{1}{D}\sum_{j} O^{}_j\equiv\overline{O},\qquad\text{and}\qquad \overline{O^{}_{mn}}=0\,\quad \text{for} \quad m\neq n. 
\end{equation}
The higher-order moments of the matrix elements $O^{}_{mn}$ depend on the type of ensemble from which the random Hamiltonian is drawn~\cite{dalessio_quantum_2016}. For complex-Hermitian or real-symmetric matrices, the random eigenvectors $\ket{\psi^{}_m}$ and $\ket{\psi^{}_n}$ are columns of matrices drawn from an ensemble of random unitary or orthogonal matrices, respectively. 

All the moments of the matrix elements $O^{}_{mn}$ can be calculated taking $c^m_j$ to be the matrix elements of a {\bf Haar-random unitary (orthogonal) matrix}, and using the {\bf Weingarten formula} for Haar ensemble averages~\cite{collins_03,collins_06,collins_22}. Haar-random unitary (orthogonal) matrices are complex (real) matrices sampled {\bf uniformly} from the unitary (orthogonal) group according to the {\bf Haar measure}, which is the unique bi-invariant probability measure. When the $c^m_j$ are drawn from a Haar-random ensemble of unitary matrices, the {\bf Haar-averages} $\overline{c^{m_1}_{j_1}\cdots c^{m_N}_{j_N}(c^{m_1'}_{j'_1})^*\cdots(c^{m'_N}_{j'_N})^*}$ read:
\vspace{-0.1cm}
\begin{equation}\label{eq:Weingarten_U(D)}
    \overline{c^{m_1}_{j_1}\cdots c^{m_N}_{j_N}(c^{m_1'}_{j'_1})^*\cdots(c^{m'_N}_{j'_N})^*}=\sum_{\sigma,\tau\in S\!^{}_N}\mathrm{Wg}^{\mathrm{U}(D)}(\sigma^{-1}\tau)\,\delta^{}_{j_1j'_{\sigma(1)}}\ldots \delta^{}_{j_Nj'_{\sigma(N)}}\delta^{}_{m_1m'_{\tau(1)}}\ldots \delta^{}_{m_Nm'_{\tau(N)}}\,,
\end{equation}
\vspace{-0.3cm}

\noindent where the different terms correspond to products of Kronecker deltas between pairs of primed and unprimed indices, where the unprimed lower (upper) indices $j_r$ ($m_r$) are matched with a permutation $\sigma$ ($\tau$) of the primed lower (upper) indices $j_r'$ ($m'_r$). The permutations $\sigma,\tau$ are elements of the symmetric group on $N$ objects, $S\!^{}_N$, and the prefactor $\mathrm{Wg}^{\mathrm{U}(D)}(\sigma^{-1}\tau)$ is the Weingarten function of the unitary group $\mathrm{U}(D)$~\cite{collins_03}. In particular, Eq.~\eqref{eq:RMT_orthonormality} corresponds to a special case of Eq.~\eqref{eq:Weingarten_U(D)} with $N=1$ and the Weingarten function $\mathrm{Wg}^{\mathrm{U}(D)}([1])=1/D$ corresponding to the identity permutation.

To compute the variance of the matrix elements of $\hat O$, one evaluates Haar-averages of the form $\overline{c^{m}_{\!j}c^{n}_{k}(c^{m'}_{j'})^*(c^{n'}_{k'})^*}$, which using the Weingarten formula in Eq.~\eqref{eq:Weingarten_U(D)} can be written as 
\begin{equation}\label{eq:Weingarten_4point_U(D)}
    \overline{c^{m}_{j}c^{n}_{k}(c^{m'}_{j'})^*(c^{n'}_{k'})^*}=\frac{1}{D^2-1}\left[(\delta^{}_{jj'}\delta^{}_{kk'}\delta^{}_{mm'}\delta^{}_{nn'} + \delta^{}_{jk'}\delta^{}_{kj'}\delta^{}_{mn'}\delta^{}_{nm'})-\frac{1}{D}\left(\delta^{}_{jj'}\delta^{}_{kk'}\delta^{}_{mn'}\delta^{}_{nm'} + \delta^{}_{jk'}\delta^{}_{kj'}\delta^{}_{mm'}\delta^{}_{nn'}\right)\right]\,,
\end{equation}
where the first two terms correspond to $\sigma=\tau$, with both permutations being either the identity $\left(\begin{smallmatrix} 1 & 2 \\ 1 & 2 \end{smallmatrix}\right)$ or the transposition $\left(\begin{smallmatrix} 1 & 2 \\ 2 & 1 \end{smallmatrix}\right)$, leading to $\sigma^{-1}\tau=\left(\begin{smallmatrix} 1 & 2 \\ 1 & 2 \end{smallmatrix}\right)$. Recall that a permutation $\sigma$ on $N$ objects can be expressed in the form $\left(\begin{smallmatrix} 1 & 2 & \ldots & N \\ \sigma(1) & \sigma(2) & \ldots & \sigma(N) \end{smallmatrix}\right)$. Since $\sigma^{-1}\tau=\left(\begin{smallmatrix} 1 & 2 \\ 1 & 2 \end{smallmatrix}\right)=\left(\begin{smallmatrix} 1 \\ 1 \end{smallmatrix}\right)\left(\begin{smallmatrix} 2 \\ 2 \end{smallmatrix}\right)$ is a product of two 1-cycles, the corresponding Weingarten function is given by $\mathrm{Wg}^{\mathrm{U}(D)}([1,1])=1/(D^2-1)$. Similarly, the last two terms contain combinations of the identity and the transposition, for which $\sigma^{-1}\tau=\left(\begin{smallmatrix} 1 & 2 \\ 2 & 1 \end{smallmatrix}\right)$ is a transposition (or one 2-cycle) with Weingarten function $\mathrm{Wg}^{\mathrm{U}(D)}([2])=-1/[D(D^2-1)]$.

Using Eq.~\eqref{eq:Weingarten_4point_U(D)}, one can calculate the fluctuations of the diagonal and off-diagonal matrix elements to the leading order in $1/D$~\cite{pappalardi_notes_25}. For the diagonal matrix elements
\begin{equation}
    \overline{(O^{}_{mm})^2}-\overline{O^{}_{mm}}^2=\sum_{j,k} O^{}_jO^{}_k \overline{c^m_j c^m_k (c^m_j)^* (c^m_k)^*}-\overline{O}^2=\sum_{j,k}O^{}_jO^{}_k\frac{1}{D^2-1}\left[1+\delta^{}_{jk}-\frac{1}{D}\left(1+\delta^{}_{jk}\right)\right]-\overline{O}^2\simeq \frac{1}{D}\left(\overline{O^2}-\overline{O}^2\right)\,,
\end{equation}
where we defined $\overline{O^2}\equiv\frac{1}{D}\sum_j O^2_j$. Similarly for the off-diagonal elements
\begin{equation}
    \overline{|O^{}_{mn}|^2}-|\overline{O^{}_{mn}}|^2=\sum_{j,k} O^{}_jO^{}_k \overline{c^n_j c^m_k (c^m_j)^* (c^n_k)^*}=\sum_{j,k}O^{}_jO^{}_k\frac{1}{D^2-1}\left[\delta^{}_{jk}-\frac{1}{D}\right]\simeq \frac{1}{D}\left(\overline{O^2}-\overline{O}^2\right)\,.
\end{equation}
Therefore, to the leading order in $1/D$, for the GUE the {\bf matrix elements} of a Hermitian operator $\hat O$ can be parametrized as 
\begin{equation}\label{eq:RMT_Result_MatrixElements}
    O^{}_{mn}=\overline{O}\,\delta^{}_{mn}+\sqrt{\frac{\overline{O^2}-\overline{O}^2}{D}} R^{O}_{mn}\,,
\end{equation}
where $R^{O}_{mn}$ is a random variable with zero mean and unit variance, i.e., $\overline{R^{O}_{mn}}=0$ and $\overline{|R^{O}_{mn}|^2}=1$.

If the $c^m_j$ are instead drawn from a Haar-random ensemble of orthogonal matrices, the Haar-averages $\overline{c^{m_1}_{j_1}\cdots c^{m_N}_{j_N}c^{m_{N+1}}_{j_{N+1}}\cdots c^{m_{2N}}_{j_{2N}}}$ can be computed using the Weingarten formula for the orthogonal group $\mathrm{O}(D)$~\cite{collins_06,collins_22}: 
\begin{equation}\label{eq:Weingarten_O(D)}
    \overline{c^{m_1}_{j_1}\cdots c^{m_N}_{j_N}c^{m_{N+1}}_{j_{N+1}}\cdots c^{m_{2N}}_{j_{2N}}}=\sum_{\sigma,\tau\in M^{}_{2N}}\mathrm{Wg}^{\mathrm{O}(D)}(\sigma^{-1}\tau)\,\delta^{}_{j_{\sigma(1)}j_{\sigma(2)}}\ldots \delta^{}_{j_{\sigma(2N-1)}j_{\sigma(2N)}}\delta^{}_{m_{\tau(1)}m_{\tau(2)}}\ldots \delta^{}_{m_{\tau(2N-1)}m_{\tau(2N)}} \,,
\end{equation}
\vspace{-0.3cm}

\noindent where the prefactor $\mathrm{Wg}^{\mathrm{O}(D)}(\sigma^{-1}\tau)$ is the Weingarten function of the orthogonal group $\mathrm{O}(D)$~\cite{collins_06}. The sum is over pairings $\sigma$ ($\tau$) of the lower (upper) indices $j_r$ ($m_r$), with $\sigma,\tau\in M^{}_{2N}\subset S^{}_{2N}$, where $M^{}_{2N}$ denotes the set of pairings of $2N$ objects. A pairing of $2N$ objects $\{1,2,\ldots,2N\}$ is defined as a partition into $N$ subsets containing two elements each. For instance, for $N=2$ with the set of $2N$ objects $\{1,2,3,4\}$, one of the three possible pairings can be written as $\pi=\{\{\pi(1),\pi(2)\},\{\pi(3),\pi(4)\}\}=\{\{1,3\},\{2,4\}\}$ associated with the permutation $\pi=\left(\begin{smallmatrix} 1 & 2 & 3 & 4 \\ 1 & 3 & 2 & 4 \end{smallmatrix}\right)$. To determine which $\mathrm{Wg}^{\mathrm{O}(D)}(\pi)$ to use for a permutation $\pi$, one considers an undirected graph $\Gamma(\pi)$ with vertex set $\{1,2,\ldots,2N\}$ and connects the vertices in the edge sets $\{\{1,2\},\ldots,\{2N-1,2N\}\}$ and $\{\{\pi(1),\pi(2)\},\ldots,\{\pi(2N-1),\pi(2N)\}\}$~\cite{collins_22}. For instance, the graphs $\Gamma$ associated with the three possible pairings $\pi_a,\pi_b$ and $\pi_c$ on $2N=4$ objects $\{1,2,3,4\}$ can be drawn as follows:
\begin{equation}\label{eq:graph_Gamma}
    \pi_{a}=\left(\begin{smallmatrix} 1 & 2 & 3 & 4 \\ 1 & 2 & 3 & 4 \end{smallmatrix}\right)\overset{\Gamma}{\longmapsto}
\vcenter{\hbox{%
\begin{tikzpicture}[scale=0.6]
  \fill (0,1) circle (3pt) node[above left] {1};
  \fill (1,1) circle (3pt) node[above right] {2};
  \fill (0,0) circle (3pt) node[below left] {4};
  \fill (1,0) circle (3pt) node[below right] {3};
  \draw (0,1) -- (1,1);
  \draw (0,0) -- (1,0);
\end{tikzpicture}}}\,,\qquad 
\pi_{b}=\left(\begin{smallmatrix} 1 & 2 & 3 & 4 \\ 1 & 3 & 2 & 4 \end{smallmatrix}\right)\overset{\Gamma}{\longmapsto}
\vcenter{\hbox{%
\begin{tikzpicture}[scale=0.6]
  \fill (0,1) circle (3pt) node[above left] {1};
  \fill (1,1) circle (3pt) node[above right] {2};
  \fill (0,0) circle (3pt) node[below left] {4};
  \fill (1,0) circle (3pt) node[below right] {3};
  \draw (0,1) -- (1,1);
  \draw (0,0) -- (1,0);
  \draw (0,1) -- (1,0);
  \draw (1,1) -- (0,0);
\end{tikzpicture}}}\,,\qquad 
\pi_{c}=\left(\begin{smallmatrix} 1 & 2 & 3 & 4 \\ 1 & 4 & 2 & 3 \end{smallmatrix}\right)\overset{\Gamma}{\longmapsto}
\vcenter{\hbox{%
\begin{tikzpicture}[scale=0.6]
  \fill (0,1) circle (3pt) node[above left] {1};
  \fill (1,1) circle (3pt) node[above right] {2};
  \fill (0,0) circle (3pt) node[below left] {4};
  \fill (1,0) circle (3pt) node[below right] {3};
  \draw (0,1) -- (1,1);
  \draw (0,0) -- (1,0);
  \draw (0,1) -- (0,0);
  \draw (1,1) -- (1,0);
\end{tikzpicture}}}\,.
\end{equation}
Coincident edges are superimposed in the graphical representation. If $\Gamma(\pi)$ consists of $r$ disconnected cycles of even lengths $2l_1,2l_2,\ldots,2l_r$ with $l_1\geq l_2 \geq \ldots \geq l_r \geq1$ then $\pi$ is said to have a coset-type $[l_1,l_2,\ldots,l_r]$ and one uses the Weingarten function $\mathrm{Wg}^{\mathrm{O}(D)}([l_1,l_2,\ldots,l_r])$ for $\pi$~\cite{collins_22}. For instance, the graph $\Gamma$ for the permutation $\pi_b$ in Eq.~\eqref{eq:graph_Gamma} (with $N=2$) consists of a single $4$-cycle (i.e., $l_1=2$) connecting edges $\{\{1,2\},\{3,4\},\{1,3\},\{2,4\}\}$ and is associated with the Weingarten function $\mathrm{Wg}^{\mathrm{O}(D)}([2])=-1/[D(D-1)(D+2)]$.

The Haar-average $\overline{c^{m}_{j}c^{n}_{k}c^{m'}_{j'}c^{n'}_{k'}}$ for $N=2$ can be evaluated using the Weingarten formula~\eqref{eq:Weingarten_O(D)} for the orthogonal group $\mathrm{O}(D)$ to obtain
\begin{align}\label{eq:Weingarten_4point_O(D)}
    \overline{c^{m}_{j}c^{n}_{k}c^{m'}_{j'}c^{n'}_{k'}}=\frac{D+1}{D(D-1)(D+2)}\Bigg(&\delta^{}_{jj'}\delta^{}_{kk'}\left[\delta^{}_{mm'}\delta^{}_{nn'} -\frac{1}{D+1}\left(\delta^{}_{mn}\delta^{}_{m'n'}+\delta^{}_{mn'}\delta^{}_{nm'}\right)\right]
    +\, \delta^{}_{jk}\delta^{}_{j'k'}\left[\delta^{}_{mn}\delta^{}_{m'n'} -\frac{1}{D+1}\left(\delta^{}_{mm'}\delta^{}_{nn'}+\delta^{}_{mn'}\delta^{}_{nm'}\right)\right] \\ 
    +\,&\delta^{}_{jk'}\delta^{}_{kj'}\left[\delta^{}_{mn'}\delta^{}_{nm'} -\frac{1}{D+1}\left(\delta^{}_{mm'}\delta^{}_{nn'}+\delta^{}_{mn}\delta^{}_{m'n'}\right)\right]\Bigg)\,,\nonumber
\end{align}
where the $3\times3=9$ terms represent the $3$ pairing choices $\{j',k,k'\}$ for the lower index $j$ and independently the $3$ pairing choices $\{m',n,n'\}$ for the upper index $m$. The first term in each square bracket corresponds to identical pairings of the lower and upper indices (i.e., $\sigma=\tau$ and $\sigma^{-1}\tau$ is the identity permutation) and results in the graph $\Gamma(\sigma^{-1}\tau)$ consisting of two disconnected 2-cycles [see $\Gamma(\pi_a)$ in Eq.~\eqref{eq:graph_Gamma}] with $\mathrm{Wg}^{\mathrm{O}(D)}([1,1])=(D+1)/[D(D-1)(D+2)]$, while for the last two terms in each square bracket $\Gamma(\sigma^{-1}\tau)$ corresponds to a single 4-cycle [see $\Gamma(\pi_b)$ and $\Gamma(\pi_c)$ in Eq.~\eqref{eq:graph_Gamma}] with $\mathrm{Wg}^{\mathrm{O}(D)}([2])=-1/[D(D-1)(D+2)]$. Using Eq.~\eqref{eq:Weingarten_4point_O(D)}, one can calculate the fluctuations of the diagonal and off-diagonal matrix elements to the leading order in $1/D$. For the diagonal elements
\begin{equation}
    \overline{(O^{}_{mm})^2}-\overline{O^{}_{mm}}^2=\sum_{j,k} O^{}_jO^{}_k \overline{c^m_j c^m_k c^m_j c^m_k}-\overline{O}^2=\sum_{j,k}O^{}_jO^{}_k\frac{D+1}{D(D-1)(D+2)}\left[2\delta^{}_{jk}+1-\frac{1}{D+1}\left(4\delta^{}_{jk}+2\right)\right]-\overline{O}^2\simeq \frac{2}{D}\left(\overline{O^2}-\overline{O}^2\right)\,,
\end{equation}
and similarly for the off-diagonal elements
\begin{equation}
    \overline{(O^{}_{mn})^2}-\overline{O^{}_{mn}}^2=\sum_{j,k} O^{}_jO^{}_k \overline{c^n_j c^m_k c^m_j c^n_k}=\sum_{j,k}O^{}_jO^{}_k\frac{D+1}{D(D-1)(D+2)}\left[\delta^{}_{jk}-\frac{1}{D+1}\left(\delta^{}_{jk}+1\right)\right]\simeq \frac{1}{D}\left(\overline{O^2}-\overline{O}^2\right)\,.
\end{equation}
Therefore, to the leading order in $1/D$, one recovers the form of Eq.~\eqref{eq:RMT_Result_MatrixElements} but with $R^{O}_{mn}$ for the GOE having zero mean and variance 1 (2) for $m\neq n$ ($m=n$). Namely, the {\bf variance} of the diagonal matrix elements is {\bf twice} that of the off-diagonal ones.

\subsection{Entanglement entropy}

Consider a system in a pure state $\ket{\psi}$ represented by the density matrix $\hat \rho =\ket{\psi}\bra{\psi}$. Given a bipartition of the system into subsystem $A$ and its complement $B$, the reduced density matrix for subsystem $A$ is obtained by taking the partial trace over its complement $B$
\begin{equation}
    \hat \rho^{}_A=\Tr^{}_{B}(\hat \rho)=\sum_{j=1}^{D_B}\bra{b^{}_j}\hat \rho \ket{b^{}_j}\,, 
\end{equation}
where $\{ \ket{b^{}_j}\}$ is a basis of states for the complement $B$. The {\bf von Neumann entanglement entropy} (we refer to it simply as the {\bf entanglement entropy} in what follows) between subsystem $A$ and its complement $B$ is defined as 
\begin{equation}
    S\!^{}_{A}(\hat \rho^{}_A)=-\Tr(\hat \rho^{}_A \ln \hat\rho^{}_A)\,.    
\end{equation}
In this chapter, we focus on lattice chains with $L$ sites, so we refer to $L$ as the volume and label the number of lattice sites in subsystems $A$ and $B$ as $L^{}_A$ and $L^{}_B=L-L^{}_A$, respectively. For our subsystem $A$, we define the {\bf subsystem fraction} as $f=L^{}_A/L$. Also, our Hamiltonian~\eqref{eq:Spin1_Hamiltonian} conserves the total magnetization, which is equivalent to particle-number conservation in itinerant models. The {\bf magnetization per site} $\m=M/L$ in the spin-$1$ model and the corresponding {\bf filling fraction} $\n=N/L$ in itinerant models are related via $\n=\m+1$. 

While initially studied mainly in the context of ground states of local many-body Hamiltonians, for which the entanglement entropy typically grows with the area of the boundary of the subsystem in gapped systems (or with the logarithm of the length in critical one-dimensional systems)~\cite{amico_fazio_08, eisert2010colloquium}, much has been learned in recent years about the volume-law entanglement of typical (highly excited) energy eigenstates of such Hamiltonians~\cite{bianchi_2022}, and the crossover from the area-law ground-state entanglement to the volume-law excited-state entanglement has also begun to be understood~\cite{qiang_barthel_21, qiang_barthel_21a, Miao2022eigenstate}. In contrast to the (nearly) {\bf maximal volume-law entanglement} exhibited by typical energy eigenstates of quantum-chaotic Hamiltonians~\cite{vidmar_rigol_2017, garrison_grover_18}, studies of many-body eigenstates of {\bf translationally invariant} quadratic (noninteracting) fermionic Hamiltonians revealed a volume-law entanglement that is qualitatively different~\cite{storms_singh_14, vidmar_17, hackl2019average}. In quantum-chaotic Hamiltonians the {\bf coefficient of the volume-law term is constant and maximal}, whereas in the quadratic fermionic Hamiltonians it depends on the ratio between the volume of the subsystem and the volume of the entire system. Furthermore, typical eigenstates of integrable interacting models were found to exhibit a qualitatively (and quantitatively) similar behavior to those of quadratic models~\cite{leblond_2019}, which motivated the conjecture in Ref.~\cite{leblond_2019} that the entanglement entropy of typical energy eigenstates of quantum many-body systems can be used as a diagnostic of quantum chaos and integrability.

The behavior of the entanglement entropy of typical energy eigenstates of quantum-chaotic and integrable Hamiltonians has been understood using RMT and, specifically, using Haar-random ensembles. For a tutorial on this topic, which also has a comprehensive list of references, see Ref.~\cite{bianchi_2022}. Here we list the nonvanishing terms that can occur as $L\rightarrow\infty$: (i) Haar-random states have a maximal volume-law entanglement, an $\O(1)$ subleading correction at $f=1/2$~\cite{page_1993}, and capture leading behavior of typical energy eigenstates of quantum-chaotic Hamiltonians; (ii) Haar-random states with a fixed number of particles have a maximal (allowed by $\n$) volume-law entanglement, can exhibit $\O(\!\sqrt{L})$ and $\O(1)$ subleading corrections depending on $f$ and $\n$~\cite{bianchi_2022, vidmar_rigol_2017, yauk_hackl_2024}, and capture leading behavior of typical energy eigenstates of quantum-chaotic Hamiltonians with particle-number conservation; (iii) Haar-random states with fixed total spin and magnetization have a maximal (allowed by $\m$ and the total spin density) volume-law entanglement, can exhibit $\O(\!\sqrt{L})$, $\O(\ln L)$, and $\O(1)$ subleading corrections depending on $f$, $\m$, and the total spin density~\cite{patil_2023, bianchi_2024, chakraborty2025}, and capture leading behavior of typical energy eigenstates of quantum-chaotic Hamiltonians with SU(2) symmetry; (iv) Haar-random fermionic Gaussian states without and with a fixed number of particles have a volume-law entanglement whose prefactor depends on $f$, can exhibit $\O(1)$ subleading corrections~\cite{bianchi_2022, lydzba2020entanglement, bianchi2021page}, and capture leading behavior of typical energy eigenstates of translationally invariant and random quadratic Hamiltonians. The differences between the entanglement entropy of the appropriate Haar-random ensembles and typical eigenstates of quantum-chaotic local Hamiltonians have been found to occur at the level of the $\O(1)$ terms~\cite{huang_21, haque_22, kliczkowski2023average, rodriguez-nieva_24, rodriguez-nieva_25}.

We discuss in more detail case (ii) above as it is relevant to our Hamiltonian~\eqref{eq:Spin1_Hamiltonian}. For Haar-random pure states with fixed total particle number $N$, the average entanglement entropy of subsystem $A$ is given by~\cite{bianchi_2019} 
\begin{equation}
\label{eq:entropy_fixed_N_exact_sum}
\braket{S\!^{}_{A}}_N=\sum_{N^{}_A}\varrho^{}_{N^{}_A}\varphi^{}_{N^{}_A}\quad\text{with}\quad\varrho^{}_{N^{}_A}=\frac{d^{}_Ad^{}_B}{d^{}_N}\,,\quad
\varphi^{}_{N^{}_A}=\Psi(d^{}_N\!+\!1)-\Psi(\max[d^{}_A,d^{}_B]\!+\!1)
-\min\left[\frac{d^{}_A-1}{2d^{}_B},\frac{d^{}_B-1}{2d^{}_A}\right]\,,
\end{equation}
where $d^{}_A=\mathcal{D}(N^{}_A,L^{}_A)$, $d^{}_B=\mathcal{D}(N-N^{}_A,L-L^{}_A)$, and $d^{}_N=\mathcal{D}(N,L)$ are the respective dimensions of the Hilbert space in subsystem $A$ (with $N^{}_A$ particles), subsystem $B$ (with $N^{}_B=N-N^{}_A$ particles), and the entire system (with $N=N^{}_A+N^{}_B$ particles), and $\Psi(x)=\frac{\Gamma'(x)}{\Gamma(x)}$ is the digamma function. As $L\rightarrow\infty$, the asymptotic scaling of the exact sum for $\braket{S\!^{}_{A}}_N$ in Eq.~\eqref{eq:entropy_fixed_N_exact_sum} was computed in Ref.~\cite{bianchi_2022} for the case of two states per site \big[for which $\mathcal{D}(N,L)=\binom{L}{N}=\frac{L!}{(L-N)!N!}$\big] and in full generality in Ref.~\cite{yauk_hackl_2024} for models with arbitrary local Hilbert space dimensions. For $f\leq1/2$, it is given by:
\begin{equation}
    \label{eq:entropy_N_asymptotic}
        \braket{S\!^{}_{A}}_N=\s(\n) fL-\frac{|\,\s'(\n)|}{\sqrt{2\pi|\,\s''(\n)|}}\sqrt{L}\,\delta^{}_{f,\frac{1}{2}}
        +\frac{1}{2}\left[f+\ln(1-f)-\delta^{}_{f,\frac{1}{2}}\delta^{}_{\n,\n^*}\right]+o(1)\,,\quad\text{where}\quad \mathcal{D}(N,L)\propto \frac{1}{\sqrt{L}}\exp[\s(\n)L].
\end{equation}
Here, the function $\s(\n)$ is the coefficient of $L$ in the exponent of the asymptotic Hilbert space dimension, which can be determined using the generating function method outlined in Ref.~\cite{yauk_hackl_2024}. The quantities $\s'(\n)$ and $\s''(\n)$ are the first- and second-order derivatives with respect to $\n$, $\n^*$ is defined using $\s'(\n^*)=0$, and $o(1)$ denotes corrections that vanish as $L\rightarrow\infty$. For our spin-$1$ model~\eqref{eq:Spin1_Hamiltonian} with three states per site, the dimension of the sector with magnetization $M$ (or, equivalently, particle number $N=M+L$) for $L$ sites can be obtained from the sum
\begin{equation}\label{eq:dimension spin-1}
    \mathcal{D}(N,L)=\sum_{k=0}^{\lfloor N/3\rfloor}(-1)^k\binom{L}{k}\binom{N-3k+L-1}{L-1}\,,
\end{equation} 
where $0\leq N\leq 2L$, with the function $\s(\n)$ given by~\cite{yauk_hackl_2024}
\begin{equation}
     \s(\n) = (\n-2)\ln(2-\n)+(\n-1)\ln(2) 
     +\ln\left(7-3\n+\sqrt{1-3\n(\n-2)}\right)  
    -\n\ln\left(\n-1+\sqrt{1-3\n(\n-2)}\right)\,. \label{eq:Beta_Spin-1} 
\end{equation}
In particular, $\s(\n)$ in Eq.~\eqref{eq:Beta_Spin-1} takes its maximum value of $\ln3$ at $\n^*=1$ (half filling or zero magnetization $\m=M/L=0$). We stress that while the analytical results are obtained for the {\bf average entanglement entropies} of Haar-random states, these averages describe the {\bf typical values} because the variances vanish exponentially fast with increasing $L$~\cite{bianchi_2022}. The latter has been verified to be the case in quantum-chaotic local Hamiltonians, while integrable ones were found to exhibit a variance that vanishes polynomially fast with increasing $L$~\cite{rafal_24}.

\begin{figure}[t]
    \centering
    \includegraphics[width=0.79\linewidth]{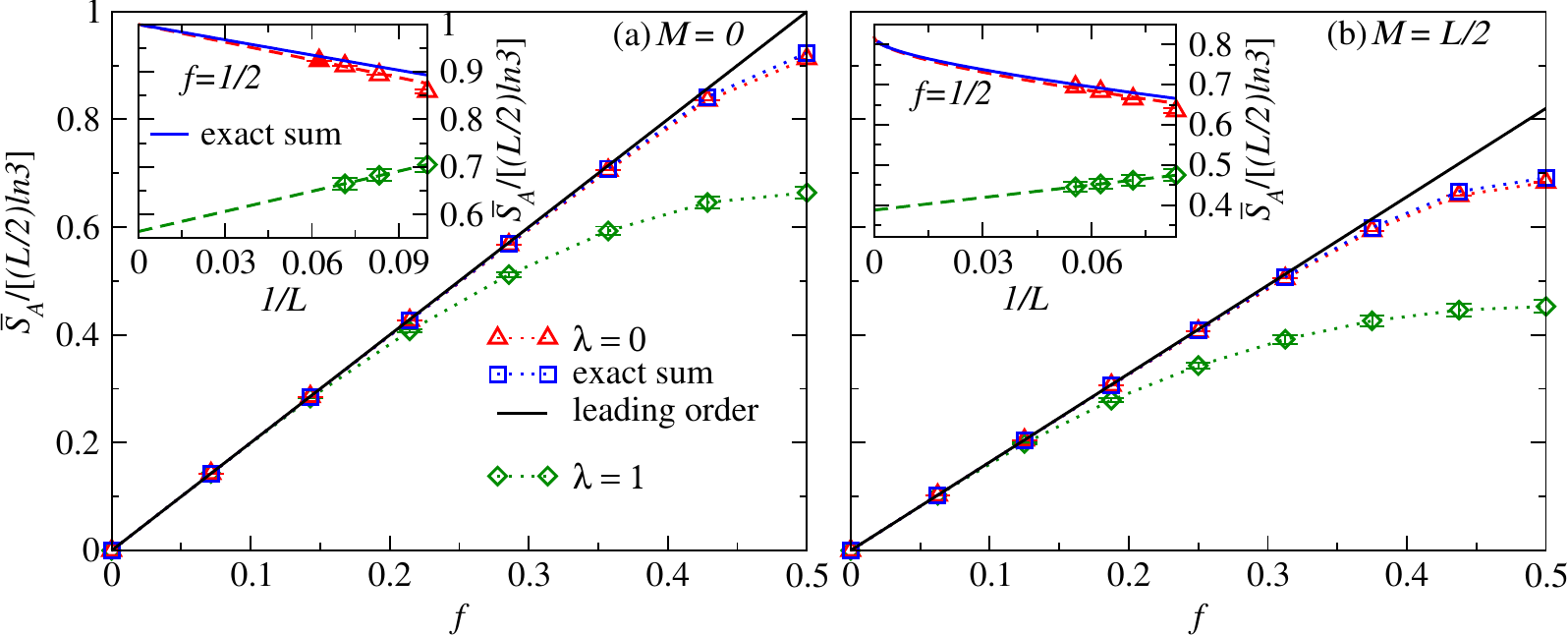}
    \caption{{\bf Entanglement entropy}: Average bipartite entanglement entropy versus the subsystem fraction $f=L^{}_A/L$ for (a) $M=0$ and (b) $M=L/2$ in chains with $L=14,16$, respectively. We show numerical results for the average $\bar S\!^{}_{A}$ obtained for 100 mid-spectrum Hamiltonian eigenstates of the spin-$1$ $XXZ$ model~\eqref{eq:Spin1_Hamiltonian} at the quantum-chaotic ($\lambda=0$) and integrable ($\lambda=1$) points. We also show results from the exact sum for random pure states $\braket{S\!^{}_{A}}_N$ in Eq.~\eqref{eq:entropy_fixed_N_exact_sum}, and the leading-order term for $\braket{S\!^{}_{A}}_N$ in Eq.~\eqref{eq:entropy_N_asymptotic} with $\s(\n)$ in Eq.~\eqref{eq:Beta_Spin-1}. The insets show the finite-size scaling of $\bar S\!^{}_{A}/[(L/2)\ln3]$ at $f=1/2$ for the quantum-chaotic and integrable Hamiltonian eigenstates, and for the exact sum in Eq.~\eqref{eq:entropy_fixed_N_exact_sum}. The dashed lines through the results for the Hamiltonian eigenstates are obtained by a fit to the three largest chains. In the quantum-chaotic case we use a single-parameter fit $c^{}_1+p/L$ (a) [$c^{}_1+c^{}_2/\sqrt{L}+p/L$ (b)], where $c^{}_1$ ($c^{}_1$ and $c^{}_2$) are set by Eq.~\eqref{eq:entropy_N_asymptotic} with $\s(\n)$ in Eq.~\eqref{eq:Beta_Spin-1} and $p$ is the fitting parameter. In the integrable case, we use a two-parameter fit $p'+p/L$. Adapted from Ref.~\cite{yauk_hackl_2024}.}
    \label{fig:page-curves}
\end{figure}

In Fig.~\ref{fig:page-curves} we show the average bipartite entanglement entropy $\bar S\!^{}_{A}$ for mid-spectrum Hamiltonian eigenstates of the spin-$1$ $XXZ$ model~\eqref{eq:Spin1_Hamiltonian} in the $M=0$ and $M=L/2$ magnetization sectors at the quantum-chaotic ($\lambda=0$) and integrable ($\lambda=1$) points. We compare the numerical results with those for the Haar-random average $\braket{S\!^{}_{A}}_N$ for: (i) the exact sum in Eq.~\eqref{eq:entropy_fixed_N_exact_sum} and (ii) the leading-order volume-law term in the asymptotic expression in Eq.~\eqref{eq:entropy_N_asymptotic} with $\s(\n)$ in Eq.~\eqref{eq:Beta_Spin-1}. The results for $\bar S\!^{}_{A}$ corresponding to the quantum-chaotic eigenstates are close to the results for the exact sum $\braket{S\!^{}_{A}}_N$ in Eq.~\eqref{eq:entropy_fixed_N_exact_sum} for Haar-random pure states, and both are well described by the leading-order term $\s(\n)fL$ in the asymptotic expression for $\braket{S\!^{}_{A}}_N$ in Eq.~\eqref{eq:entropy_N_asymptotic} for $f\lesssim0.35$. This agreement is remarkable given the relatively small system sizes considered. On the other hand, $\bar S\!^{}_{A}$ for the integrable Hamiltonian eigenstates deviates from the exact sum $\braket{S\!^{}_{A}}_N$ as $f$ increases from $f=0$. The difference between $\bar S\!^{}_{A}$ for the quantum-chaotic and the integrable Hamiltonian eigenstates can be seen more clearly in the insets of Fig.~\ref{fig:page-curves} where we show finite-size scaling analyses for $f=1/2$. $\bar S\!^{}_{A}$ for the quantum-chaotic Hamiltonian eigenstates is consistent with $\braket{S\!^{}_{A}}_N$ for Haar-random states with deviations at the level of the subleading $\O(1)$ corrections, while $\bar S\!^{}_{A}$ for the integrable Hamiltonian eigenstates already differs from $\braket{S\!^{}_{A}}_N$ at the level of the leading volume-law term. This provides evidence supporting the expectation that entanglement entropy can serve as a diagnostic of quantum chaos and integrability, and shows that RMT captures the leading behavior of the results obtained for Hamiltonian eigenstates even for the small system sizes accessible to full exact diagonalization.

\section{Eigenstate Thermalization Hypothesis (ETH)}\label{sec: ETH}

Thermalization refers to the fact that, after equilibration, the expectation values of observables (which are represented by Hermitian operators) are described by traditional statistical mechanics ensembles. As mentioned before, thermalization in generic isolated quantum many-body systems is understood to be a consequence of eigenstate thermalization~\cite{deutsch_1991, srednicki_1994, rigol_2008}, whereby highly excited energy eigenstates appear thermal from the perspective of physical observables. The corresponding ETH ansatz for the matrix elements $O^{}_{mn}\equiv \langle \psi^{}_m|\hat O| \psi^{}_n \rangle$ of a few-body intensive observable $\hat O$ in the energy eigenstates $\hat H\ket{\psi^{}_m}=E^{}_m\ket{\psi _m}$ within each symmetry sector reads~\cite{dalessio_quantum_2016, srednicki_99}:
\vspace{-0.1cm}
\begin{equation}\label{eq:ETH_ansatz}
 O^{}_{mn}=O(E^{}_m)\delta^{}_{mn}\,+\,\frac{1}{\sqrt{\Omega(\bar E^{}_{mn})}}f^{}_O(\bar E^{}_{mn},\omega^{}_{mn})R^{O}_{mn}\,,
 \end{equation}
\vspace{-0.3cm}

\noindent where $\bar E^{}_{mn}=(E^{}_m+E^{}_n)/2$ is the average energy, $\omega^{}_{mn}=E^{}_m-E^{}_n$ is the energy difference, $O(E^{}_m)$ and $f^{}_O(\bar E^{}_{mn},\omega^{}_{mn})$ are smooth functions, $\Omega^{}(\bar E^{}_{mn})\propto \exp[S(\bar E^{}_{mn})]$ is the density of states with $S(\bar E^{}_{mn})$ the thermodynamic entropy at the energy $\bar E^{}_{mn}$, and $R^{O}_{mn}$ are random numbers with the same first and second moments we found in Sec.~\ref{sec:matelemobsRMT}, as determined by the symmetries of $\hat H$ and $\hat O$. The ETH ansatz in Eq.~\eqref{eq:ETH_ansatz} captures the first two moments of the distribution of $O^{}_{mn}$. There exist higher-order correlations between the $O^{}_{mn}$ that are required to describe quantities such as the out-of-time-order correlation functions. They have been recently included in an extended ETH; see Sec.~\ref{sec:higher}. 

Note that the ETH ansatz~\eqref{eq:ETH_ansatz} resembles the expression for the matrix elements of Hermitian operators in RMT~\eqref{eq:RMT_Result_MatrixElements}. The diagonal matrix elements $O^{}_{mm}$ in the ETH encode the mean value $O(E^{}_m)$ with fluctuations about the mean that vanish as $\Omega^{-1/2}(E^{}_{m})$, in analogy with $\bar O$ and its fluctuations that vanish as $D^{-1/2}$ in RMT. Similarly, the magnitude of the off-diagonal matrix elements in the ETH vanishes as $\Omega^{-1/2}(\bar E^{}_{mn})$ in analogy with $D^{-1/2}$ in RMT. The ETH ansatz extends the RMT expression to account for structure in the energy spectrum of physical Hamiltonians, which remarkably for the first two moments of the distribution of $O^{}_{mn}$ is all encoded in the smooth functions $O(E^{}_m)$, $\Omega(\bar E^{}_{mn})$, and $f^{}_O(\bar E^{}_{mn},\omega^{}_{mn})$. Only in a vanishingly small energy window in which the functions $O(E^{}_m)$, $\Omega(\bar E^{}_{mn})$, and $f^{}_O(\bar E^{}_{mn},\omega^{}_{mn})$ are constant does one recover the structureless RMT behavior, see, e.g.,~Refs.~\cite{brenes_pappalardi_21, Dymarsky_2022, wang_lamann_22}.  

\subsection{Thermalization}

To see how {\bf statistical mechanics} behavior emerges in an {\bf isolated quantum system}~\cite{dalessio_quantum_2016}, we consider a system prepared in an initial state $\ket{\Psi^{}_0}$ that evolves unitarily under a time-independent Hamiltonian $\hat H$, $\hat H\ket{\psi^{}_m}=E^{}_m\ket{\psi^{}_m}$:
\vspace{-0.1cm}
\begin{equation}
    \ket{\Psi(t)}=\sum_{m=1}^Dc^{}_me^{-iE^{}_mt}\ket{\psi^{}_m},\quad\text{with}\quad c^{}_m=\braket{\psi^{}_m|\Psi^{}_0},\quad\bar E\equiv\braket{\Psi^{}_0|\hat H|\Psi^{}_0}= \O(L),\quad\text{and}\quad \delta E^{}_0\equiv\sqrt{\braket{\Psi^{}_0|\hat H^2|\Psi^{}_0}-\braket{\Psi^{}_0|\hat H|\Psi^{}_0}^2} =\O(\sqrt{L}).
\end{equation}
\vspace{-0.2cm}

\noindent The fact that, for physically relevant initial states, the energy fluctuations $\delta E^{}_0$ are in general {\bf subextensive} and $\O\!\left(\!\sqrt{L}\right)$ can be shown on general grounds (see the Supplementary Discussion in Ref.~\cite{rigol_2008}). The expectation value of an observable $\hat O$ at time $t$ can then be written as
\vspace{-0.1cm}
\begin{equation}\label{eq:O(t)}
    O(t)\equiv\braket{\Psi(t)|\hat O|\Psi(t)}=\sum_{m=1}^D|c^{}_m|^2O^{}_{mm}+\underset{m \neq n}{\sum_{n,m=1}^D} c^*_mc^{}_n e^{i(E^{}_m-E^{}_n)t}O^{}_{mn}\,,
\end{equation}
\vspace{-0.3cm}

\noindent where $O^{}_{mn}\equiv \langle \psi^{}_m|\hat O| \psi^{}_n \rangle$ are the matrix elements in the energy eigenstates. The observable $\hat O$ is said to {\bf thermalize} if $O(t)$ equilibrates at the value predicted by statistical mechanics. By {\bf equilibration} we mean that $O(t)$ relaxes to its {\bf infinite-time average} $\overline{O(t)}$ and remains close to it at most times. For a generic $\hat H$ with no degeneracies (beyond accidental ones) after resolving all the symmetries, $\overline{O(t)}$ can be written as:
\vspace{-0.1cm}
\begin{equation}\label{eq:O(t)_diagonal_ensemble}
    \overline{O(t)}\equiv\lim_{t\rightarrow\infty} \frac{1}{t}\int_0^{t} O(t')dt'\simeq\sum_{m=1}^D |c^{}_m|^2 O^{}_{mm}=\Tr\left[\hat \rho^{}_\text{DE}\!\left(c^{}_1,c^{}_2,\dots,c^{}_D\right) \hat O\right]\,,\quad\text{where}\quad \hat \rho^{}_\text{DE}\!\left(c^{}_1,c^{}_2,\dots,c^{}_D\right)\equiv\sum_{m=1}^D |c^{}_m|^2\ket{\psi^{}_m}\bra{\psi^{}_m} 
\end{equation}
\vspace{-0.2cm}

\noindent is the so-called {\bf diagonal ensemble} (DE)~\cite{rigol_2008}. To obtain this result we used that the second sum in Eq.~\eqref{eq:O(t)} averages to zero (up to accidental degeneracies) due to the oscillating phase factors. Thermalization, as defined via the microcanonical ensemble (ME), requires that 
\vspace{-0.1cm}
\begin{equation}\label{eq:thermalizationTL}
\Tr\left[\hat \rho^{}_\text{DE}\!\left(c^{}_1,c^{}_2,\dots,c^{}_D\right) \hat O\right]\simeq\Tr\left[\hat \rho^{}_\text{ME}(\bar E) \hat O\right],\qquad \text{where}\qquad \hat \rho^{}_\text{ME}(\bar E)=\frac{1}{\mathcal{N}^{}_\text{ME}}\sum_{m=1}^{\mathcal{N}^{}_\text{ME}}\ket{\psi^{}_m}\bra{\psi^{}_m}
\end{equation}
\vspace{-0.2cm}

\noindent is the density matrix of the microcanonical ensemble, in which the $\sum_m$ runs over $\mathcal{N}^{}_\text{ME}$ eigenstates with $E^{}_m\simeq\bar E$, chosen such that $\Tr\left[\hat \rho^{}_\text{ME}(\bar E)\hat H\right]=\bar E$ and the energy fluctuations $\delta E^{}_\text{ME}$ are subextensive. We can see that thermalization is not to be taken for granted under unitary dynamics because $\hat \rho^{}_\text{DE}\!\left(c^{}_1,c^{}_2,\dots,c^{}_D\right)$ depends on the initial state through the coefficients $c^{}_m$ while $\hat \rho^{}_\text{ME}(\bar E)$ only depends on it through $\bar E$.

Using the ETH in Eq.~\eqref{eq:ETH_ansatz}, we can calculate the leading terms in the diagonal and microcanonical averages:
\vspace{-0.1cm}
\begin{align}\label{eq:traceOinDE}
\vspace{-0.1cm}
    \Tr\left[\hat \rho^{}_\text{DE}\!\left(c^{}_1,c^{}_2,\dots,c^{}_D\right) \hat O\right]&\simeq\sum_{m=1}^D |c^{}_m|^2 O(E^{}_m)=\sum_{m=1}^D |c^{}_m|^2 \Bigg[O(\bar E)+\left.\frac{\partial O}{\partial E}\right\lvert_{\bar E}\!\left(E^{}_m-\bar E\right)+\left.\frac{1}{2!}\frac{\partial^2 O}{\partial E^2}\right\lvert_{\bar E}\!\left(E^{}_m-\bar E\right)^2 +\ldots \Bigg]\,\approx O(\bar E)+\frac{1}{2}O''(\bar E)\delta E_0^2\,,\\ \label{eq:traceOinME}
    \Tr\left[\hat \rho^{}_\text{ME}(\bar E) \hat O\right]&=\frac{1}{\mathcal{N}^{}_\text{ME}}\sum_{m=1}^{\mathcal{N}^{}_\text{ME}} O^{}_{mm}\simeq\frac{1}{\mathcal{N}^{}_\text{ME}}\sum_{m=1}^{\mathcal{N}^{}_\text{ME}} O(E^{}_m)\approx O(\bar E)+ \frac{1}{2}O''(\bar E)\delta E_\text{ME}^2\,,
\end{align}
\vspace{-0.2cm}

\noindent where, to obtain the final result in Eq.~\eqref{eq:traceOinME}, we used the same expansion as in Eq.~\eqref{eq:traceOinDE}. Noting that the energy is extensive, we define its intensive counterpart $\epsilon\equiv E/L$ and the corresponding $o(\epsilon)\equiv O(L\epsilon)$. Using $\epsilon$ and $o(\epsilon)$, we find that:
\begin{equation}\label{eq:DerivativeScaling}
O''(\bar E)=\left.\frac{\partial^2 O(E)}{\partial E^2}\right|_{\bar E}=\left.\frac{\partial^2 o(\epsilon)}{\partial \epsilon^2}\right|_{\bar \epsilon}\frac{1}{L^2}=\O\left(L^{-2}\right)\,, 
\end{equation}
\vspace{-0.2cm}

\noindent for a smooth $o(\epsilon)$. Therefore, we see that for subextensive $\delta E_0^{}=\O(\sqrt{L})$ and $\delta E_\text{ME}^{}=\O(\sqrt{L})$ the second terms in Eqs.~\eqref{eq:traceOinDE} and~\eqref{eq:traceOinME} are subleading so that Eq.~\eqref{eq:thermalizationTL} holds thanks to the ETH. It then follows that differences between the l.h.s.~and the r.h.s.~of Eq.~\eqref{eq:thermalizationTL} are $\O(1/L)$ in finite systems, like the differences between different ensembles (microcanonical, canonical, and grand-canonical) in statistical mechanics.

Furthermore, $O(t)$ remains close to $\overline{O(t)}$ at most times since the mean-square temporal fluctuations
\begin{align}
    \lim_{t\rightarrow\infty} \frac{1}{t}\int_0^{t} dt'[O(t')]^2-\overline{O(t)}^2&= \lim_{t\rightarrow\infty} \frac{1}{t} \int_0^{t}dt' \sum_{m,n,p,q} c^*_mc^{}_nc^*_pc^{}_qe^{i(E^{}_m-E^{}_n+E^{}_p-E^{}_q)t'}O^{}_{mn}O^{}_{pq}- \sum_{m,p} |c^{}_m|^2|c^{}_p|^2O^{}_{mm}O^{}_{pp}\,\nonumber\\&\simeq \sum_{m,n\neq m} |c^{}_m|^2|c^{}_n|^2 |O^{}_{mn}|^2\sim |O^{}_{mn}|^2_\text{typ} \sum_{m,n\neq m} |c^{}_m|^2 |c^{}_n|^2\simeq |O^{}_{mn}|^2_\text{typ} \sim \Omega^{-1}(\bar E),
\end{align}
\vspace{-0.2cm}

\noindent vanish exponentially with the system size (neglecting accidental gap resonances). One can also see, by extending this analysis, that the contribution of terms due to accidental degeneracies to $\overline{O(t)}$ in Eq.~\eqref{eq:O(t)_diagonal_ensemble} is exponentially small and can be neglected.

\subsection{Diagonal ETH}\label{sec:diagonal-ETH}

\begin{figure}[!t]
    \centering
    \includegraphics[width=\linewidth]{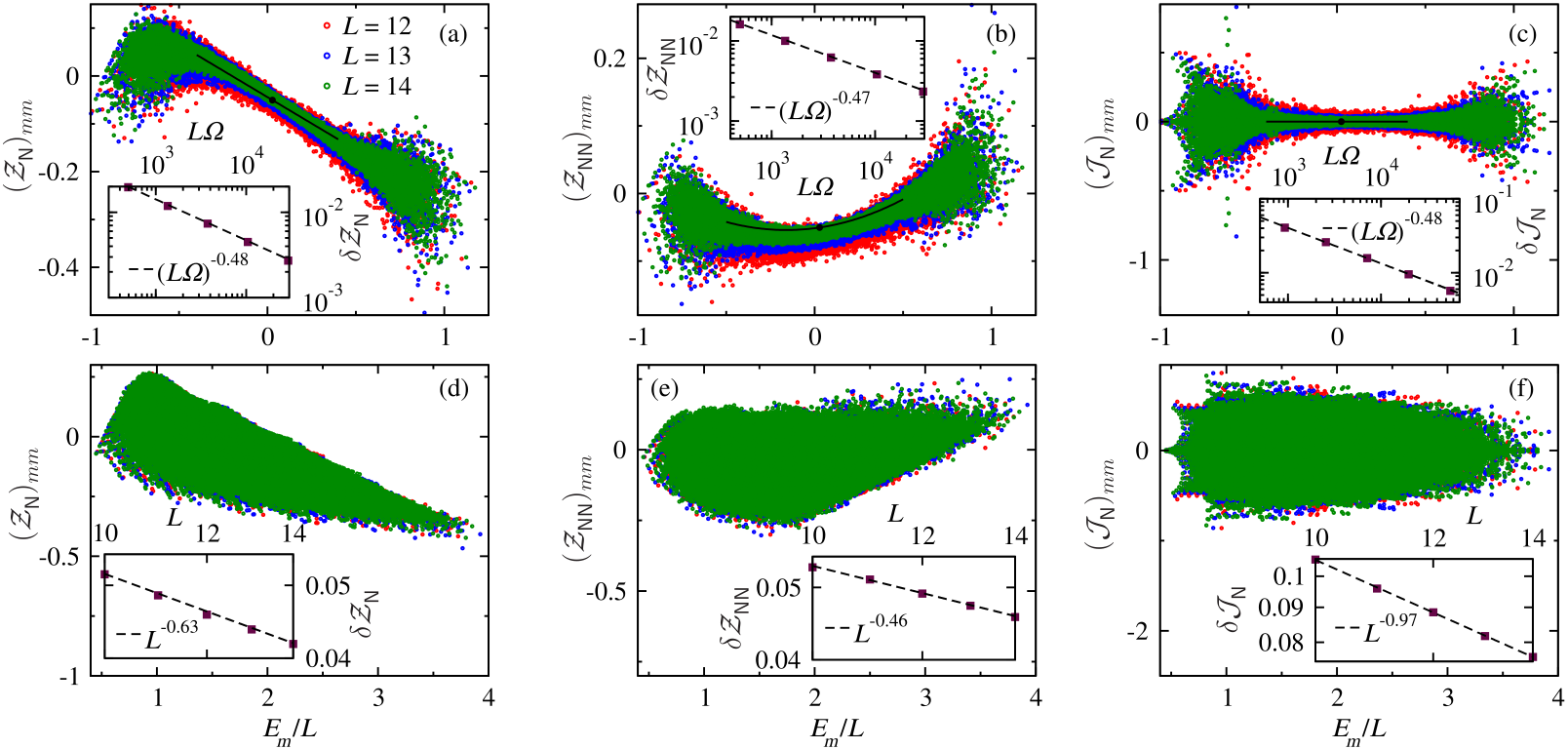}
    \caption{{\bf Diagonal matrix elements}: $\tO^{}_{mm}$ for the observables $\hat \tO=\hat \tZ^{}_{\Nt},\hat \tZ^{}_{\Nt\Nt}$, and $\hat \tJ^{}_{\Nt}$ in the energy eigenstates of the Hamiltonian~\eqref{eq:Spin1_Hamiltonian} within sectors with magnetization $M=0$ and $\mathbb{Z}^{}_2=\pm1$ ($\hat \tZ^{}_{\Nt},\hat \tZ^{}_{\Nt\Nt}$) and $M=1$ ($\hat \tJ^{}_{\Nt}$) and total quasimomenta $k\neq0,\pi$ plotted as functions of energy density $E^{}_m/L$ at the nonintegrable [(a)--(c)] and integrable [(d)--(f)] points. In the nonintegrable case, the leading-order behavior of the microcanonical function $\tO(E^{}_m)$ for $E^{}_m\approx \E$ is depicted by the black dots corresponding to the mean $\tO(\E)$ and the black lines centered at the black dots corresponding to the leading (as well as subleading in the case of $\hat \tZ^{}_{\Nt\Nt}$) order energy dependence. The insets show the finite-size scaling of the fluctuations $\delta \tO=\braket{|\tO^{}_{mm}-\overline{\tO^{}_{mm}}|}$ about the mean $\overline{\tO^{}_{mm}}$ calculated as a running average $\overline{(\ldots)}$ over $50$ states centered at $E^{}_m$, averaged $\braket{\ldots}$ over the central $50\%$ of the spectrum, plotted as a function of $L\Omega$ ($L$) in the nonintegrable (integrable) case where $\Omega$ is the average density of states in the central $50\%$ of the spectrum. The dashed lines in the nonintegrable (integrable) case are power-law fits $\propto (L\Omega)^{-\gamma}$ [$\propto L^{-\gamma}$] to the data with fitting parameter $\gamma$ reported in the legends.}
    \label{fig:diagonal-matrix-elements}
\end{figure}

In this section, we study the diagonal matrix elements of the observables $\hat \tO=\hat \tZ^{}_{\Nt},\hat \tZ^{}_{\Nt\Nt},$ and $\hat \tJ^{}_{\Nt}$ [see Eq.~\eqref{eq:observables}] within the energy eigenstates of $\hat H$~\eqref{eq:Spin1_Hamiltonian} in total quasimomentum sectors with $k\neq0,\pi$ belonging to total magnetization sectors with $M=0$ (and $\mathbb{Z}^{}_2=\pm1$) for $\hat \tZ^{}_{\Nt},\hat \tZ^{}_{\Nt\Nt}$ and $M=1$ for $\hat \tJ^{}_{\Nt}$.\footnote{We take $M=1$ instead of $M=0$ for $\hat \tJ^{}_{\Nt}$ because $(\tJ^{}_{\Nt})^{}_{mm}$ vanishes in the $M=0$ sector since $\hat \tJ^{}_{\Nt}$ is odd under the $\mathbb{Z}^{}_2$ spin inversion symmetry.} In Figs.~\ref{fig:diagonal-matrix-elements}(a)--\ref{fig:diagonal-matrix-elements}(c) we plot the diagonal matrix elements $\tO^{}_{mm}$ vs $E^{}_m/L$ within the energy eigenstates of the Hamiltonian~\eqref{eq:Spin1_Hamiltonian} at the {\bf nonintegrable point} ($\lambda=0$). The {\bf support} of the matrix elements {\bf shrinks rapidly} with increasing system size at fixed $E^{}_m/L$ consistent with the emergence of the smooth microcanonical function $\tO(E^{}_m)$ predicted by the ETH ansatz~\eqref{eq:ETH_ansatz}. On the other hand, at the {\bf integrable point} ($\lambda=1$) the support of \{$\tO^{}_{mm}$\} in Figs.~\ref{fig:diagonal-matrix-elements}(d)--\ref{fig:diagonal-matrix-elements}(f) {\bf does not shrink} with increasing system size, indicating a violation of the ETH.

At the nonintegrable point, the microcanonical function $\tO(E^{}_m)$ in the sector with magnetization $M$, for $E^{}_m$ close to the mean energy at infinite temperature $\E=\braket{\hat H}_M=\Tr^{}_M(\hat H)/\Tr^{}_M(\mathbbm{1})$ can be written as [using $\epsilon\equiv E/L$ and $o(\epsilon)\equiv \tO(L\epsilon)$ in the second expression]
\begin{equation}\label{eq:Taylor-expansion-diagonal}
    \tO(E^{}_m)= \tO(\E)+\left.\frac{\partial \tO(E^{}_m)}{\partial E^{}_m}\right\lvert_{E^{}_m=\E} \!\!\!(E^{}_m-\E)+\ldots\,\quad\implies\quad o(\epsilon^{}_m)= o(\epsilon^{}_{\bb=0})+\left.\frac{\partial o(\epsilon^{}_m)}{\partial \epsilon^{}_m}\right\lvert_{\epsilon^{}_m=\epsilon^{}_{\bb=0}}\!\!\!(\epsilon^{}_m-\epsilon^{}_{\bb=0})+\ldots\,,
\end{equation}
where $\tO(\E)=\braket{\hat \tO}_M=\Tr^{}_M(\hat \tO)/\Tr^{}_M(\mathbbm{1})$ is the infinite-temperature ($\bb=0$) expectation value of $\hat \tO$. For $o(\epsilon^{}_m)$, the first nonvanishing expansion coefficient of $(\epsilon^{}_m-\epsilon^{}_{\bb=0})^n$, for $n>0$, can be written in terms of joint cumulants $\braket{\ldots}_c$ evaluated at $\bb=0$~\cite{capizzi_poletti_25}:
\begin{equation}\label{eq:linear-in-energy coefficient}
    \frac{1}{n!}\left.\frac{\partial^n o(\epsilon^{}_m)}{\partial \epsilon_m^n}\right\lvert_{\epsilon^{}_m=\epsilon^{}_{\bb=0}}=\frac{1}{n!}\frac{\braket{\hat H^n\hat \tO}_c}{(\braket{\hat H^2}_c/L)^n}\,,\qquad\text{which for $n=1$ reads:}\qquad\frac{\braket{\hat H\hat \tO}_c}{(\braket{\hat H^2}_c/L)}=\frac{\braket{\hat H\hat \tO}_M-\braket{\hat H}_M\braket{\hat \tO}_M}{\braket{\hat H^2}_M-\braket{\hat H}^2_M}L\,.
\end{equation}

To gain intuition into the leading-order behavior of the microcanonical function $\tO(E^{}_m)$ in a sector with fixed $M$ with dimension $D_M^L\equiv \mathcal{D}(M+L,L)$~\eqref{eq:dimension spin-1}, it is instructive to first consider the calculation of the leading behavior of $\tO(E^{}_m)$ over the entire $3^L$-dimensional Hilbert space. In the entire Hilbert space, the operators $\hat \tO=\hat \tZ^{}_{\Nt},\hat \tZ^{}_{\Nt\Nt},\hat \tJ^{}_{\Nt}$ as well as the Hamiltonian $\hat H$ (with $\lambda=0$) are traceless, therefore $\E=\braket{\hat H}=0$ and $\tO(\E)=0$. Since $\hat \tZ^{}_{\Nt}$ is part of $\hat H$, it overlaps with $\hat H$ at the linear order, i.e.:
\begin{equation}\label{eq:full_cumulant_ZN}
    \braket{\hat H \hat \tZ^{}_{\Nt}}_c=\braket{\hat H \hat \tZ^{}_{\Nt}}-\braket{\hat H}\braket{\hat \tZ^{}_{\Nt}}=-\frac{\Delta}{L}\sum_{j=1}^L\braket{(\hat S^z_{\!j}\hat S^z_{\!j+1})(\hat S^z_{\!j}\hat S^z_{\!j+1})}=-\frac{\Delta}{L}L \frac{\Tr[(\hat S^z)^2\otimes(\hat S^z)^2\otimes\mathbbm{1}^{\otimes L-2}]}{3^L}=-\Delta \frac{2\cdot2\cdot3^{L-2}}{3^L}=-\Delta \left(\frac{2}{3}\right)^2\,,  
\end{equation}
where we used that $\braket{\hat H},\braket{\hat \tZ^{}_{\Nt}}=0$ and the traces $\Tr[(\hat S^z)^2]=2$ and $\Tr[\mathbbm{1}]=3$ on a single site. The operator $\hat \tZ^{}_{\Nt\Nt}$ does not overlap with $\hat H$ at the linear order, but does so at the second order, i.e.:
\begin{align}\label{eq:full_cumulant_ZNN}
    \braket{\hat H^2 \hat \tZ^{}_{\Nt\Nt}}_c&=\braket{\hat H^2 \hat \tZ^{}_{\Nt\Nt}}-\braket{\hat H^2}\braket{\hat \tZ^{}_{\Nt\Nt}}-2\braket{\hat H}\braket{\hat H\hat \tZ^{}_{\Nt\Nt}}+2\braket{\hat H}^2\braket{\hat \tZ^{}_{\Nt\Nt}}=\frac{2\Delta^2}{L}\sum_{j=1}^L \braket{(\hat S^z_{\!j}\hat S^z_{\!j+1})(\hat S^z_{\!j+1}\hat S^z_{\!j+2})(\hat S^z_{\!j}\hat S^z_{\!j+2})}\,\nonumber\\&=2\Delta^2\frac{\Tr[(\hat S^z)^2\otimes(\hat S^z)^2\otimes(\hat S^z)^2\otimes\mathbbm{1}^{\otimes L-3}]}{3^L}= 2\Delta^2\left(\frac{2}{3}\right)^3\,,
\end{align}
where only $\braket{\hat H^2 \hat \tZ^{}_{\Nt\Nt}}$ contributes because $\braket{\hat H}=\braket{\hat \tZ^{}_{\Nt\Nt}}=0$. Since the operator $\hat \tJ^{}_{\Nt}$ is odd under parity $\hat P$ (space-reflection), i.e., $\hat P\hat \tJ^{}_{\Nt}\hat P=-\hat \tJ^{}_{\Nt}$ and $\hat H$ commutes with $\hat P$, one has that for any integer power $n\ge0$,
\begin{equation}\label{eq:full_cumulant_JN}
     \braket{\hat H^n \hat \tJ^{}_{\Nt}}=\braket{\hat H^n \hat P\hat P\hat \tJ^{}_{\Nt}\hat P\hat P}=\frac{\Tr[(\hat P\hat H^n \hat P)(\hat P\hat \tJ^{}_{\Nt}\hat P)]}{3^L}=-\frac{\Tr[\hat H^n \hat \tJ^{}_{\Nt}]}{3^L}=-\braket{\hat H^n \hat \tJ^{}_{\Nt}}=0\,,
\end{equation}
which implies that $\braket{\hat H^n \hat \tJ^{}_{\Nt}}_c=0$. Lastly, the variance of $\hat H$ is given by
\begin{equation}\label{eq:full_cumulant_H2}
     \braket{\hat H^2}_c=\braket{\hat H^2}-\braket{\hat H}^2=\sum_{j=1}^L\braket{(\hat S^x_{\!j}\hat S^x_{\!j+1})^2+(\hat S^y_{\!j}\hat S^y_{\!j+1})^2+\Delta^2(\hat S^z_{\!j}\hat S^z_{\!j+1})^2}=L(1+1+\Delta^2) \frac{2\cdot2\cdot3^{L-2}}{3^L}=L(2+\Delta^2) \left(\frac{2}{3}\right)^2\,.
\end{equation}
Combining the results in Eqs.~\eqref{eq:full_cumulant_ZN}--\eqref{eq:full_cumulant_H2} using Eq.~\eqref{eq:linear-in-energy coefficient}, the leading-order behavior of $\tO(E^{}_m)$~\eqref{eq:Taylor-expansion-diagonal} over the entire $3^L$-dimensional Hilbert space can be written as
\begin{equation}
    \tZ^{}_{\Nt}(E^{}_m)\approx -\frac{\Delta}{2+\Delta^2}\left(\frac{E^{}_m}{L}\right)\,,\quad \tZ^{}_{\Nt\Nt}(E^{}_m)\approx \frac{3\Delta^2}{2(2+\Delta^2)^2}\left(\frac{E^{}_m}{L}\right)^2\,,\quad \tJ^{}_{\Nt}(E^{}_m)=0\,.
\end{equation}

Returning to our setup of interest with magnetization conservation, the leading-order behavior of $\tO(E^{}_m)$ with $E^{}_m\approx \E$ in a sector with fixed total $M$ can be obtained as follows. We first note two recursion relations that allow one to express $D^{L}_{M}$ in terms of $D^{L-1}_{M}$ and $D^{L-1}_{M\pm1}$:
\begin{equation}\label{eq:recursion_sum_dim}
    D_M^L=\sum_{m_i=-1}^1 D^1_{m_i}D^{L-1}_{M-m_i}=D^{L-1}_{M+1}+D^{L-1}_M+D^{L-1}_{M-1}\,, \qquad \frac{M}{L}D^L_M=\Tr^{}_M(\hat S^z_{\!i})=\sum_{m_i=-1}^1 m_iD^1_{m_i}D^{L-1}_{M-m_i}=D_{M-1}^{L-1}-D_{M+1}^{L-1}\,.
\end{equation}
The first one follows from $\Tr^{}_M(\mathbbm{1})=D_M^L$, where we evaluated $\Tr^{}_M(\mathbbm{1})$ by partitioning the system into a single site $i$ with magnetization $m_i\in\{-1,0,1\}$ and the complement comprising $L-1$ sites with magnetization $M-m_i$. The second one follows from the definition of magnetization, i.e., $M D^L_M=\Tr^{}_M(\hat M)\equiv\Tr^{}_M(\sum_i\hat S^z_{\!i})=L\Tr^{}_M(\hat S^z_{\!i})$.

The expectation value $\tO(\E)$ for $\hat \tO=\hat \tZ^{}_{\Nt}$ ($\hat \tZ^{}_{\Nt\Nt}$) corresponds to $\braket{\hat S^z_{\!i}\hat S^z_{\!j}}_M$ with $j=i+1$ ($i+2$). $\Tr^{}_M(\hat S^z_{\!i}\hat S^z_{\!j})$ for $j\neq i$ is given by
\begin{equation}
    \sum_{m_i,m_j=-1}^1m_im_jD^{L-2}_{M-(m_i+m_j)}=(D^{L-2}_{M-2}-D^{L-2}_{M})-(D^{L-2}_{M}-D^{L-2}_{M+2}) =\frac{M-1}{L-1}D^{L-1}_{M-1}-\frac{M+1}{L-1}D^{L-1}_{M+1}=\frac{M^2}{L(L-1)}D_M^L-\frac{1}{L-1}(D^{L}_M-D^{L-1}_M)\,,
\end{equation}
where we used the second (first) relation from Eq.~\eqref{eq:recursion_sum_dim} in the last two steps (last step). Hence, $\tO(\E)$ for $\hat \tO=\hat \tZ^{}_{\Nt},\hat \tZ^{}_{\Nt\Nt}$ can be written as
\begin{equation}\label{eq:expectation Z_N Z_NN}
    \tZ^{}_{\Nt}(\E)=\tZ^{}_{\Nt\Nt}(\E)=\frac{M^2}{L(L-1)}- \frac{1}{L-1}\left(1-\frac{D_M^{L-1}}{D_M^{L}}\right)\,,
\end{equation}
where $D_M^{L-1}/D_M^{L}<1$. Note that the expectation values $\tZ^{}_{\Nt}(\E)$ and $\tZ^{}_{\Nt\Nt}(\E)$ vanish (in the thermodynamic limit) unless $M=\O(L)$ corresponding to a finite magnetization density. Similarly, the mean energy $\E$ is given by
\begin{equation}\label{eq:expectation H}
    \E=\braket{\hat H}_M=-\Delta\sum_{i=1}^L\braket{\hat S^z_{\!i}\hat S^z_{\!i+1}}_M=-\Delta L\left[\frac{M^2}{L(L-1)}- \frac{1}{L-1}\left(1-\frac{D_M^{L-1}}{D_M^{L}}\right)\right].
\end{equation}
Since $\hat \tJ^{}_{\Nt}$ is odd under parity, repeating the same argument as in Eq.~\eqref{eq:full_cumulant_JN}, one finds $\braket{\hat H^n\hat \tJ^{}_{\Nt}}_M=0$ which implies that $\tJ^{}_{\Nt}(E^{}_m)=0$ is structureless in energy.  
To obtain the coefficient of the linear-in-energy-density term in Eq.~\eqref{eq:Taylor-expansion-diagonal}, given by Eq.~\eqref{eq:linear-in-energy coefficient}, one needs to evaluate the expectation values $\braket{\hat H\hat \tO}_M$ and $\braket{\hat H^2}_M$ in addition to $\braket{\hat \tO}_M=\tO(\E)$ and $\braket{\hat H}_M=\E$ given above. The expectation values $\braket{\hat H\hat \tZ^{}_{\Nt}}_M$, $\braket{\hat H\hat \tZ^{}_{\Nt\Nt}}_M$ and $\braket{\hat H^2}_M$ can be written as
\begin{align}
    \braket{\hat H\hat \tZ^{}_{\Nt}}_M&=-\frac{\Delta}{L}\left\langle\sum_i \hat S^z_{\!i} \hat S^z_{\!i+1}\sum_j \hat S^z_{\!j} \hat S^z_{\!j+1}\right\rangle_M=-\frac{\Delta}{L}\left[L(L-3)\braket{\hat S^z_{\!i} \hat S^z_{\!i+1} \hat S^z_{\!j} \hat S^z_{\!j+1}}_M+2L\braket{\hat S^z_{\!i} (\hat S^z_{\!i+1})^2 \hat S^z_{\!i+2}}_M+L\braket{(\hat S^z_{\!i})^2 (\hat S^z_{\!i+1})^2}_M\right]\,,\label{eq:overlap <HZ_N>}\\
    \braket{\hat H\hat \tZ^{}_{\Nt\Nt}}_M&=-\frac{\Delta}{L}\left\langle\sum_i \hat S^z_{\!i} \hat S^z_{\!i+1}\sum_j \hat S^z_{\!j} \hat S^z_{\!j+2}\right\rangle_M=-\frac{\Delta}{L}\left[L(L-4)\braket{\hat S^z_{\!i} \hat S^z_{\!i+1} \hat S^z_{\!j} \hat S^z_{\!j+2}}_M+4L\braket{\hat S^z_{\!i} (\hat S^z_{\!i+1})^2 \hat S^z_{\!i+3}}_M\right]\,,\label{eq:overlap <HZ_NN>}\\
    \braket{\hat H^2}_M&=\Delta^2\left\langle\sum_i \hat S^z_{\!i} \hat S^z_{\!i+1}\sum_j \hat S^z_{\!j} \hat S^z_{\!j+1}\right\rangle_M+\frac{1}{4}\left\langle\left(\sum_i\hat S^+_{\!i}\hat S^-_{\!i+1} +\mathrm{H.c.}\right)^2\right\rangle_M=-\Delta L\braket{\hat H\hat \tZ^{}_{\Nt}}_M+\frac{L}{4}\braket{(\hat S^+_{\!i}\hat S^-_{\!i+1} +\mathrm{H.c.})^2}_M\,,\label{eq:overlap <H^2>} 
\end{align} 
given in terms of the following expectation values for distinct sites,
\begin{align}
  \braket{\hat S^z_{\!i} \hat S^z_{\!j} \hat S^z_{\!k} \hat S^z_{\!\ell} }_M&=\frac{1}{D^L_M}[D^{L-4}_{M+4}+D^{L-4}_{M-4}-4(D^{L-4}_{M+2}+D^{L-4}_{M-2})+6D^{L-4}_{M}]\,,\qquad 
  \braket{\hat S^z_{\!i} (\hat S^z_{\!j})^2 \hat S^z_{\!k} }_M=\frac{1}{D^L_M}[D^{L-3}_{M+3}+D^{L-3}_{M-3}-(D^{L-3}_{M+1}+D^{L-3}_{M-1})]\,,\\
  \braket{(\hat S^z_{\!i})^2 (\hat S^z_{\!j})^2}_M&=\frac{1}{D^L_M}(D^{L-2}_{M+2}+D^{L-2}_{M-2}+2D^{L-2}_M)\,, \hspace{2.1cm}
  \frac{1}{4}\braket{(\hat S^+_{\!i}\hat S^-_{\!j} +\mathrm{H.c.})^2}_M=\frac{1}{D^L_M}[2(D^{L-2}_{M+1}+D^{L-2}_{M-1})+4D^{L-2}_M]\,.
\end{align}
Therefore, the coefficient of the linear-in-energy-density term in Eq.~\eqref{eq:linear-in-energy coefficient} for the observables $\hat \tO=\hat \tZ^{}_{\Nt},\hat \tZ^{}_{\Nt\Nt}$ can be obtained by combining the expectation values $\braket{\hat H\hat \tZ^{}_{\Nt}}_M$, $\braket{\hat H\hat \tZ^{}_{\Nt\Nt}}_M$ and $\braket{\hat H^2}_M$ in Eqs.~\eqref{eq:overlap <HZ_N>}--\eqref{eq:overlap <H^2>} as well as the expectation values $\braket{\hat \tZ^{}_{\Nt}}_M\equiv \tZ^{}_{\Nt}(\E)$, $\braket{\hat \tZ^{}_{\Nt\Nt}}_M\equiv \tZ^{}_{\Nt\Nt}(\E)$ and $\braket{\hat H}_M\equiv \E$ in Eqs.~\eqref{eq:expectation Z_N Z_NN}--\eqref{eq:expectation H}. Combining these results, to the leading order in $1/L$, the coefficients can be written as
\begin{align}
    \left.\frac{\partial \tZ^{}_{\Nt}(\epsilon^{}_m)}{\partial \epsilon^{}_m}\right\lvert_{\epsilon^{}_m=\epsilon^{}_{\bb=0}}&\simeq\frac{-\Delta\left[(\mathfrak{m}^2-1)^2+2(\mathfrak{m}^2-1)\frac{D^{L-1}_M}{D^{L}_M}+\frac{D^{L-2}_M}{D^{L}_M}\right]}{\left[\Delta^2(\mathfrak{m}^2-1)^2+2[1+\Delta^2(\mathfrak{m}^2-1)]\frac{D^{L-1}_M}{D^{L}_M}+(2+\Delta^2)\frac{D^{L-2}_M}{D^{L}_M}\right]}\,,\label{eq:linear-order Z_N}\\
    \left.\frac{\partial \tZ^{}_{\Nt\Nt}(\epsilon^{}_m)}{\partial \epsilon^{}_m}\right\lvert_{\epsilon^{}_m=\epsilon^{}_{\bb=0}}&\simeq\frac{1}{L}\frac{\Delta\left[(2\mathfrak{m}^4-5\mathfrak{m}^2+3)+(4\mathfrak{m}^2-6+\frac{D^{L-1}_M}{D^{L}_M})\frac{D^{L-1}_M}{D^{L}_M}-2\frac{D^{L-2}_M}{D^{L}_M}\right]}{\left[\Delta^2(\mathfrak{m}^2-1)^2+2[1+\Delta^2(\mathfrak{m}^2-1)]\frac{D^{L-1}_M}{D^{L}_M}+(2+\Delta^2)\frac{D^{L-2}_M}{D^{L}_M}\right]}\,,\label{eq:linear-order Z_NN}
\end{align}
where $\mathfrak{m}=M/L$ is the magnetization density and $D_M^{L-2}/D_M^{L}< D_M^{L-1}/D_M^{L}<1$. The linear coefficient for $\hat \tZ^{}_{\Nt}$ is nonzero, whereas the one for $\hat \tZ^{}_{\Nt\Nt}$ vanishes in the thermodynamic limit. To obtain the leading-order (nonvanishing for $L\rightarrow\infty$) energy dependence of $\hat \tZ^{}_{\Nt\Nt}$ one needs to calculate the coefficient for the quadratic-in-energy-density term in Eq.~\eqref{eq:Taylor-expansion-diagonal} which, to leading order, is given by Eq.~\eqref{eq:linear-in-energy coefficient} with $n=2$:
\begin{align}\label{eq:quadratic-order Z_NN}
    \frac{1}{2!}\left.\frac{\partial^2 \tZ^{}_{\Nt\Nt}(\epsilon^{}_m)}{\partial \epsilon_m^2}\right\lvert_{\epsilon^{}_m=\epsilon^{}_{\bb=0}}&\simeq\frac{1}{2!}\frac{\braket{\hat H^2\hat \tZ^{}_{\Nt\Nt}}_c}{(\braket{\hat H^2}_c/L)^2}=\frac{L^2}{2}\frac{\braket{\hat H^2\hat \tZ^{}_{\Nt\Nt}}_M-\braket{\hat H^2}_M\braket{\hat \tZ^{}_{\Nt\Nt}}_M-2\braket{\hat H \hat \tZ^{}_{\Nt\Nt}}_M\braket{\hat H}_M+2\braket{\hat H}_M^2\braket{\hat \tZ^{}_{\Nt\Nt}}_M}{\left[\braket{\hat H^2}_M-\braket{\hat H}^2_M\right]^2}\,,\\ \text{where}\quad\ \
    \braket{\hat H^2\hat \tZ^{}_{\Nt\Nt}}_M&=\frac{\Delta^2}{L}\left\langle\sum_i \hat S^z_{\!i} \hat S^z_{\!i+1}\sum_j \hat S^z_{\!j} \hat S^z_{\!j+1}\sum_k \hat S^z_{\!k} \hat S^z_{\!k+2}\right\rangle_M+\frac{1}{4L}\left\langle\left(\sum_i\hat S^+_{\!i}\hat S^-_{\!i+1} +\mathrm{H.c.}\right)^2\sum_j \hat S^z_{\!j} \hat S^z_{\!j+2}\right\rangle_M\,.
\end{align}
In Figs.~\ref{fig:diagonal-matrix-elements}(a)--\ref{fig:diagonal-matrix-elements}(c) we show the leading-order behavior of the microcanonical function $\tO(E^{}_m)$ close to $E^{}_m \approx \E$ for the observables $\hat \tO= \hat \tZ^{}_{\Nt},\hat \tZ^{}_{\Nt\Nt}$ and $\hat \tJ^{}_{\Nt}$, with the mean values in Eq.~\eqref{eq:expectation Z_N Z_NN} and $\tJ^{}_{\Nt}(\E)=0$, respectively, shown as black dots. The solid black curves centered on the black dots show the leading-order linear (linear and quadratic) scaling for $\hat \tZ^{}_{\Nt}$ ($\hat \tZ^{}_{\Nt\Nt}$), while $\tJ^{}_{\Nt}$ is structureless in energy. 

At this point, we should emphasize that the ETH ansatz, with the normalization convention used in Eq.~\eqref{eq:ETH_ansatz}, assumes operators $\hat O$ that are (i) intensive, so that the microcanonical function $O(E^{}_m)=\O(1)$, and (ii) properly normalized, \ie $||\hat O||=\sqrt{\Tr(\hat O^\dagger \hat O)/\Tr^{}(\mathbbm{1})}=\O(1)$, so that the smooth function $|f^{}_O(\bar E^{}_{mn},\omega^{}_{mn})|^2=\O(1)$ at fixed $\omega=\O(1)$. While local operators $\hat O\equiv\hat \lO^j$ with finite support near site $j$ satisfy both (i) and (ii), for the translationally invariant intensive sums of traceless local operators $\hat \tO\equiv\sum_j\hat \lO^j/L$ considered here [see Eq.~\eqref{eq:observables}], $||\hat \tO||\propto1/\sqrt{L}$, and the ETH in Eq.~\eqref{eq:ETH_ansatz} for the diagonal matrix elements needs to include an additional factor of $1/\sqrt{L}$ to account for the operator normalization~\cite{mierzejewski_vidmar_20, patrycja_rafal_24}:
\vspace{-0.1cm}
\begin{equation}\label{eq:diagonal-ETH}
    \tO^{}_{mm}=\tO(E^{}_m)\,+\,\frac{1}{\sqrt{L\Omega(E_m)}}f^{}_\tO(E^{}_m,0)R^{\tO}_{mm}\,.
\end{equation}
\vspace{-0.25cm}

\noindent To probe the relevance of the last term in Eq.~\eqref{eq:diagonal-ETH} in our model, we compute the average magnitude $\delta \tO$ of the fluctuations $\Delta \tO^{}_{mm}$:
\begin{equation}\label{eq:diagonal-fluctuations}
    \delta \tO\equiv\langle |\Delta \tO^{}_{mm}|\rangle\,, \qquad \Delta \tO^{}_{mm}\equiv\tO^{}_{mm}-\overline{\tO^{}_{mm}}
\end{equation}
where $\overline{(\ldots)}$ corresponds to a running average of $\tO^{}_{mm}$ over 50 states centered at $E^{}_m$, and $\langle\ldots\rangle$ is obtained by averaging over the central $50\%$ of the spectrum. Using the diagonal ETH in Eq.~\eqref{eq:diagonal-ETH} (with $\overline{R^{\tO}_{mm}}=0$), one expects the fluctuations $\delta \tO$ in Eq.~\eqref{eq:diagonal-fluctuations} to decay {\bf exponentially} with system size~\cite{leblond_2019, beugeling_2014}, i.e., $\delta \tO\propto [L\Omega(E_m)]^{-1/2}$ [recall that $\Omega^{}(E^{}_{m})\propto \exp[S(E^{}_{m})]$ grows exponentially with system size]. Indeed, the results shown in the insets in Figs.~\ref{fig:diagonal-matrix-elements}(a)--\ref{fig:diagonal-matrix-elements}(c) at the nonintegrable point are consistent with $\delta \tO\propto (L\Omega)^{-\gamma}$ and $\gamma\approx1/2$ about the center of the energy spectrum. At the integrable point, on the other hand, $\delta \tO$ [insets in Figs.~\ref{fig:diagonal-matrix-elements}(d)--\ref{fig:diagonal-matrix-elements}(f)] exhibits a slower {\bf algebraic} decay, $\delta \tO \propto L^{-\gamma}$ with $\gamma\gtrsim1/2$. The latter result is consistent with the expectation that for the translationally invariant observables considered here, $\delta \tO$ over the entire Hilbert space decays at least as fast as $L^{-1/2}$~\cite{patrycja_rafal_24, biroli_10, alba_2015}.

\begin{figure}[t]
    \centering
    \includegraphics[width=0.8\linewidth]{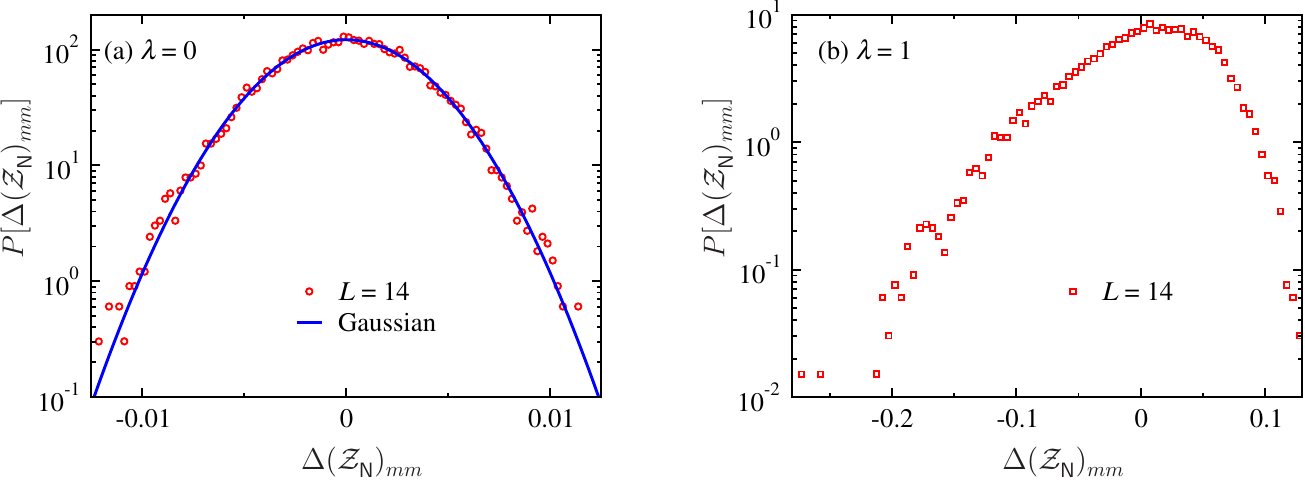}
    \caption{{\bf Diagonal matrix elements}: Distribution of $\Delta(\tZ^{}_{\Nt})^{}_{mm}$ in the central $5\%$ of the spectrum at the (a) nonintegrable ($\lambda=0$) and (b) integrable ($\lambda=1$) points of the Hamiltonian~\eqref{eq:Spin1_Hamiltonian}, for the largest chain ($L=14$) considered. The solid line in (a) shows a Gaussian fit.}
    \label{fig:histogram-diagonal-elements}
\end{figure}

We conclude this section on the diagonal ETH by discussing the {\bf distribution of the fluctuations} $\Delta \tO^{}_{mm}$, which are described at the nonintegrable point by the $R^{\tO}_{mm}$ in Eq.~\eqref{eq:diagonal-ETH}. In Fig.~\ref{fig:histogram-diagonal-elements} we show the distribution of $\Delta(\tZ^{}_{\Nt}){}_{mm}$ for the observable $\hat \tO=\hat \tZ^{}_{\Nt}$ for the energy eigenstates in a narrow energy window (central $5\%$ of the spectrum) of the Hamiltonian~\eqref{eq:Spin1_Hamiltonian} at both the nonintegrable and integrable points considered. At the {\bf nonintegrable point} [Fig.~\ref{fig:histogram-diagonal-elements}(a)], the distribution of $\Delta(\tZ^{}_{\Nt})^{}_{mm}$ is {\bf approximately Gaussian}, while at the integrable point [Fig.~\ref{fig:histogram-diagonal-elements}(b)] it is an asymmetric, non-Gaussian distribution.

\subsection{Off-diagonal ETH}

In terms of the frequency-resolved variance $\overline{|O^{}_{mn}|^2}$, obtained as a running average of $|O^{}_{mn}|^2$ with average energy $\bar E^{}_{mn}\approx\bar E$ and $\omega^{}_{mn}$ within a narrow window of width $\delta\omega$ about $\omega$, the spectral function for an observable $\hat O$ in a nonintegrable system can be written as [see Eq.~\eqref{eq:ETH_ansatz}]:
\begin{equation}\label{eq:f_var}
    |f^{\mathrm{var}}_O(\bar E,\omega)|^2=\Omega(\bar E)\, \overline{|O^{}_{mn}|^2}, \qquad \text{for}\qquad \bar E^{}_{mn}\in\text{ME},
\end{equation}
where we stress that this expression is valid only when one calculates the average $\overline{|O^{}_{mn}|^2}$ for $\bar E^{}_{mn}$ within a microcanonical window centered at $\bar E$. Note that we use the superscript `var' to make explicit that the spectral function $|f^{\mathrm{var}}_O(\bar E,\omega)|^2\simeq |f^{}_O(\bar E,\omega)|^2$ is obtained using the variance of the matrix elements. $|f^{\mathrm{var}}_O(\bar E,\omega)|^2$ as defined above is also a meaningful quantity for integrable systems for which the variance $\overline{|O^{}_{mn}|^2}$ decays exponentially with system size~\cite{leblond_2019, brenes_low_freq_2020, leblond_2020, zhang_22}, although the matrix elements $O^{}_{mn}$ do not comply with the ETH.

\paragraph{Spectral functions from autocorrelation functions}

An alternative way to define the spectral function is in terms of the connected autocorrelation function. In an energy eigenstate $\ket{\psi^{}_m}$, the autocorrelation function for an observable $\hat O$ can be written as 
\begin{equation}\label{eq:single-eigenstate-autocorrelation}
    C^{}_O(E^{}_m,t)=\braket{\psi^{}_m|\hat O(t)\hat O|\psi^{}_m}-\braket{\psi^{}_m|\hat O|\psi^{}_m}^2=\sum_{n\neq m} |O^{}_{mn}|^2e^{i\omega^{}_{mn}t}\,,
\end{equation}
\vspace{-0.2cm}

\noindent where $\hat O(t)=e^{i\hat Ht}\hat Oe^{-i\hat Ht}$ and $\omega^{}_{mn}=E^{}_m-E^{}_n$. The Fourier transform of $C^{}_O(E^{}_m,t)$ is the spectral density $S\!^{}_O(E^{}_m,\omega)$ corresponding to the eigenstate $\ket{\psi^{}_m}$:
\vspace{-0.2cm}
\begin{equation}\label{eq:single-eigenstate-spectral-density}
    S\!^{}_O(E^{}_m,\omega)=\frac{1}{2\pi}\int dt e^{-i\omega t} C^{}_O(E^{}_m,t)=\sum_{n\neq m}|O^{}_{mn}|^2\delta(\omega-\omega^{}_{mn})\,.
\end{equation}
One is often interested in the autocorrelation function in a thermal equilibrium state corresponding to an ensemble of statistical mechanics, like the canonical ensemble, with density matrix $\hat \rho=e^{-\upbeta\hat H}/\Z$ where $\Z=\Tr e^{-\upbeta\hat H}$, for which the autocorrelation function reads~\cite{dalessio_quantum_2016,brenes_pappalardi_21}
\begin{align}\label{eq:thermal-autocorrelation}
    C^\upbeta_O(t)&=\Tr\left(\hat \rho \hat O(t)\hat O\right)-\Tr\left(\hat \rho \hat O\right)^2=\sum_{m}\frac{e^{-\upbeta E^{}_m}}{\Z} \sum_{n}|O^{}_{mn}|^2e^{i\omega^{}_{mn}t}-\left(\sum_{m} \frac{e^{-\upbeta E^{}_m}}{\Z}O^{}_{mm}\right)^2\\
    &=\sum_{m}\frac{e^{-\upbeta E^{}_m}}{\Z}\sum_{n}|O^{}_{mn}|^2e^{i\omega^{}_{mn}t}-\sum_{m} \frac{e^{-\upbeta E^{}_m}}{\Z}O^2_{mm}+\underbrace{\sum_{m} \frac{e^{-\upbeta E^{}_m}}{\Z}O^2_{mm}  -\left(\sum_{m} \frac{e^{-\upbeta E^{}_m}}{\Z}O^{}_{mm}\right)^2}_{\text{Var}^{}_\upbeta(O^{}_{mm})}=\sum_{m}\frac{e^{-\upbeta E^{}_m}}{\Z}\underbrace{\sum_{n\neq m}|O^{}_{mn}|^2e^{i\omega^{}_{mn}t}}_{C^{}_O(E^{}_m,t)}+\mathscr{O}\left(\frac{1}{L}\right)\,,\nonumber
\end{align}
where in the last step, we assumed that $O^{}_{mm}$ is a smooth function of the energy density $E^{}_m/L$ with corrections that vanish exponentially in $L$ (which is the case if $O^{}_{mm}$ follows the ETH) to obtain the (time-independent) statistical variance
\begin{equation}\label{eq:thermal-variance}
    \text{Var}^{}_\upbeta(O^{}_{mm})\simeq\int dE\frac{e^{S(E)-\upbeta E}}{\Z} O^2(E)-\left(\int dE\frac{e^{S(E)-\upbeta E}}{\Z} O(E)\right)^2\simeq \left(\left.\frac{\partial O}{\partial E}\right\vert_{E\!^{}_\upbeta}\right)^2\int dE\frac{e^{S(E)-\upbeta E}}{\Z}(E-E\!^{}_\upbeta)^2\equiv\left(\left.\frac{\partial O}{\partial E}\right\vert_{E\!^{}_\upbeta}\right)^2 \delta E^2 = \mathscr{O}\left(\frac{1}{L}\right)\,,\nonumber
\end{equation}
by Taylor expanding $O(E)$ around the saddle point $E=E\!^{}_\upbeta$ with $E\!^{}_\upbeta=\Tr(\hat \rho \hat H)$, and using that $\partial O/\partial E=\mathscr{O}(L^{-1})$ with subextensive energy fluctuations $\delta E=\mathscr{O}(\sqrt{L})$. Therefore, up to a time-independent correction that vanishes in the thermodynamic limit, the thermal autocorrelation function $C^\upbeta_O(t)$ in Eq.~\eqref{eq:thermal-autocorrelation} is equivalent to the single-eigenstate result $ C^{}_O(E^{}_m,t)$ in Eq.~\eqref{eq:single-eigenstate-autocorrelation} averaged over the Gibbs probability distribution associated with the thermal state $\hat \rho=e^{-\upbeta\hat H}/\Z$. The Fourier transform of $C^\upbeta_O(t)$ corresponds to the spectral density
\begin{equation}\label{eq:spectral-density-ETH}
    S\!^\upbeta_O(\omega)\simeq\sum_{m}\frac{e^{-\upbeta E^{}_m}}{\Z} S\!^{}_O(E^{}_m,\omega)= \sum_{m,n\neq m}\frac{e^{-\upbeta E^{}_m}}{\Z} |O^{}_{mn}|^2\delta(\omega-\omega^{}_{mn})\simeq\sum_{m,n\neq m}\frac{e^{-\upbeta E^{}_m}}{\Z} \frac{|f^{}_O(\bar E^{}_{mn},\omega^{}_{mn})|^2}{\Omega(\bar E^{}_{mn})} \delta(\omega-\omega^{}_{mn})\,,
\end{equation}
where the last approximate equality follows after substituting the ETH ansatz for the matrix elements of $\hat O$ from Eq.~\eqref{eq:ETH_ansatz} and coarse graining over nearby matrix elements, using $\overline{|R^{O}_{mn}|^2}=1$. Replacing the sums over $E^{}_m$ and $E^{}_n$ by integrals, one gets
\begin{align}\label{eq:spectral-density-and-spectral-function}
    S\!^\upbeta_O(\omega)&\simeq\!\int\! dE e^{S(E)}\frac{e^{-\upbeta E}}{\Z}\!\int\! dE' e^{S(E')} \frac{\Big|f^{}_O\Big(\frac{E+E'}{2},E\!-\!E'\Big)\Big|^2}{\exp\Big[S\Big(\frac{E+E'}{2}\Big)\Big]}\delta\big(\omega\!-\!(E\!-\!E')\big)\,\nonumber=\frac{1}{\Z}\!\int\! dE \exp\!\Big[S(E)\!+\!S(E-\omega)\!-\!S\big(E-\tfrac{\omega}{2}\big)\!-\!\upbeta E\Big] \big|f^{}_O(E\!-\!\tfrac{\omega}{2},\omega)\big|^2\,\nonumber\\&\simeq\frac{1}{\Z}\!\int\! dE \exp\!\Big[S(E)-\frac{\partial S}{\partial E}\frac{\omega}{2}+\frac{3}{8}\frac{\partial^2S}{\partial E ^2}\omega^2-\upbeta E\Big]\bigg[\big|f^{}_O(E,\omega)\big|^2-\frac{\partial|f^{}_O(E,\omega)|^2}{\partial E}\frac{\omega}{2}\bigg]\simeq e^{-\upbeta\omega/2}\,\big|f^{}_O(E\!^{}_\upbeta,\omega)\big|^2\,, 
\end{align}
where, in the second line, we Taylor expanded about small $\omega=\mathscr{O}(1)$ and used the saddle-point approximation about $E\!^{}_\upbeta=\Tr(\hat \rho \hat H)$, so that $\tfrac{\partial S}{\partial E}|_{E\!^{}_\upbeta}=\upbeta$. Therefore, $S\!^\upbeta_O(\omega)$, which is the Fourier transform of the thermal autocorrelation function $C_O^\upbeta(t)$, is related to the ETH spectral function $|f^{}_O(E\!^{}_\upbeta,\omega)|^2$ up to the prefactor $e^{-\upbeta\omega/2}$. Our result in Eq.~\eqref{eq:spectral-density-and-spectral-function} is consistent with the Kubo-Martin-Schwinger relation~\cite{kubo_57,martin_schwinger_59,schonle_autocorrelation_2021}
\vspace{-0.1cm}
\begin{equation}
 \hat O = \hat O^{\dagger} \quad\implies\quad |f^{}_O(E,\omega)|^2=|f^{}_O(E,-\omega)|^2  \quad\implies\quad  \frac{S\!^\upbeta_O(\omega)}{S\!^\upbeta_O(-\omega)}=e^{-\upbeta\omega}\,.
\end{equation}
\vspace{-0.2cm}

Using Eqs.~\eqref{eq:spectral-density-and-spectral-function} and~\eqref{eq:spectral-density-ETH}, we can write the infinite-temperature spectral function, \ie at the average energy $E\!^{}_{\upbeta=0}=\Tr(\hat H)/D$, as
\begin{equation}\label{eq:f_corr}
    |f^{\mathrm{corr}}_O(\E,\omega)|^2=\frac{1}{D}\sum_{m,n\neq m} |O^{}_{mn}|^2\delta(\omega-\omega^{}_{mn})\,, 
\end{equation}
\vspace{-0.2cm}

\noindent where we added a superscript `corr' to make explicit that the spectral function $|f^{\mathrm{corr}}_O(\E,\omega)|^2\simeq |f^{}_O(\E,\omega)|^2$ is obtained using the autocorrelation function. In Eq.~\eqref{eq:f_corr}, $D=\Tr(\mathbbm{1})$ is the Hilbert space dimension. $|f^{\mathrm{corr}}_O(\E,\omega)|^2$ is the average of $S\!^{}_O(E^{}_m,\omega)$ over all the energy eigenstates. To compute $|f^{\mathrm{corr}}_O(\E,\omega)|^2$ numerically, we regularize the $\delta$-function by replacing $\delta(x)\rightarrow \tfrac{1}{\sqrt{2\pi \sigma^2}}\exp\big[-\tfrac{x^2}{2\sigma^2}\big]$ with the broadening parameter $\sigma$. We take $\sigma=0.1\omega^{}_H$, where $\omega^{}_H$ is the average level spacing in the central 10\% of the energy spectrum~\cite{abdelshafy_25}.

Next, we relate the two definitions of the spectral functions $|f^{\mathrm{var}}_O(\E,\omega)|^2$ [cf.~Eq.~\eqref{eq:f_var}] and $|f^{\mathrm{corr}}_O(\E,\omega)|^2$ [cf.~Eq.~\eqref{eq:f_corr}] coarse-grained within a narrow frequency window of width $\delta\omega$:
\vspace{-0.1cm}
\begin{align}\label{eq:f_corr coarse-grained}
    |f^{\mathrm{corr}}_O(\E,\omega)|^2\simeq \frac{1}{\delta\omega}\int_{\omega-\frac{\delta\omega}{2}}^{\omega+\frac{\delta\omega}{2}} |f^{\mathrm{corr}}_O(\E,\omega')|^2 d\omega'&=\frac{1}{D\delta\omega}\sum_{m=1}^{D}\underset{|\omega^{}_{mn}-\omega|<\frac{\delta\omega}{2}}{\sum^D_{n\neq m}}|O^{}_{mn}|^2\,\nonumber=\frac{1}{D}\underbrace{\left(\frac{N^{}_{\delta\omega}(\omega)}{\delta\omega}\right)}_{\rho(\omega)}\Bigg(\frac{1}{N^{}_{\delta\omega}(\omega)}\sum_{m=1}^{D}\!\!\!\underset{\quad|\omega^{}_{mn}-\omega|<\frac{\delta\omega}{2}}{\sum^D_{n\neq m}}|O^{}_{mn}|^2\Bigg)\,\nonumber\\
    &=\frac{\rho(\omega)}{D\W}\left(\W\overline{|O^{}_{mn}|^2}\right)\simeq\frac{\rho(\omega)}{D\W}\sqrt{2}\ |f^{\mathrm{var}}_O(\E,\omega)|^2\,, 
\end{align}
where $N^{}_{\delta\omega}(\omega)$ counts the number of pairs of eigenstates with energy difference $\omega^{}_{mn}$ within a window of width $\delta\omega$ centered at $\omega$, $\rho(\omega)$ is the density of energy differences, and we stress that $\overline{|O^{}_{mn}|^2}$ is computed over the entire spectrum [unlike the microcanonical average in Eq.~\eqref{eq:f_var}, see Eq.~\eqref{eq:f_var calc fullspectrum}]. Therefore, $|f^{\mathrm{corr}}_O(\E,\omega)|^2$ coarse-grained within a narrow frequency window is related to $|f^{\mathrm{var}}_O(\E,\omega)|^2$ through a ratio between the density of energy differences $\rho(\omega)$ and the density of states at the mean energy $\W$. Since the density of states $\Omega(E)$ of local Hamiltonians is generally Gaussian (see Fig.~\ref{fig:DOS}), \ie $\Omega(E)\simeq\tfrac{D}{\sqrt{2\pi \sigma_E^2}}\exp\Big[-\tfrac{(E-\E)^2}{2\sigma_E^2}\Big]$ with $\sigma_E^2$ quantifying the energy variance, one has $\W\simeq D/\sqrt{2\pi\sigma_E^2}$ and the density of energy differences has the form:
\begin{equation}\label{eq:rho_omega}
    \rho (\omega)=\int dE\, \Omega(E) \,\Omega(E-\omega)\simeq\frac{D^2}{2\pi \sigma_E^2}\int dE \exp\left[-\frac{(E-\E)^2}{2\sigma_E^2}\right]\exp\left[-\frac{(E-\E-\omega)^2}{2\sigma_E^2}\right]
    =\frac{D^2}{\sqrt{4\pi \sigma_E^2}} \exp\left(-\frac{\omega^2}{4\sigma_E^2}\right)\,,
\end{equation}
which is a Gaussian with twice the energy variance $2\sigma_E^2$. Since the variance $\overline{|O^{}_{mn}|^2}$ in Eq.~\eqref{eq:f_corr coarse-grained} is obtained using the entire energy spectrum, in contrast to the one in Eq.~\eqref{eq:f_var} calculated within a microcanonical energy window, the former evaluates to
\begin{align}\label{eq:f_var calc fullspectrum}
&\W \overline{|O_{mn}|^2}=\frac{\W}{N_{\delta\omega}(\omega)}\sum_{m=1}^{D}\!\underset{|\omega^{}_{mn}-\omega|<\frac{\delta\omega}{2}}{\sum^D_{n\neq m}}\!\!\!|O^{}_{mn}|^2\simeq\frac{\W}{N_{\delta\omega}(\omega)}\sum_{m=1}^{D}\!\underset{|\omega^{}_{mn}-\omega|<\frac{\delta\omega}{2}}{\sum^D_{n\neq m}}\!\!\!\frac{|f^{\mathrm{var}}_O(\bar E_{mn},\omega_{mn})|^2}{\Omega(\bar E_{mn})}\simeq\frac{\W\,\delta \omega}{N_{\delta\omega}(\omega)}\int dE\, \frac{\Omega(E+\frac{\omega}{2}) \Omega(E-\frac{\omega}{2})}{\Omega(E)} |f^{\mathrm{var}}_O(E,\omega)|^2\nonumber\\
&\simeq\frac{D^2}{\sqrt{2\pi\sigma_E^2}\,\rho(\omega)} \exp\left(-\frac{\omega^2}{4\sigma_E^2}\right)\int\!  \frac{dE}{\sqrt{2\pi\sigma_E^2}}\exp\left(-\frac{(E-\E)^2}{2\sigma_E^2}\right)\left[|f^{\mathrm{var}}_O(\E,\omega)|^2+\left.\frac{\partial |f^{\mathrm{var}}_O(E,\omega)|^2}{\partial E}\right|_{E=\E}\!\!\!(E-\E)+\ldots\right]\nonumber\\&\simeq \frac{D^2}{\sqrt{2\pi\sigma_E^2}\,\rho(\omega)}\exp\left(-\frac{\omega^2}{4\sigma_E^2}\right)|f^\mathrm{{var}}_O(\E,\omega)|^2=\sqrt{2}\,|f^\mathrm{{var}}_O(\E,\omega)|^2\qquad \implies \qquad |f^{\mathrm{var}}_O(\E,\omega)|^2\simeq\frac{\W\, \overline{|O^{}_{mn}|^2}}{\sqrt{2}},
\end{align} 
where, in the second line, we Taylor expanded the smooth $|f^{\mathrm{var}}_O(E,\omega)|^2$ about the peak of the Gaussian at $E=\E$ to then use the subextensivity of the energy width $\sigma_E\propto \sqrt{L}$ with arguments similar to Eqs.~\eqref{eq:traceOinDE}--\eqref{eq:DerivativeScaling}, which allowed us to obtain the leading-order contribution to the integral, $|f^\mathrm{{var}}_O(\E,\omega)|^2+\O(1/L)$. Therefore, when evaluating the variance $\overline{|O^{}_{mn}|^2}$ over the entire spectrum, the spectral function $|f^{\mathrm{var}}_O(\E,\omega)|^2$ is obtained using Eq.~\eqref{eq:f_var calc fullspectrum} [as opposed to Eq.~\eqref{eq:f_var}]. This is what we do here numerically considering the relevant symmetry sectors. Substituting the expressions for $\W$ and $\rho(\omega)$ in Eq.~\eqref{eq:f_corr coarse-grained}, one finds that:
\begin{equation}\label{eq:f_corr vs f_var}
     |f^{\mathrm{corr}}_O(\E,\omega)|^2\simeq\exp\left(-\frac{\omega^2}{4\sigma_E^2}\right)\, \big|f^{\mathrm{var}}_O(\E,\omega)\big|^2\,.
\end{equation}
Therefore, $|f^{\mathrm{corr}}_O(\E,\omega)|^2$ is modulated by an additional Gaussian decay relative to $|f^{\mathrm{var}}_O(\E,\omega)|^2$. In the thermodynamic limit for $\omega=\mathscr{O}(1)$, since $\sigma_E^2\propto L$, the Gaussian factor reduces to unity and $|f^{\mathrm{corr}}_O(\E,\omega)|^2$ is the same as $ |f^{\mathrm{var}}_O(\E,\omega)|^2$.

\paragraph{Numerical results for translationally invariant intensive observables}

\begin{figure}[!t]
    \centering
    \includegraphics[width=0.95\linewidth]{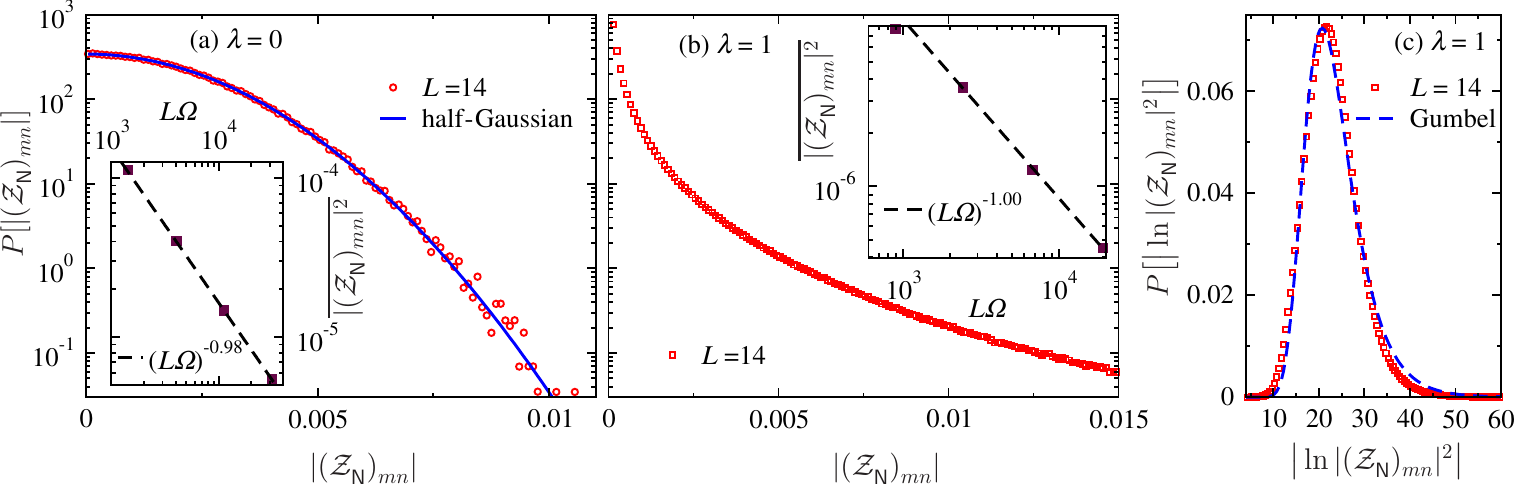}
    \caption{{\bf Off-diagonal matrix elements}: Distribution of $|(\tZ^{}_{\Nt})^{}_{mn}|$ [$|\ln|(\tZ^{}_{\Nt})^{}_{mn}|^2|$] with $\bar E^{}_{mn}$ in the central $5\%$ of the spectrum at (a) the nonintegrable point ($\lambda=0$ and $|\omega^{}_{mn}|<0.01$), and (b) [(c)] the integrable point ($\lambda=1$ and no constraint on $\omega^{}_{mn}$), for chains with $L=14$. The solid (dashed) line in (a) [(c)] shows a half-Gaussian (Gumbel) fit to the data points. The insets show the variance $\overline{|(\tZ^{}_{\Nt})^{}_{mn}|^2}$ vs $L\Omega$ (for $|\omega^{}_{mn}|<0.01$) in chains with $L=11$--$14$, where $\Omega$ is the average density of states in the central $5\%$ of the spectrum. The dashed lines show power-law fits $\propto (L\Omega)^{-\gamma}$ to the data for the three largest chains.}
    \label{fig:histogram-offdiagonal-elements}
\end{figure}

Next, we numerically test the off-diagonal ETH and our results for the spectral functions considering the observables in Eq.~\eqref{eq:observables}. As for the diagonal ETH in Eq.~\eqref{eq:diagonal-ETH}, we need to account for the proper operator norm because the observables~\eqref{eq:observables} are translationally invariant intensive sums of traceless local observables, so that the off-diagonal ETH in Eq.~\eqref{eq:ETH_ansatz} and the spectral functions in Eqs.~\eqref{eq:f_var calc fullspectrum} and~\eqref{eq:f_corr} are written as
\vspace{-0.1cm}
\begin{equation}\label{eq:offdiagonal-ETH}
     \tO^{}_{mn}=\,\frac{1}{\sqrt{L\Omega(\bar E^{}_{mn})}}f^{}_\tO(\bar E^{}_{mn},\omega^{}_{mn})R^{\tO}_{mn}\,,\quad |f^{\mathrm{var}}_\tO(\E,\omega)|^2\simeq\frac{L\W \overline{|\tO^{}_{mn}|^2}}{\sqrt{2}} \quad \text{and}\quad |f^{\mathrm{corr}}_\tO(\E,\omega)|^2=\frac{L}{D}\sum_{m,n\neq m} |\tO^{}_{mn}|^2\delta(\omega-\omega^{}_{mn}).
\end{equation}
\vspace{-0.2cm}

We first examine the distribution of the off-diagonal matrix elements $\tO^{}_{mn}$ in the sectors with $M=0$, $k\neq0,\pi$ and $\mathbb{Z}^{}_2=\pm1$ [$(\mathbb{Z}^{}_2)_m=-(\mathbb{Z}^{}_2)^{}_n$ for $(\tJ^{}_{\Nt})^{}_{mn}$]. In Fig.~\ref{fig:histogram-offdiagonal-elements} we show the distribution of $|\tO^{}_{mn}|$ for the observable $\hat \tO=\hat \tZ^{}_{\Nt}$ within the eigenstates of the Hamiltonian~\eqref{eq:Spin1_Hamiltonian} with $\bar E^{}_{mn}$ in a narrow window about $\E$ at both the nonintegrable and integrable points considered. At the nonintegrable point [Fig.~\ref{fig:histogram-offdiagonal-elements}(a)], the distribution of $|\tO^{}_{mn}|$ is approximately half-Gaussian as expected~\cite{leblond_2019, brenes_low_freq_2020, leblond_2020, beugeling_offdiag_2015}, while at the integrable point [Fig.~\ref{fig:histogram-offdiagonal-elements}(b)] it is clearly non-Gaussian~\cite{leblond_2019, brenes_low_freq_2020, leblond_2020, zhang_22}. Instead, at integrability, the transformed off-diagonal matrix elements $x=\big|\ln|(\tZ^{}_{\Nt})^{}_{mn}|^2\big|$ are well described by a Gumbel distribution~\cite{essler_2024, rottoli_alba_26}
\vspace{-0.1cm}
\begin{equation}
    P^{}_{\mu,\,\sigma}(x)=\frac{1}{\sigma}\exp\left[-\left(\frac{x-\mu}{\sigma}\right)-\exp\left(-\frac{x-\mu}{\sigma}\right)\right]\,,
\end{equation}
\vspace{-0.2cm}

\noindent as shown in Fig.~\ref{fig:histogram-offdiagonal-elements}(c). Nevertheless, irrespective of integrability or chaos, the variance $\overline{|(\tZ^{}_{\Nt})^{}_{mn}|^2}$ decays exponentially with increasing system size as $(L\Omega)^{-\gamma}$ with $\gamma\approx1$ as shown in the insets of Figs.~\ref{fig:histogram-offdiagonal-elements}(a) and~\ref{fig:histogram-offdiagonal-elements}(b)~\cite{leblond_2019, jansen_2019}.

\begin{figure}[!t]
    \centering
    \includegraphics[width=\linewidth]{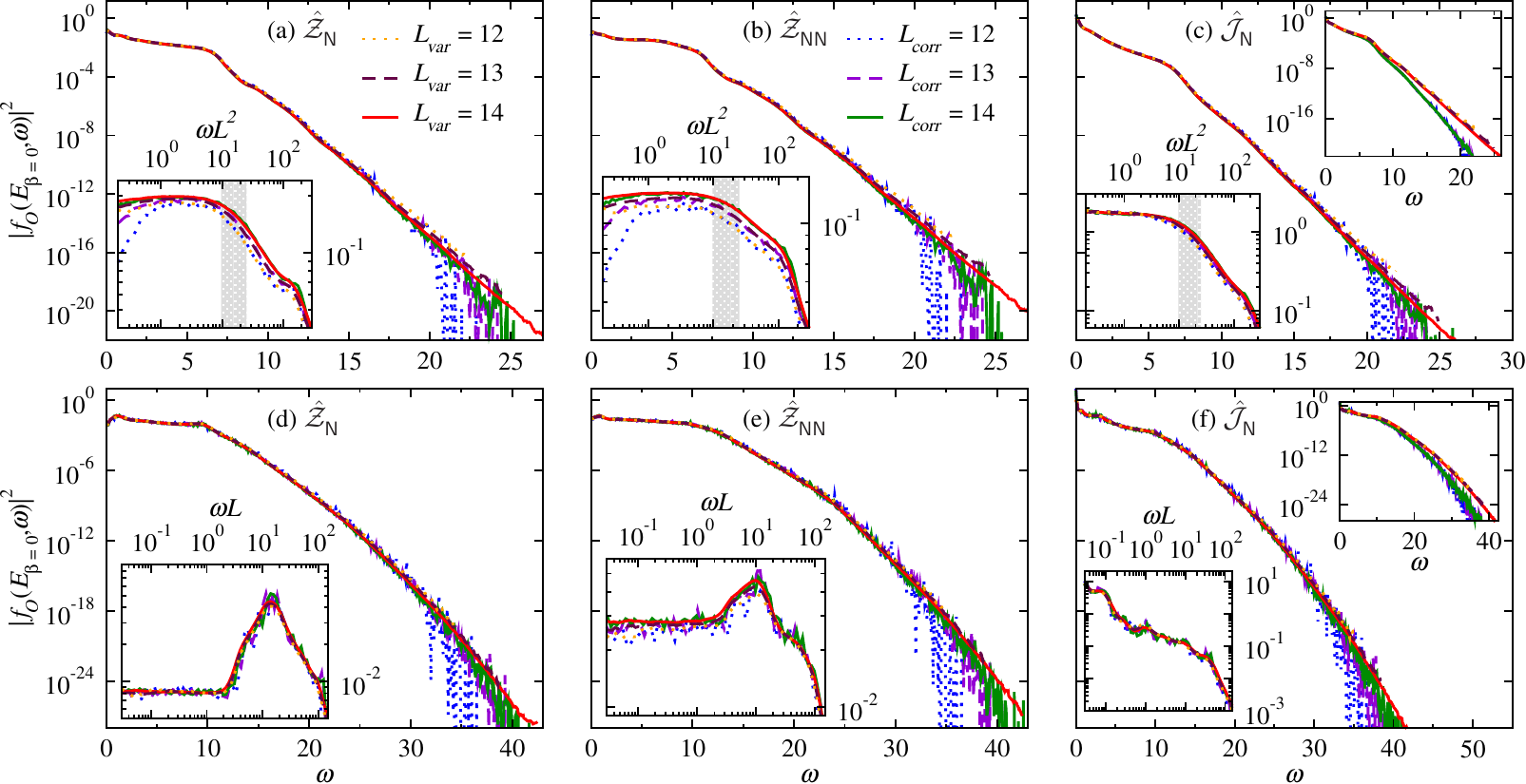}
    \vspace{-0.3cm}
    \caption{{\bf Spectral functions}: $|f^{\mathrm{var}}_\tO(\E,\omega)|^2$ and $|f^{\mathrm{resc}}_\tO(\E,\omega)|^2\equiv\exp[\omega^2/(4\sigma_E^2)]|f^{\mathrm{corr}}_\tO(\E,\omega)|^2$ [see Eqs.~\eqref{eq:offdiagonal-ETH} and~\eqref{eq:f_corr vs f_var}] evaluated at infinite temperature for the observables $\hat \tO=\hat \tZ^{}_{\Nt},\hat \tZ^{}_{\Nt\Nt}$, and $\hat \tJ^{}_{\Nt}$ with the Hamiltonian~\eqref{eq:Spin1_Hamiltonian} at the nonintegrable [(a)--(c)] and integrable [(d)--(f)] points. Results are shown for the three largest chains with $L=12,13,14$, labeled as $L^{}_{\mathrm{var}}$ ($L^{}_{\mathrm{corr}}$) corresponding to $|f^{\mathrm{var}}_\tO(\E,\omega)|^2$ [$|f^{\mathrm{resc}}_\tO(\E,\omega)|^2$] in the legends. The bottom-left insets show the same results plotted on a log-log scale to highlight the low-frequency regime. At the nonintegrable point [(a)--(c)], we use a gray shaded band to indicate the range of $\omega^{}_{\text{SFF}}L^2=2\pi L^2/t^{}_{\text{SFF}}$ corresponding to the range of $t^{}_{\text{SFF}}/L^2$ identified in Fig.~\ref{fig:level-statistics}(c), and we rescale the $x$-axes by a factor of $L^2$, while at the integrable point [(d)--(f)], we rescale the $x$-axes by a factor of $L$. The top-right insets in (c) and (f) contrast the bare functions $|f^{\mathrm{var}}_\tO(\E,\omega)|^2$ and $|f^{\mathrm{corr}}_\tO(\E,\omega)|^2$.}
    \label{fig:spectral-function-highfreq-PBC-Trans}
\end{figure}

In the main panels in Fig.~\ref{fig:spectral-function-highfreq-PBC-Trans}, we plot our numerical results for $|f^{\mathrm{var}}_\tO(\E,\omega)|^2$ and for the rescaled spectral function $|f^{\mathrm{resc}}_\tO(\E,\omega)|^2\equiv \exp\big(\tfrac{\omega^2}{4\sigma_E^2}\big)|f^{\mathrm{corr}}_\tO(\E,\omega)|^2$, which exhibit data collapse. The top-right insets in Figs.~\ref{fig:spectral-function-highfreq-PBC-Trans}(c) and~\ref{fig:spectral-function-highfreq-PBC-Trans}(f) show a direct comparison between $|f^{\mathrm{var}}_{\tJ^{}_{\Nt}}(\E,\omega)|^2$ and $|f^{\mathrm{corr}}_{\tJ^{}_{\Nt}}(\E,\omega)|^2$, which makes apparent that $|f^{\mathrm{corr}}_{\tJ^{}_{\Nt}}(\E,\omega)|^2$ decays faster than and departs from $|f^{\mathrm{var}}_{\tJ^{}_{\Nt}}(\E,\omega)|^2$ with increasing $\omega$ both at the nonintegrable and integrable points, as anticipated in Eq.~\eqref{eq:f_corr vs f_var}. The log-linear scale used for the plots makes apparent that the spectral functions at the nonintegrable point [Figs.~\ref{fig:spectral-function-highfreq-PBC-Trans}(a)--\ref{fig:spectral-function-highfreq-PBC-Trans}(c)] exhibit an approximately exponential decay in frequency for $\omega\gtrsim 10$, preceded by a slower, observable-dependent decay for $\omega\lesssim 10$. In the integrable case [Figs.~\ref{fig:spectral-function-highfreq-PBC-Trans}(d)--\ref{fig:spectral-function-highfreq-PBC-Trans}(f)], the spectral functions exhibit a faster-than-exponential decay for $\omega\gtrsim 10$. Exponential and faster-than-exponential (e.g., Gaussian) decays of the spectral functions have been found to be robust in a wide range of models~\cite{leblond_2019, brenes_low_freq_2020, leblond_2020, zhang_22, schonle_autocorrelation_2021, brenes_2020, richter2020, fadingvidmar2024, fadingvidmar2025}.

In the bottom-left insets in Fig.~\ref{fig:spectral-function-highfreq-PBC-Trans}, we replot the results in the main panels on a log-log scale to highlight the properties of the spectral functions in the low-frequency regime. In the nonintegrable case, we rescale the $x$-axes $\omega\rightarrow\omega L^2$ to underscore the low-frequency plateau in the spectral functions below a frequency $\omega^{}_{\text{ETH}}$ that is consistent with $\omega^{}_{\text{ETH}}\propto\tfrac{1}{L^2}$, as expected from diffusion~\cite{dalessio_quantum_2016}. In our model and for the chain sizes considered, we find that the plateaus in the spectral functions end close to the range of $\omega^{}_{\text{SFF}}L^2=2\pi L^2/t^{}_{\text{SFF}}$ [corresponding to the range of $t^{}_{\text{SFF}}/L^2$ in Fig.~\ref{fig:level-statistics}(c)] indicated by the gray shaded bands, i.e., there appears to be consistency between the frequencies associated with RMT behavior in the SFF and in the matrix elements of observables; see Ref.~\cite{ceven_patil_26} for recent related results in spin ladders. For the observables $\hat \tZ^{}_{\Nt}$ and $\hat \tZ^{}_{\Nt\Nt}$, which have a finite overlap with powers of the Hamiltonian (as discussed in Sec.~\ref{sec:diagonal-ETH}), the values of the spectral functions in the plateaus increase with increasing system size, while for the observable $\hat \tJ^{}_{\Nt}$, which does not overlap with any power of the Hamiltonian, the value of the spectral function does not scale with the system size~\cite{leblond_2019, capizzi_poletti_25, patil_2025}. On the other hand, for the integrable case, the $x$-axes are rescaled by a factor of $L$ to underscore a ballistic regime in which there are features in the spectral function at frequencies $\omega\propto \tfrac{1}{L}$, as also seen in the integrable spin-$\tfrac12$ $XXZ$ model in one dimension~\cite{leblond_2020}.

\section{Outlook: Symmetries and Higher-Order Correlations}\label{sec:outlook}

\subsection{ETH and symmetries}\label{sec:ETHSymmetries}

As mentioned in Sec.~\ref{sec: ETH}, the ETH ansatz in Eq.~\eqref{eq:ETH_ansatz} captures the first two moments of the observable matrix elements $O^{}_{mn}$ written in the energy eigenstates within symmetry-resolved blocks (subspaces) of quantum-chaotic Hamiltonians. Those blocks are labeled by the quantum numbers associated with the symmetries and so are the corresponding ETH smooth functions. For simplicity, we did not include these labels in Eq.~\eqref{eq:ETH_ansatz}. There are important considerations to have in mind when dealing with symmetries, which we highlight next. 

Models of interest can exhibit {\bf continuous symmetries} with associated {\bf extensive conserved quantities}, e.g., U(1), which results in the conservation of total magnetization for $\hat H$ in Eq.~\eqref{eq:Spin1_Hamiltonian}. In such cases, sectors with different densities of the conserved quantities can have exponentially different dimensions of the Hilbert space, see for instance the asymptotic form of $\mathcal{D}(N,L)$ in Eq.~\eqref{eq:entropy_N_asymptotic}, and the smooth functions $O(E^{}_m)$, $\Omega(\bar E^{}_{mn})$, and $f^{}_O(\bar E^{}_{mn},\omega^{}_{mn})$ in Eq.~\eqref{eq:ETH_ansatz} depend on the sector considered, as expected on physical grounds. Furthermore, if the observable of interest {\bf breaks a continuous symmetry} of the Hamiltonian, then it can have nonvanishing off-diagonal matrix elements between energy eigenstates that belong to symmetry-resolved blocks with different quantum numbers corresponding to the broken symmetry. The ETH can be generalized to describe observables that respect as well as those that break the symmetries of the Hamiltonian. For example, when only total magnetization is conserved (in addition to the total energy) one can write for a generic observable~\cite{patil_2025}:
\begin{equation}\label{eq:gen-AbelianETH}
 O^{}_{mn}=O(E^{}_m,M^{}_m)\delta^{}_{mn} + \frac{1}{\sqrt{\Omega(\bar E^{}_{mn},\bar M^{}_{mn})}}\,f^{}_O(\bar E^{}_{mn},\omega^{}_{mn};\bar M^{}_{mn},\nu_{mn})\,R^{O}_{mn}\,,
\end{equation}
where $\bar M^{}_{mn}=(M^{}_m+M^{}_n)/2$ is the average magnetization and $\nu_{mn}=M^{}_m-M^{}_n$ is the corresponding magnetization difference. In our calculations so far, we have focused on specific magnetization sectors of $\hat H$ in Eq.~\eqref{eq:Spin1_Hamiltonian}. The ansatz in Eq.~\eqref{eq:gen-AbelianETH} can be straightforwardly extended to account for additional mutually commuting extensive conserved quantities. Remarkably, it can also be generalized to account for noncommuting extensive conserved quantities, such as those appearing in the presence of non-Abelian symmetries~\cite{patil_2025, murthy_nonabelian_2023, noh_2023, lasek_24}.

Another important set of symmetries that can be present in models of interest is {\bf discrete symmetries}, which {\bf lack associated extensive conserved quantities}. Examples of those symmetries are lattice translations and other space symmetries (e.g., specific rotations and/or reflections, space inversion, etc.), and spin inversion. Our model Hamiltonian~\eqref{eq:Spin1_Hamiltonian} exhibits several of these symmetries, as discussed in Sec.~\ref{sec:Model_and_observables}. Discrete symmetries result in symmetry-resolved subspaces that roughly have similar Hilbert space dimensions. Remarkably, early studies of the ETH in models with lattice translational invariance showed that $\tO(E^{}_m)$ does not depend on the total quasimomentum sector~\cite{rigol_09a,rigol_09b, mondaini_2016}, and similar results were found for $f^{}_\tO(\bar E^{}_{mn},\omega^{}_{mn})$~\cite{leblond_2019, leblond_2020}. This is why in our calculations we can average over different subsectors associated with discrete symmetries to improve the statistics in the finite-size calculations.

An important question in the context of eigenstate thermalization in quantum-chaotic models with discrete symmetries is what happens to observables that break such symmetries, such as {\bf local operators in models that are invariant under lattice translations}. In that case, in contrast to the translationally invariant observables that we have considered so far in this chapter, there are off-diagonal matrix elements between sectors with different total quasimomentum. In Ref.~\cite{leblond_2020} it was shown that the distributions of $|\lO^j_{mn}|$ of local operators for $k_m\neq k_n$ are dense and close to half-Gaussian, like the one shown in Fig.~\ref{fig:histogram-offdiagonal-elements}(a) for a translationally invariant observable for $k_m= k_n$, and that those local operators exhibit spectral functions that are qualitatively similar to those of their translationally invariant counterparts. A physical understanding of the contrast between local and translationally invariant operators in translationally invariant chains was recently provided in Ref.~\cite{patil_2026}, along with a generalization of the ETH for such systems. We review the main findings of Ref.~\cite{patil_2026} in the context of our model Hamiltonian~\eqref{eq:Spin1_Hamiltonian} and the next-nearest-neighbor $\hat{S}^z$-$\hat{S}^z$ correlations considered in earlier sections. Namely, focusing on sectors with zero total magnetization ($M=0$), we consider the translationally invariant operator $\hat \tO=\hat \tZ^{}_{\Nt\Nt}$ and the corresponding local operator $\hat \lO^j=\hat \lZ^{j}_{\Nt\Nt}$, see Eq.~\eqref{eq:observables}, for which the normalization prefactors in the spectral functions differ by a factor of $L$ because of the Hilbert-Schmidt normalization~\cite{mierzejewski_vidmar_20, patrycja_rafal_24}:
\vspace{-0.1cm}
\begin{equation}\label{eq:local-vs-trans-spectral-functions}
    |f^{\mathrm{corr}}_{\tO}(\E,\omega)|^2=\frac{L}{D}\sum_{m,n\neq m} |\tO^{}_{mn}|^2\delta(\omega-\omega^{}_{mn})\,, \quad\text{and}\quad |f^{\mathrm{corr}}_{\lO^j}(\E,\omega)|^2=\frac{1}{D}\sum_{m,n\neq m} |\lO^j_{mn}|^2\delta(\omega-\omega^{}_{mn})\,. 
\end{equation}
\vspace{-0.2cm}

\begin{figure}[!t]
    \centering
    \includegraphics[width=0.98\linewidth]{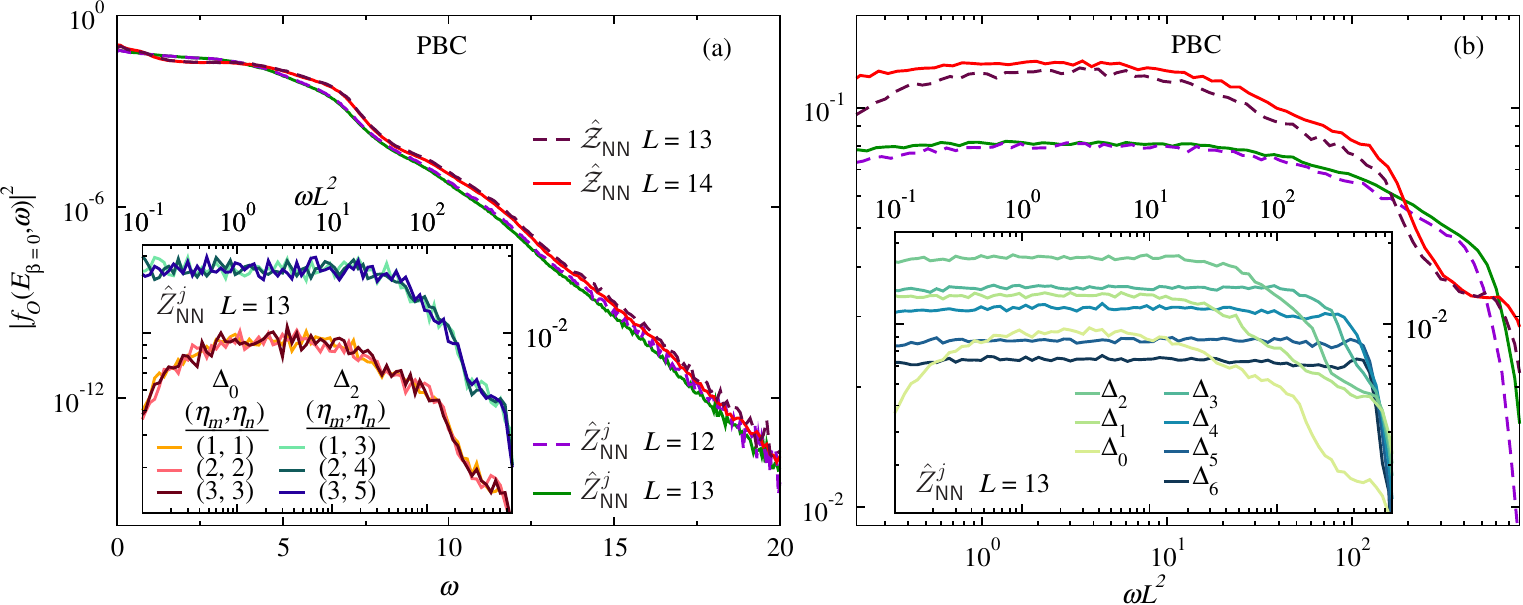}
    \caption{{\bf Spectral functions (PBCs):} Rescaled spectral functions $|f^{\mathrm{resc}}_O(\E,\omega)|^2=\exp\left[\omega^2/(4\sigma_E^2)\right]|f^{\mathrm{corr}}_O(\E,\omega)|^2$ in the (a) high- and (b) low-frequency regimes for the translationally invariant $\hat\tO=\hat \tZ^{}_{\Nt\Nt}$ [see Eq.~\eqref{eq:observables}] and its local counterpart $\hat\lO^j=\hat \lZ^{j}_{\Nt\Nt}$ in nonintegrable spin-$1$ $XXZ$ chains [$\lambda=0$ in Eq.~\eqref{eq:Spin1_Hamiltonian}] with PBCs. The inset in (b) shows the different $\Delta_\ell$ contributions (with $\ell=0,\ldots,6$ for $L=13$) to the rescaled function $|f^{\mathrm{resc}}_{\lO^{j}}(\E,\omega)|^2=\exp\left[\omega^2/(4\sigma_E^2)\right]|f^{\mathrm{corr}}_{\lO^j}(\E,\omega)|^2$ in Eq.~\eqref{eq:PBC-local-vs-trans-spectral-functions}. The inset in (a) shows the $(k_m,k_n)$-resolved contributions from the quasimomentum blocks $(k_m,k_n)=\tfrac{2\pi}{L}(\eta_m,\eta_n)$ with $(\eta_m,\eta_n)=(1,1),(2,2)$ and $(3,3)$ [$(\eta_m,\eta_n)=(1,3),(2,4)$ and $(3,5)$] to the $\Delta_0$ ($\Delta_2$) curve in the inset in (b) times $L$ ($2L$) to account for the number of $(k_m,k_n)$ blocks contributing to $\Delta_0$ ($\Delta_2$).}
    \label{fig:PBCs-loc-vs-trans-spectral-function}
\end{figure}

\noindent In Fig.~\ref{fig:PBCs-loc-vs-trans-spectral-function}(a), we show numerical results for the rescaled spectral functions $|f^{\mathrm{resc}}_\tO(\E,\omega)|^2$ and $|f^{\mathrm{resc}}_{\lO^j}(\E,\omega)|^2$ [Eq.~\eqref{eq:local-vs-trans-spectral-functions} rescaled by $\exp(\omega^2/(4\sigma_E^2))$] for two different chain sizes in each case. One can see that $|f^{\mathrm{resc}}_\tO(\E,\omega)|^2$ and $|f^{\mathrm{resc}}_{\lO^j}(\E,\omega)|^2$ are qualitatively and quantitatively similar, but they are not identical. Figure~\ref{fig:PBCs-loc-vs-trans-spectral-function}(b) highlights the differences at low frequency. To understand the origin of their difference we need to discuss the effect of the invariance of the Hamiltonian under lattice translations and of the invariance or lack thereof of the observables also under lattice translations.

As we mentioned in Sec.~\ref{sec:Model_and_observables}, the eigenstates of the Hamiltonian $\hat H$~\eqref{eq:Spin1_Hamiltonian} are simultaneous eigenstates of the (single-site) lattice translation operator $\hat{\mathcal{T}}$, i.e., $\hat H\ket{E_m,k_m}=E_m\ket{E_m,k_m}$ and $\hat{\mathcal{T}}\ket{E_m,k_m}=e^{-ik_m}\ket{E_m,k_m}$, with energy $E_m$ and quasimomentum $k_m=\tfrac{2\pi}{L}\eta_m$ (we set the lattice spacing $a=1$). Consequently, as we used in all our earlier numerical calculations in this chapter, for translationally invariant observables the matrix $\tO_{mn}\equiv\bra{E_m,k_m}\hat \tO\ket{E_n,k_n}$ has $L^2$ quasimomentum-resolved blocks of about the same size (with $\sim\frac{D^2}{L^2}$ matrix elements) defined by the values of $k_m$ and $k_n$, in which the matrix elements vanish when $k_m \neq k_n$. For local operators, on the other hand, they need not vanish when $k_m \neq k_n$. Using that $\hat \lO^l=(\hat{\mathcal{T}}^\dagger)^{(j-l)} \hat \lO^j(\hat{\mathcal{T}})^{(j-l)}$ in a periodic chain, the matrix elements of local operators in the energy eigenbasis of translationally invariant Hamiltonians are related via:
\vspace{-0.1cm}
\begin{equation}\label{eq:different-sites-phase-relation}
    \lO^l_{mn}=\lO^j_{mn}e^{i(j-l)(k_m-k_n)}\,.
\end{equation}
Equation~\eqref{eq:different-sites-phase-relation} implies that: (i) for $k_m=k_n$, $\lO^j_{mn}=\lO^l_{mn}\ \forall\, j,\, l$ so that $\tO_{mn}=\lO^j_{mn}\ \forall\, j$, and (ii) $\tO_{mn}=0$ for $k_m \neq k_n$.

Using these insights, we can rewrite the spectral functions in Eq.~\eqref{eq:local-vs-trans-spectral-functions} for energy eigenstates that are simultaneous eigenstates of the lattice translation operator by resolving their quasimomenta $k^{}_m$ and $k^{}_n$:
\vspace{-0.1cm}
\begin{equation}\label{eq:PBC-local-vs-trans-spectral-functions}
    |f^{\mathrm{corr}}_{\tO}(\E,\omega)|^2=\frac{L}{D}\underset{k^{}_m=k^{}_n}{\sum_{m,n\neq m}} |\tO^{}_{mn}|^2\delta(\omega-\omega^{}_{mn})\,, \qquad\text{and}\qquad |f^{\mathrm{corr}}_{\lO^j}(\E,\omega)|^2=\frac{1}{D}\sum_{\ell=0}^{\lfloor L/2\rfloor}\underset{k^{}_m=k^{}_n\pm\Delta_\ell}{\sum_{m,n\neq m}} |\lO^j_{mn}|^2\delta(\omega-\omega^{}_{mn})\,,
\end{equation}
\vspace{-0.2cm}

\noindent where the equalities between quasimomenta are understood modulo $2\pi$, with $k^{}_m,k^{}_n\in(-\pi,\pi]$, and $\Delta_\ell=\tfrac{2\pi}{L}\ell$ denotes the quasimomentum difference with $\ell=0,1,\ldots,\lfloor\tfrac{L}{2}\rfloor$. Note that $|f^{\mathrm{corr}}_{\lO^j}(\E,\omega)|^2$ in Eq.~\eqref{eq:PBC-local-vs-trans-spectral-functions} is independent of the site $j$ considered because $|\lO^l_{mn}|=|\lO^j_{mn}|\ \forall\, j,\, l$, see Eq.~\eqref{eq:different-sites-phase-relation}. Next we note that while $|f^{\mathrm{corr}}_{\tO}(\E,\omega)|^2$ in Eq.~\eqref{eq:PBC-local-vs-trans-spectral-functions} has contributions from matrix elements $\tO^{}_{mn}$ that only connect energy eigenstates with equal quasimomentum $k_m=k_n$, $|f^{\mathrm{corr}}_{\lO^j}(\E,\omega)|^2$ contains additional terms connecting different quasimomenta $k_m\neq k_n$. Furthermore, since $\tO_{mn}=\lO^j_{mn}$ within sectors with $k_m=k_n$, we realize that the $\Delta_{\ell=0}$ contribution to $|f^{\mathrm{corr}}_{\lO^j}(\E,\omega)|^2$ equals $\tfrac{1}{L}|f^{\mathrm{corr}}_{\tO}(\E,\omega)|^2$, which vanishes as $L\rightarrow \infty$. Therefore, unless the remaining terms with $\Delta_{\ell\neq0}$ add to give $L-1$ times the contribution from $\Delta_{\ell=0}$, $|f^{\mathrm{corr}}_{\lO^j}(\E,\omega)|^2$ and $|f^{\mathrm{corr}}_{\tO}(\E,\omega)|^2$ are expected to differ in general.

The inset in Fig.~\ref{fig:PBCs-loc-vs-trans-spectral-function}(b) shows the (rescaled) contributions from terms with a given $\Delta_{\ell=0,\ldots,6}$ to $|f^{\mathrm{corr}}_{\lO^j}(\E,\omega)|^2$ in Eq.~\eqref{eq:PBC-local-vs-trans-spectral-functions}. The $\Delta_{\ell=0}$ curve for $L=13$ corresponds to a $\tfrac{1}{L}$-rescaled version of $|f^{\mathrm{resc}}_{\tO}(\E,\omega)|^2$ in the main panel in Fig.~\ref{fig:PBCs-loc-vs-trans-spectral-function}(b). The curves corresponding to $\Delta_{\ell\neq0}$ contributions differ from the one for $\Delta_{\ell=0}$, resulting in different $|f^{\mathrm{resc}}_{\lO^{j}}(\E,\omega)|^2$ and $|f^{\mathrm{resc}}_{\tO}(\E,\omega)|^2$ as highlighted in the main panel in Fig.~\ref{fig:PBCs-loc-vs-trans-spectral-function}(b). Furthermore, the $\Delta_{\ell\neq 0}$ contributions exhibit a physically meaningful behavior, \ie increasing $\Delta_{\ell}$ corresponds to a shortening length scale $\propto 1/\Delta_\ell$, and the associated low-frequency plateaus start to emerge at higher frequencies $\omega$ (or shorter time scales). In other words, the contribution from the largest quasimomentum difference $\Delta_{\ell=\lfloor L/2\rfloor}$ (shortest length scale) relaxes in the shortest time, while the contribution from the smallest finite difference $\Delta_{\ell=1}$ (longest finite length scale) takes the longest to relax. The $\Delta_{\ell=0}$ contribution, which is the only contribution in $|f^{\mathrm{corr}}_{\tO}(\E,\omega)|^2$, does not have a length scale associated with it as $1/\Delta_{\ell=0}$ diverges. 

Motivated by the previous observations, in Ref.~\cite{patil_2026} an ETH ansatz was introduced for local observables in the presence of lattice-translational symmetry:
\vspace{-0.2cm}
\begin{equation}\label{eq:TranslationalETH}
    \braket{E^{}_m,k^{}_m|\hat \lO^j|E^{}_n,k^{}_n}= \lO^j(E^{}_m)\delta^{}_{mn}+\frac{1}{\sqrt{\Omega(\bar E^{}_{mn})}}f^{}_{\lO^j}(\bar E^{}_{mn},\omega^{}_{mn},\kappa^{}_{mn})R^{\lO^j}_{mn}\,,
\end{equation}
\vspace{-0.2cm}

\noindent where $\kappa^{}_{mn}=|k^{}_m-k^{}_n|$ denotes the shortest quasimomentum distance between $k^{}_m$ and $k^{}_n$ modulo $2\pi$, with $0 \leq \kappa^{}_{mn} \leq \pi$. Interestingly, unlike the ETH for continuous symmetries [\eg Eq.~\eqref{eq:gen-AbelianETH} for U(1)], since the lattice-translational symmetry is discrete, the conserved quasimomentum $k^{}_m$ is nonextensive and hence does not enter the microcanonical function $\lO^{j}(E^{}_m)$ encoding the thermodynamic prediction, nor does it affect the thermodynamic entropy $S(\bar E^{}_{mn})$, which (as $L$ increases) is the same in all quasimomentum sectors contributing to the density of states $\Omega(\bar E^{}_{mn})\propto \exp[S(\bar E^{}_{mn})]$. The total quasimomentum $k^{}_m$ only enters the function $f^{}_{\lO^j}(\bar E^{}_{mn},\omega^{}_{mn},\kappa^{}_{mn})$ through $\kappa^{}_{mn}$, implying a lack of finer structure at the level of individual $k^{}_m$. The inset in Fig.~\ref{fig:PBCs-loc-vs-trans-spectral-function}(a) shows the quasimomentum-resolved contributions to the $\Delta_{\ell=0}$ and $\Delta_{\ell=2}$ results [shown in the inset in Fig.~\ref{fig:PBCs-loc-vs-trans-spectral-function}(b)] from various quasimomentum blocks labeled by $(k_m,k_n)$. The collapse of the different curves supports the expectation that $\kappa^{}_{mn}=\Delta_{\ell}$ is the relevant quantity describing the different $\Delta_{\ell}$ results.

While the focus of this chapter has been on the ETH to understand quantum dynamics under unitary time evolution involving time-independent Hamiltonians, so that the total energy is conserved, we should mention that quantum dynamics have also been extensively studied for time-periodically driven or {\bf Floquet} lattice models, which for finite-dimensional Hilbert spaces generically heat up to featureless infinite-temperature states~\cite{dalessio_rigol_2014, lazarides_2014}. The random matrix theory connections also run deep in such systems~\cite{mark_24}, and one can formulate an equivalent of the ETH ansatz as follows. In a Floquet system, in which the unitary time-evolution operator after $q$ cycles can be written as $\hat U^{}_q=(\hat U^{}_T)^q$, the ETH for an observable $\hat O$ in the quasienergy eigenstates defined by the unitary evolution operator over one period, $\hat U^{}_T\ket{\varepsilon_m}=e^{-i\varepsilon_m}\ket{\varepsilon_m}$ (we set $T=1$) with $\varepsilon_m\in(-\pi,\pi]$, can be written as
\vspace{-0.1cm}
\begin{equation}
    \braket{\varepsilon^{}_m|\hat O|\varepsilon_n}=\bar O\,\delta_{mn}+\frac{1}{\sqrt{\Omega}} f^{}_O(\omega_{mn})\,R^{O}_{mn}\,,
\end{equation}
\vspace{-0.2cm}

\noindent where $\bar O=\Tr(\hat O)/D$ is the infinite-temperature average, $\Omega=D/(2\pi)$ is the density of states, $D=\Tr(\mathbbm{1})$ is the Hilbert space dimension, and $\omega_{mn}=\varepsilon_m-\varepsilon_n\ (\text{mod}\ 2\pi)$ with $\omega_{mn}\in(-\pi,\pi]$. Since the quasienergies $\varepsilon_m$ are nonextensive, $\bar O$ is structureless in the quasienergy spectrum, whose density of states is uniform in generic quantum-chaotic Floquet systems. In such systems, since the density of states is uniform, $|f^{\mathrm{var}}_{O}|^2$ obtained from the variance of the matrix elements and $|f^{\mathrm{corr}}_{O}|^2$ obtained from the autocorrelation function are identical~\cite{prosen_2021}.

\subsection{ETH and higher-order correlations}\label{sec:higher}

The symmetry considerations in Sec.~\ref{sec:ETHSymmetries} go beyond the formulation of the ETH as an extension of random matrix theory for quantum systems in the absence of symmetries~\cite{deutsch_1991, srednicki_1994, rigol_2008}. Another important direction in which the ETH~\eqref{eq:ETH_ansatz} has been extended is to account for correlations between the matrix elements of observables. Higher-order correlations among the matrix elements $O^{}_{mn}$~\cite{foini_kurchan_19, pappalardi_foini_kurchan_22} are needed to describe higher moments of observables, such as out-of-time-order correlators~\cite{brenes_pappalardi_21, wang_lamann_22, foini_kurchan_19, chan_deluca_19, murthy_srednicki_19, pappalardi_fritzsch_25}, as well as correlators involving different observables~\cite{dalessio_quantum_2016, foini_kurchan_19, pappalardi_foini_kurchan_22, noh_20}. These correlations have been connected to the topic of free probability~\cite{pappalardi_foini_kurchan_22, fava_25}. In what follows, we briefly discuss these developments and their consequences in the context of local vs average operator equivalents of translationally invariant observables in systems with open boundary conditions (OBCs).

In Sec.~\ref{sec:Chaos}, we discussed the structure of matrix elements $O^{}_{mn}$ of a Hermitian operator $\hat O$ expressed in the eigenstates of an RMT ensemble, where the rotational invariance of the ensemble enabled the calculation of the mean and the variance of $O^{}_{mn}$ [Eq.~\eqref{eq:RMT_Result_MatrixElements}] using the Weingarten calculus. The rotational invariance also leads to higher-order correlations (subleading in $1/D$) among the eigenstates given by the Weingarten formula [Eqs.~\eqref{eq:Weingarten_U(D)} and~\eqref{eq:Weingarten_O(D)}], which can be used to calculate higher moments of $O^{}_{mn}$ beyond the leading predictions in Eq.~\eqref{eq:RMT_Result_MatrixElements}. In a similar spirit, since the ETH is expected to reduce to the RMT results in a small enough energy window, Ref.~\cite{foini_kurchan_19} extended the ETH to incorporate higher-order correlations using a typicality argument, \ie using that the statistics of $O^{}_{mn}$ are invariant under local unitary rotations involving energy eigenstates in a small energy window. The extended ETH ansatz reads~\cite{foini_kurchan_19,pappalardi_foini_kurchan_22}:
\begin{align}
   \overline{O_{m_1m_2}^{}\ldots O_{m_{r-1}m_{r}}^{}O_{m_{r}m_{r+1}}^{} \dots O_{m_{\Nm}m_1}^{}}&=\left[\Omega(\bar E)\right]^{-(\Nm-1)}\,F_O^{(\Nm)}(\bar E^{},\omega^{}_{m_1m_2},\omega^{}_{m_2m_3},\ldots,\omega^{}_{m_{\Nm-1}m_\Nm})\,,\quad \text{for}\quad m_r\neq m_{r'}\, \forall\, r\neq r'\,,\label{eq:extend-ETH-ansatz-A}\\
    \overline{O_{m_1m_2}^{}\ldots O_{m_{r-1}m_{1}}^{}O_{m_{1}m_{r+1}}^{} \dots O_{m_{\Nm}m_1}^{}}&\simeq \overline{O_{m_1m_2}^{}\ldots O_{m_{r-1}m_{1}}^{}}\,\,\,\overline{O_{m_{1}m_{r+1}}^{} \dots O_{m_{\Nm}m_1}^{}}\,,\label{eq:extend-ETH-ansatz-B}
\end{align}
where $\overline{(\ldots)}$ represents an average of the matrix elements, $\bar E=(E^{}_{m_1}+\ldots +E^{}_{m_\Nm})/\Nm$ is the average energy, $\omega^{}_{m_rm_{r+1}}=E^{}_{m_{r}}-E^{}_{m_{r+1}}$ are the energy differences, and $F^{(\Nm)}_O$ is a smooth $\O(1)$ function at fixed $\omega^{}_{m_rm_{r+1}}=\O(1)$. The extended ETH ansatz in Eq.~\eqref{eq:extend-ETH-ansatz-A} associates the average of a product of $\Nm$ matrix elements with distinct indices $\{m_1,m_2,\ldots, m_\Nm\}$ to a smooth $\O(1)$ function $F_O^{(\Nm)}$ times $\Nm-1$ factors of $1/\Omega(\bar E)$ characterizing the decay corresponding to the density of states, $\Omega(\bar E)\propto\exp[S(\bar E)]$. In addition, Eq.~\eqref{eq:extend-ETH-ansatz-B} states that the average in the thermodynamic limit factorizes for repeated indices, \eg for $m_r=m_1$ in this case. Note that the ETH ansatz in Eq.~\eqref{eq:ETH_ansatz} is a special case of the extended ETH in Eq.~\eqref{eq:extend-ETH-ansatz-A} for $\Nm=1$ and $\Nm=2$, and can be recast as:
\begin{equation}
    \overline{O^{}_{mm}}=F^{(1)}_O(E^{}_m)\,, \qquad \overline{O^{}_{mn}O^{}_{nm}}=\frac{1}{\Omega(\bar E)}\,F^{(2)}_O(\bar E,\omega^{}_{mn})\quad\text{for}\quad m\neq n\,,
\end{equation}
where $F^{(1)}_O(E^{}_m)\equiv O(E^{}_m)$, $F^{(2)}_O(\bar E,\omega^{}_{mn})\equiv |f^{}_O(\bar E^{}_{mn},\omega^{}_{mn})|^2$ and $\bar E\equiv \bar E^{}_{mn}$. In addition to the higher-order correlations encoded in Eq.~\eqref{eq:extend-ETH-ansatz-A}, the extended ETH also includes correlations between matrix elements of two different observables $\hat A$ and $\hat B$~\cite{pappalardi_foini_kurchan_22}, advanced in Ref.~\cite{dalessio_quantum_2016}:
\begin{equation}\label{eq:ETH-cross-correlations}
    \overline{A^{}_{mn}B^{}_{nm}}=\frac{1}{\Omega(\bar E)}\,F^{(2)}_{A,B}(\bar E,\omega^{}_{mn})\quad\text{for}\quad m\neq n\,.
\end{equation}

We now discuss the consequences of the correlations in Eq.~\eqref{eq:ETH-cross-correlations} when comparing the predictions for the local operator $\hat \lO^j=\hat \lZ^{j}_{\Nt\Nt}$ in the bulk of the chain (\ie at site $j=\lfloor L/2\rfloor$) and the ``average'' operator $\hat \tO=\hat{\bar \tZ}^{}_{\Nt\Nt}$ counterpart of the translationally invariant $\hat \tZ^{}_{\Nt\Nt}$~\eqref{eq:observables},
\vspace{-0.1cm}
\begin{equation}\label{eq:TransInvObservableInOBC}
\hat \tO=\hat{\bar \tZ}^{}_{\Nt\Nt}\equiv\frac{1}{L-2}\sum_{j=1}^{L-2} \hat{S}^z_{\!j} \hat{S}^z_{\!j+2},    
\end{equation}
in nonintegrable spin-$1$ $XXZ$ chains with OBCs:
\vspace{-0.15cm}
\begin{equation}\label{eq:OBCs_Spin1_Hamiltonian}
    \hat H= -\sum_{j=1}^{L-1} \left(\hat{S}^x_{\!j} \hat{S}^x_{\!j+1} + \hat{S}^y_{\!j} \hat{S}^y_{\!j+1} +\Delta\hat{S}^z_{\!j} \hat{S}^z_{\!j+1}\right)+h^{(1)}_z\, \hat{S}^z_{\!1}\,,
\end{equation}
\vspace{-0.2cm}

\noindent where $\Delta=0.55$ and a $z$-field of strength $h^{(1)}_z=0.1$ is added to break the space-reflection and spin-inversion symmetries of the model, and we focus on the sector with zero total magnetization ($M=0$).

\begin{figure}[t]
    \centering
    \includegraphics[width=0.98\linewidth]{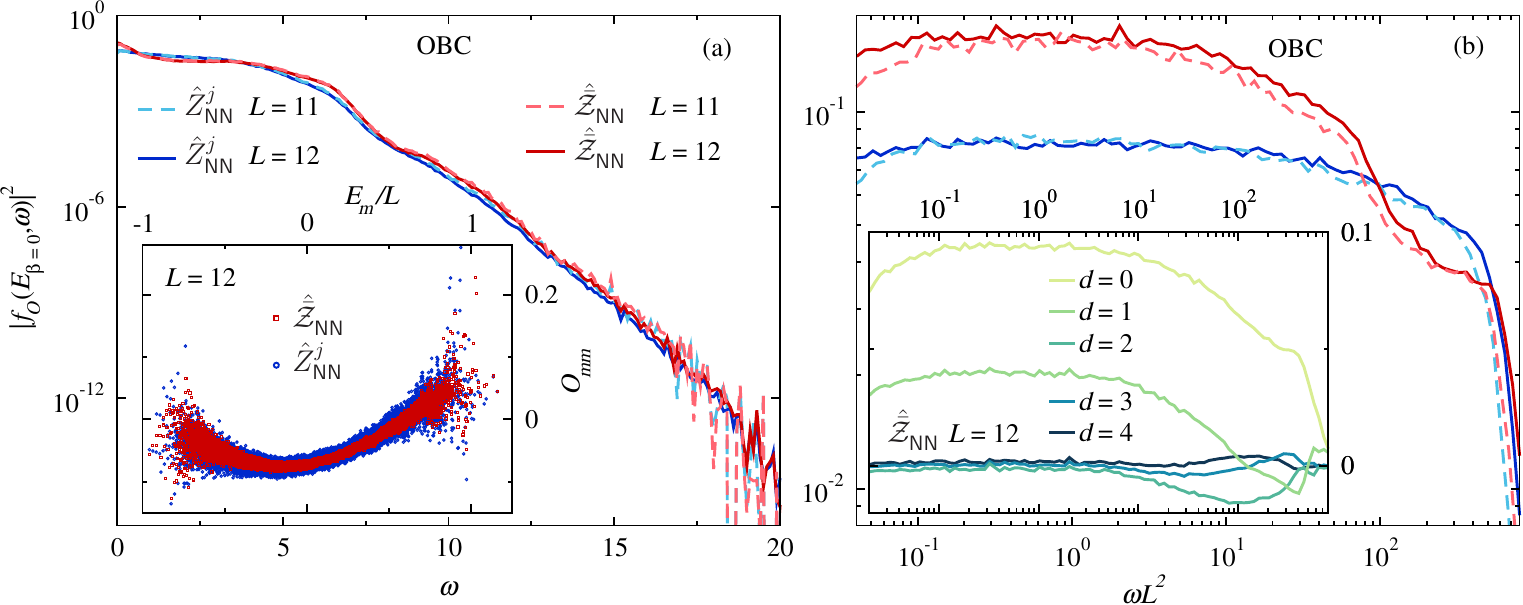}
    \caption{{\bf Spectral functions (OBCs)}: Rescaled spectral functions $|f^{\mathrm{resc}}_O(\E,\omega)|^2= \exp\left[\omega^2/(4\sigma_E^2)\right]|f^{\mathrm{corr}}_O(\E,\omega)|^2$ in the (a) high- and (b) low-frequency regimes for the average operator $\hat \tO=\hat{\bar \tZ}^{}_{\Nt\Nt}$ and its local counterpart $\hat \lO^j=\hat \lZ^{j}_{\Nt\Nt}$ (at site $j=\lfloor L/2\rfloor$) in the nonintegrable spin-$1$ $XXZ$ chains with OBCs; see Eq.~\eqref{eq:OBCs_Spin1_Hamiltonian}. The inset in (a) shows the diagonal matrix elements $O^{}_{mm}$ vs the energy density $E^{}_m/L$ for $\hat \lO^{j}$ and $\hat{\tO}$ in the chain with $L=12$. The inset in (b) shows the contributions to $|f^{\mathrm{resc}}_{\tO}(\E,\omega)|^2$ in Eq.~\eqref{eq:local-vs-trans-spectral-functions} (with $L\rightarrow L-2$) from terms with different values of $d=0,\ldots,4$ in Eq.~\eqref{eq:trans-inv-observable-matrix-elements-decompostion} in the chain with $L=12$.} 
    \label{fig:OBCs-loc-vs-trans-spectral-function}
\end{figure}

In the inset of Fig.~\ref{fig:OBCs-loc-vs-trans-spectral-function}(a), we plot the diagonal matrix elements of $\hat{\tO}^{}$ and $\hat \lO^{j}$ in a chain with $L=12$ and OBCs. Even though the two operators exhibit different eigenstate-to-eigenstate fluctuations in the finite chain considered, the fluctuations decay exponentially fast with increasing system size for both of them (not shown). Furthermore, in the thermodynamic limit, the two operators provide identical predictions for the expectation values of the next-nearest-neighbor $\hat{S}^z$-$\hat{S}^z$ correlations in the bulk of the system in thermal equilibrium~\cite{patil_2026, rigol_shastry_08, iyer_15}. This need not be the case for dynamical quantities, as we discuss next.

In the main panels of Figs.~\ref{fig:OBCs-loc-vs-trans-spectral-function}(a) and~\ref{fig:OBCs-loc-vs-trans-spectral-function}(b), we plot the rescaled spectral functions $|f^{\mathrm{resc}}_{\tO}(\E,\omega)|^2$ and $|f^{\mathrm{resc}}_{\lO^{j}}(\E,\omega)|^2$ corresponding to $|f^{\mathrm{corr}}_{\tO}(\E,\omega)|^2$ and $|f^{\mathrm{corr}}_{\lO^{j}}(\E,\omega)|^2$ in Eq.~\eqref{eq:local-vs-trans-spectral-functions}\footnote{For $\hat{\bar \tZ}^{}_{\Nt\Nt}$ in Eq.~\eqref{eq:TransInvObservableInOBC}, we replace the prefactor $L\rightarrow L-2$ for $|f^{\mathrm{corr}}_{\tO}(\E,\omega)|^2$ in Eq.~\eqref{eq:local-vs-trans-spectral-functions} to account for the appropriate Hilbert-Schmidt norm.} rescaled by $\exp[\omega^2/(4\sigma_E^2)]$, in the high- and low-frequency regimes, respectively, for the two operators $\hat \tO=\hat{\bar \tZ}^{}_{\Nt\Nt}$ and $\hat\lO^j=\hat \lZ^{j}_{\Nt\Nt}$. The spectral functions $|f^{\mathrm{resc}}_{\tO}(\E,\omega)|^2$ and $|f^{\mathrm{resc}}_{\lO^j}(\E,\omega)|^2$ exhibit the same differences we observed in chains with PBCs in Fig.~\ref{fig:PBCs-loc-vs-trans-spectral-function}. As a matter of fact, the results for each observable are very close in chains with PBCs and OBCs, as also observed in spin-$1$ tilted-field Ising chains in Ref.~\cite{patil_2026}. To understand the origin of the differences in the chains with OBCs, the $|\tO^{}_{mn}|^2$ in $|f^{\mathrm{corr}}_{\tO}(\E,\omega)|^2$~\eqref{eq:local-vs-trans-spectral-functions} can be expanded in terms of matrix elements $\lO^j_{mn}$ and $\lO^l_{mn}$ of the local operators at different sites as
\begin{equation}\label{eq:trans-inv-observable-matrix-elements-decompostion}
    |\tO^{}_{mn}|^2=\frac{1}{(L-2)^2}\Bigg(\sum_{d=0}^{L-3}\underset{|j-l|=d}{\sum_{j,l=1}^{L-2}}\lO^j_{mn}(\lO^l_{mn})^*\Bigg)\,,
\end{equation}
\vspace{-0.2cm}

\noindent with $d=0,1,\ldots,L-3$ being the distance between local operators at sites $j,l$. If the random matrix elements $R^{\lO^j}_{mn}$ and $R^{\lO^l}_{mn}$ appearing in the ETH ansatz~\eqref{eq:ETH_ansatz} for the local operators at sites $j\neq l$ are assumed to be uncorrelated, then the corresponding average of the product of matrix elements $\overline{\lO^j_{mn}(\lO^l_{mn})^*}$ vanishes for $d=|j-l|\neq0$. In addition, if $\overline{|\lO^j_{mn}|^2}\approx \overline{|\lO^l_{mn}|^2}$ for all $j,l$ in the bulk of the system, then $\overline{|\tO^{}_{mn}|^2}\approx \overline{|\lO^j_{mn}|^2}/(L-2)$, which from Eq.~\eqref{eq:local-vs-trans-spectral-functions} implies that $|f^{\mathrm{corr}}_{\tO}(\E,\omega)|^2\approx|f^{\mathrm{corr}}_{\lO^j}(\E,\omega)|^2$. Therefore, deviations between the two spectral functions point towards the existence of correlations between $\lO^{j}_{mn}$ at different sites in chains with OBCs, as advanced by Eq.~\eqref{eq:ETH-cross-correlations}. The inset in Fig.~\ref{fig:OBCs-loc-vs-trans-spectral-function}(b), which shows the contribution to $|f^{\mathrm{resc}}_{\tO}(\E,\omega)|^2$ in Eq.~\eqref{eq:local-vs-trans-spectral-functions} from terms with different values of $d$ in Eq.~\eqref{eq:trans-inv-observable-matrix-elements-decompostion}, makes it apparent that correlations exist between the matrix elements of observables at different lattice sites, as the terms with $d\neq0$ are nonzero. 

The findings discussed in this section connect a symmetry (lattice translations) with correlations between matrix elements of observables in the absence of that symmetry. Understanding the broader origin and implications of this connection, and whether similar relations arise for other discrete symmetries, remains an interesting direction for future work. Other important directions include understanding how quantum chaos and eigenstate thermalization emerge as a system is driven away from integrability~\cite{santos_rigol_10a, abdelshafy_25, rigol_09a, pandey_claeys_20, leblond_sels_21} or many-body localization~\cite{suntajs_bonca_20, sierant_25, zakrzewski2026}, as well as exploring novel paradigms such as Hilbert-space fragmentation and quantum many-body scars~\cite{Moudgalya_2022}.

\begin{ack}[Acknowledgments]

This work was supported by the National Science Foundation (NSF) Grants No.~PHY-2309146 and~PHY-2607767. The computations were done in the Institute for Computational and Data Sciences Roar supercomputer at Penn State. 
\end{ack}

% \seealso{article title article title}

\bibliographystyle{JHEP}%
\bibliography{references}

\end{document}